\input harvmac
\input epsf
\noblackbox


\overfullrule=0pt

\def\cC{{\cal C}}
\def\IP{{\bf P}}

\def\bfone{\relax{\rm 1\kern-.35em 1}}
\def\inbar{\vrule height1.5ex width.4pt depth0pt}
\def\IC{\relax\,\hbox{$\inbar\kern-.3em{\mss C}$}}
\def\ID{\relax{\rm I\kern-.18em D}}
\def\IF{\relax{\rm I\kern-.18em F}}
\def\IH{\relax{\rm I\kern-.18em H}}
\def\II{\relax{\rm I\kern-.17em I}}
\def\IN{\relax{\rm I\kern-.18em N}}
\def\IQ{\relax\,\hbox{$\inbar\kern-.3em{\rm Q}$}}
\def\us#1{\underline{#1}}
\def\IR{\relax{\rm I\kern-.18em R}}
\font\cmss=cmss10 \font\cmsss=cmss10 at 7pt
\def\ZZ{\relax\ifmmode\mathchoice
{\hbox{\cmss Z\kern-.4em Z}}{\hbox{\cmss Z\kern-.4em Z}}
{\lower.9pt\hbox{\cmsss Z\kern-.4em Z}}
{\lower1.2pt\hbox{\cmsss Z\kern-.4em Z}}\else{\cmss Z\kern-.4em
Z}\fi}

\def\cC{{\cal C}}

\def\cJ{{\cal J}} 
 \def\cM{{\cal M}}
 \def\cO{{\cal O}}

\def\gg{{r}}
\def\ag{{g}}

\def\nup#1({Nucl.\ Phys.\ $\us {B#1}$\ (}
\def\plt#1({Phys.\ Lett.\ $\us  {B#1}$\ (}
\def\cmp#1({Comm.\ Math.\ Phys.\ $\us  {#1}$\ (}
\def\prp#1({Phys.\ Rep.\ $\us  {#1}$\ (}
\def\prl#1({Phys.\ Rev.\ Lett.\ $\us  {#1}$\ (}
\def\prv#1({Phys.\ Rev.\ $\us  {#1}$\ (}
\def\mpl#1({Mod.\ Phys.\ Let.\ $\us  {A#1}$\ (}
\def\ijmp#1({Int.\ J.\ Mod.\ Phys.\ $\us{A#1}$\ (}
\def\tit#1|{{\it #1},\ }

\def\Coe#1.#2.{{#1\over #2}}

\def\coe#1.#2.{\relax{\textstyle {#1 \over #2}}\displaystyle}
\def\half{{1 \over 2}}

\def\dd{\rm d}
\def\MS#1#2{ {{\bar {\cal M}}_{#1,#2}} }
\def\frac#1#2{{#1\over #2}}
\def\hc#1{\cC^{(#1)}}

\lref\ghovaf{D.~Ghoshal and C.~Vafa,
``c=1 String as the Topological Theory of the Conifold,''
Nucl. Phys. $\us {B453}$ (1995) 121.}

\lref\quot{I. Ciocan-Fontanine and M. M. Kapranov, ``Derived Quot Schemes,''
math.AG/9905174.}

\lref\kthy{E. Witten, ``D-Branes and K-Theory,'' hep-th/9810188.}

\lref\marinomoore{M.~Marino and G.~Moore,
``Counting Higher Genus Curves in a Calabi-Yau Manifold,''
Nucl. Phys. $\us {B543}$ (1999) 592--614, hep-th/9808131.}

\lref\hst{S. Hosono, M.-H. Saito, and A. Takahashi, ``Holomorphic
Anomaly Equation and BPS State Counting of Rational Elliptic
Surface,'' hep-th/9901151.}

\lref\swrefs{
W. Lerche, 
``Introduction to Seiberg-Witten Theory and its Stringy Origin,'' 
Nucl. Phys. Proc. Suppl. {\bf 55B} (1997) 83-117, hep-th/9611190; 
A. Klemm, ``On the Geometry behind N=2 Supersymmetric Effective Actions in 
Four-Dimensions'', Proceedings of the Trieste Summer School on 
High Energy Physics and Cosmology {\bf 96}, World Scientic, Singapore (1997) 
120-242, 
hep-th/9705131; P. Mayr, ``Geometric Construction of N=2 Gauge Theories,'' 
Fortsch. Phys. {\bf 47} (1999) 39-63, hep-th/9807096.}

\lref\kappriv{M. M. Kapranov, private communication.}

\lref\aspdon{P. S. Aspinwall and R. Y. Donagi, ``The Heterotic
String, the Tangent Bundle, and Derived Categories,''
Adv. Theor. Math. Phys. {\bf 2} (1998) 1041-1074;
hep-th/9806094.}

\lref\earlymulticover{P. Aspinwall and D. Morrison, ``Topological Field
Theory and Rational Curves,'' \cmp51(1993)245, 
Yu. I. Manin, ``Generating Functions in Algebraic Geometry and Sums 
over Trees'' in ``The moduli space of curves''
(Texel Island, 1994), Progr. Math. $\us{129}$, 
Birkh\"auser, Boston (1995) 401--417; alg-geom/9407005, 
C.\ Voisin, {\sl A Mathematical Proof of a formula of Aspinwall 
and Morrison}, Comp. Math. $\us{104}$ (1996) 135 .}

\lref\vafaoog{H. Ooguri and C. Vafa, ``Summing up Dirichlet
Instantons,''
Phys. Rev. Lett. {\bf 77} (1996) 3296-3298.}

\lref\cfkm{P. Candelas, A. Font, S. Katz, and D. R. Morrison
    ``Mirror Symmetry for Two Parameter Models -- II,''
      Nucl. Phys. {\bf B429} (1994) 626-674.}

\lref\cdfkm{P. Candelas, X. de la Ossa, A. Font, S. Katz, and D. R.
Morrison
    ``Mirror Symmetry for Two Parameter Models -- I,''
      Nucl.Phys. {\bf B416} (1994) 481-538.}

\lref\morvaf{D. R. Morrison and C. Vafa, ``Compactifications
of F-Theory on Calabi-Yau threefolds -- II,''
Nucl. Phys. {\bf B476} (1996) 437-469, hep-th/9603161.}

\lref\vain{I. Vainsencher, ``Enumeration of n-fold Tangent
Hyperplanes to a Surface,'' J. Algebraic Geom. $\us{4}$,
1995, 503-526, math.AG/9312012.}

\lref\fultoni{W. Fulton, ``{\it Intersection Theory},'' Second edition,
Springer-Verlag (Berlin) 1998.}

\lref\faber{C. Faber, ``Algorithm for Computing Intersection Numbers
on Moduli Spaces of Curves, with an Application to the Class of
the Locus of the Jacobians,'' math.AG/9706006.}

\lref\penglu{P. Lu, ``Special Lagrangian Tori on a
Borcea-Voisin Threefold,'' math.DG/9902063.}

\lref\vafagop{R.\ Gopakumar and C.\ Vafa, ``M-theory and
Topological Strings-I \& II,'' hep-th/9809187, hep-th/9812127.}

\lref\litian{J. Li and G. Tian, ``Virtual Moduli Cycles
and Gromov-Witten Invariants of Algebraic Varieties,''
J. Amer. Math. Soc. {\bf 11} (1998), no. 1, 119-174.}

\lref\pandh{R. Pandharipande, ``Hodge Integrals and
Degenerate Contributions,'' math.AG/9811140.}

\lref\behfan{K. Behrend and B. Fantechi, ``The Intrinsic
Normal Cone,'' math.AG/9601010.}

\lref\locvir{T. Graber and R. Pandharipande, ``Localization
of Virtual Classes,'' math.AG/9708001.}

\lref\fp{C.~Faber and R.~Pandharipande, 
{\sl Hodge Integrals and Gromov-Witten Theory}, math.AG/9810173.} 

\lref\kontloc{M. Kontsevich, ``Enumeration of Rational Curves
via Torus Actions,'' in {\sl The Moduli Space of Curves,}
Dijkgraaf et al eds., Progress in Mathematics {\bf 129},
Birkh\"auser (Boston) 1995.}

\lref\kontdgrav{M. Kontsevich, ``Intersection theory on the moduli space of curves 
and the matrix Airy function'', Comm. Math. Phys. $\us{147}$ (1992) 1-23 .}

\lref\mm{M. Mari\~no and G.\ Moore, 
``Counting higher genus curves in a Calabi-Yau Manifold,'' 
Nucl. Phys. $\us {B543}$ (1999) 592, hep-th/9808131. }

\lref\gasd{ C. Vafa, ``Gas of D-Branes and Hagedorn Density of BPS
States,''
Nucl. Phys. {\bf B463} (1996) 415-419; hep-th/9511088.}

\lref\bsv{ M. Bershadsky, V. Sadov, and C. Vafa, ``D-Branes and
Topological
Field Theories,'' Nucl. Phys. {\bf B463} (1996) 420-434;
hep-th/9511222.}

\lref\yz{S.-T.~Yau and E.~Zaslow, ``BPS States, String Duality, and
Nodal Curves on K3,'' Nucl. Phys. $\us{B471}$ (1996) 503-512; hep-th/9512121.}

\lref\yztwo{ S.-T. Yau
and E. Zaslow, ``BPS States as Symplectic Invariants from String
Theory,''  in {\sl Geometry and Physics,} Proceedings of the
Special Session on Geometry and Physics, Aarhus, Denmark, 1996.}

\lref\cdgp{ P. Candelas, X. C. De La Ossa, P. Green, and L Parkes,
``A Pair of C-Y Manifolds as an Exactly Soluble Superconformal
Theory,'' Nucl. Phys. $\us {B359}$ (1991) 21--74.}

\lref\mcl{R. McLean, ``Deformations of Calibrated Submanifolds,"
Duke preprint 96-01:  www.math.duke.edu/preprints/1996.html.}

\lref\hl{F. R. Harvey and H. B. Lawson, ``Calibrated Geometries,''
Acta Math. {\bf 148} (1982) 47;
F. R. Harvey, {\sl Spinors and Calibrations,} Academic Press,
New York, 1990.}

\lref\ooguri{ K. Becker, M. Becker, D. R. Morrison, H. Ooguri,
Y. Oz, and Z. Yin, ``Supersymmetric Cycles in Exceptional Holonomy
Manifolds and Calabi-Yau Fourfolds,'' Nucl. Phys. {\bf B480} (1996)
225-238.}

\lref\vafmir{ C. Vafa, ``Extending Mirror Conjecture to Calabi-Yau
with Bundles,'' hep-th/9804131.}

\lref\kach{ S. Kachru, A. Klemm, W. Lerche, P. Mayr, C. Vafa,
``Nonperturbative Results on the Point Particle Limit of N=2 Heterotic
String,''
Nucl. Phys. {\bf B459} (1996) 537.}

\lref\kkv{ S. Katz, A. Klemm, and C. Vafa,
``Geometric Engineering of Quantum Field Theories,''
Nucl. Phys. ${\us B497}$ (1997) 173-195.}

\lref\wittdgrav{E.~Witten,
``Two-dimensional Gravity and Intersection Theory on Moduli Space,''
Surveys in Differential Geometry $\us{1}$ (1991) 243-310.}

\lref\lly{ B. Lian, K. Liu, and S.-T. Yau, ``Mirror Principle I,''
Asian J. of Math. Vol. {\bf 1} No. 4 (1997) 729-763; math.AG/9712011.}

\lref\llyt{ B. Lian, K. Liu, and S.-T. Yau, ``Mirror Principle II,''
in preparation.}

\lref\givental{ A. Givental, ``A Mirror Theorem for Toric Complete
Intersections,''
{\it Topological Field Theory, Primitive Forms and Related Topics
(Kyoto,
1996)}, Prog. Math. {\bf 160}, 141-175; math.AG/9701016.}

\lref\syz{ A. Strominger,
S.-T. Yau, and E. Zaslow, ``Mirror Symmetry is T-Duality,''
Nuclear Physics {\bf B479} (1996) 243-259; hep-th/9606040.}

\lref\slag{ D. Morrison, ``The Geometry Underlying Mirror Symmetry,''
math.AG/9608006;
M. Gross and P. Wilson, ``Mirror Symmetry via 3-tori for a Class of
Calabi-Yau
Threefolds,'' to appear in Math. Ann., math.AG/9608009;  B. Acharya, ``A

Mirror Pair of Calabi-Yau Fourfolds in Type II String Theory,'' Nucl.
Phys.
{\bf B524} (1998) 283-294, hep-th/9703029;
N. C. Leung and C. Vafa, ``Branes and Toric Geometry,'' Adv. Theor.
Math.
Phys. {\bf 2} (1998) 91-118, hep-th/9711013;
N. Hitchin, ``The Moduli Space of Special Lagrangian Submanifolds,''
math.DG/9711002.}

\lref\kontsevich{ M. Kontsevich, ``Homological Algebra of Mirror
Symmetry,''
Proceedings of the 1994 International Congress of Mathematicians {\bf
I},
Birk\"auser, Z\"urich, 1995, p. 120;
math.AG/9411018.}

\lref\kkv{ S. Katz, A. Klemm, and C. Vafa, ``Geometric Engineering of
Quantum Field Theories,'' Nucl. Phys. $\us{B497}$ (1997) 173-195,
hep-th/9609239.}

\lref\guzz{ D. Guzzetti, ``Stokes Matrices and Monodromy for the
Quantum Cohomology of Projective Spaces,'' preprint SISSA 87/98/FM.}

\lref\dub{ B. Dubrovin,
{\sl Geometry of 2D Topological Field Theories,}
Lecture Notes in Math {\bf 1620} (1996) 120-348.}

\lref\ttstar{ S. Cecotti and C. Vafa, ``Topological Anti-Topological
Fusion,'' Nucl. Phys. {\bf B367} (1991) 359-461.}

\lref\class{ S. Cecotti and C. Vafa, ``On Classification of
N=2 Supersymmetric Theories,'' Commun. Math. Phys. {\bf 158} (1993)
569-644.}

\lref\dubconj{ ``Geometry and Analytic Theory of Frobenius
Manifolds,'' math.AG/9807034.}

\lref\ad{ P. Aspinwall and R. Dongagi,
``The Heterotic String, the Tangent Bundle,
and Derived Categories,'' hep-th/9806094.}

\lref\batyrev{V.\ Batyrev, ``Dual Polyhedra and Mirror Symmetry for
Calabi-Yau Hypersurfaces
in Toric Varieties,'' J. Algebraic Geom. {\bf 3} (1994) 493-535.}

\lref\batyrevII{V.\ Batyrev, ``Variations of the Mixed Hodge Structure
of Affine Hypersurfaces in Algebraic Tori,''
Duke Math. Jour. {\bf 69}, 2 (1993) 349}

\lref\hkty{ S.\  Hosono, A.\  Klemm, S.\ Theisen and S.T. Yau,
``Mirror Symmetry, Mirror Map and Applications to Calabi-Yau
Hypersurfaces,''
\cmp167(1995)301--350, hep-th/9308122;
and ``Mirror Symmetry, Mirror Map and Applications to Calabi-Yau
Hypersurfaces,'' Nucl. Phys. $\us{B433}$ (1995) 501-554, hep-th/9406055.}

\lref\cox{D.\ Cox, ``The Homogeneous Coordinate Ring of a Toric
Variety,''
J.\ Alg. Geom $\us{4}$ (1995) 17, math.AG/9206011.}

\lref\danilov{V. I. Danilov, {\sl The Geometry of Toric Varieties,}
Russian Math. Surveys, {\bf  33} (1978) 97.}

\lref\oda{T.\ Oda,  {\sl Convex Bodies and Algebraic Geometry, An
Introduction to the
Theory of Toric Varieties,} Ergennisse der Mathematik und ihrer
Grenzgebiete, 3. Folge, Bd. {\bf 15},
Springer-Verlag (Berlin) 1988.}

\lref\fulton{W.\ Fulton, {\sl Introduction to Toric Varieties,}
Princeton Univ. Press {\bf 131} (Princeton) 1993}

\lref\batyrevborisov{V.\ Batyrev and L.\ Borisov, ``On Calabi-Yau
Complete Intersections in
Toric Varieties,'' in {\sl Higher-dimensional Complex Varieties,}
(Trento, 1994), 39-65, de Gruyter  (Berlin) 1996.}

\lref\secondaryfan{I. M. Gel'fand, M. Kapranov, A. Zelevinsky, {\sl
Multidimensional Determinants,
Discriminants and Resultants}, Birkh\"auser (Boston) 1994;
L. Billera, J. Filiman and B.\ Sturmfels, ``
Constructions and Complexity of Secondary Polytopes,'' Adv. Math. {\bf
83}
(1990),  155-17.}

\lref\kmv{A.~Klemm, P.~Mayr, and C.~Vafa, ``BPS States of Exceptional
Non-Critical Strings,'' Nucl. Phys. $\us{B58}$
(Proc. Suppl.) (1997) 177-194; hep-th/9607139.}

\lref\ganor{O.\ Ganor,  ``A Test Of The Chiral E8 Current Algebra 
               On A 6D Non-Critical String'' Nucl.Phys. {\bf B479} (1996) 197-217, 
               hep-th/607020}

\lref\morrison{D.\ Morrison,
``Where is the large Radius Limit?'' Int. Conf. on Strings 93,
Berkeley; hep-th/9311049}
\lref\griffith{P. Griffiths, ``On the Periods of certain
Rational Integrals,'' Ann. Math. {\bf 90} (1969) 460.}

\lref\linearsigmamodel{E. Witten, ``Phases of $N=2$ Theories in Two
Dimensions,'' Nucl. Phys. {\bf B403} (1993) 159.}

\lref\lv{ N. C. Leung and C. Vafa, ``Branes and Toric Geometry,'' Adv.
Theor.Math.
Phys. {\bf 2} (1998) 91-118, hep-th/9711013}

\lref\ln{A. Lawrence and N. Nekrasov, ``Instanton sums and
five-dimensional theories,''
Nucl. Phys. $\us{B513}$ (1998) 93.}

\lref\ckyz{T.-M. Chiang, A. Klemm, S.-T. Yau, and E. Zaslow,
``Nonlocal Mirror Symmetry:  Calculations and Interpretations,''
hep-th/9903053.}

\lref\bcovI{M. Bershadsky, S. Cecotti, H. Ooguri, and C. Vafa,
``Holomorphic Anomalies in Topological Field Theories,''
Nucl. Phys. $\us{B405}$ (1993) 279.}

\lref\bcovII{M. Bershadsky, S. Cecotti, H. Ooguri, and C. Vafa,
``Kodaira-Spencer Theory of Gravity and Exact
Results for Quantum String Amplitudes,''
Commun. Math. Phys. {\bf 165} (1994) 311-428.}

\lref\kz{A.\ Klemm and E.\ Zaslow, ``Local Mirror Symmetry at Higher Genus'', 
hep-th/9906046.}  

\lref\ck{D.A. Cox and S. Katz, 
{\sl Mirror Symmetry and Algebraic Geometry},
Mathematical Surveys and Monographs vol.\ $\us {68}$, Amer.\ Math.\ Soc., Providence,
RI 1999.}   

\lref\klm{A.\ Klemm, P.\ Mayr and W.\ Lerche, hep-th/9506122,\plt357(1995)313}

\lref\gms{Ori J. Ganor, David R. Morrison, Nathan Seiberg, Nucl.Phys. B487 (1997) 
93-127}

\lref\lmw{W. Lerche, P. Mayr, N.P. Warner, Nucl.Phys. $\us{B499}$ (1997) 125--148.}

\lref\mnw{J.A.~Minahan, D.~Nemechansky and N.P.~ Warner, 
``Partition function for the BPS States of the Non-Critical $E_8$ String,'' 
Adv. Theor. Math.Phys. $\us { 1}$ (1998) 167-183; hep-th/9707149, 
J.A.~Minahan, D.~Nemechansky, C.~Vafa and N.P.~Warner,
``$E$-Strings and $N=4$ Topological Yang-Mills Theories,''
Nucl. Phys. $\us {B527}$ (1998) 581-623.}

\lref\mirrorproofs{ 
M. Kontsevich, ``Enumeration of Rational Curves
via Torus Actions,'' in {\sl The Moduli Space of Curves,}
Dijkgraaf et al eds., Progress in Mathematics $\us {129}$,
Birkh\"auser (Boston) 1995,
A. Givental, ``A Mirror Theorem for Toric Complete Intersections,''
{\it Topological Field Theory, Primitive Forms and Related Topics
(Kyoto,1996)}, Prog. Math. $\us{160}$, 141-175; math.AG/9701016, 
B. Lian, K. Liu, and S.-T. Yau, ``Mirror Principle I\& II,''
Asian J. of Math. Vol. $\us { 1}$  (1997) 729-763; 
math.AG/9712011 \& math.AG/9905006. }

\lref\ckyz{
T.-M. Chiang, A. Klemm, S.-T. Yau, and E. Zaslow,
``Local Mirror Symmetry:  Calculations and Interpretations,''
hep-th/9903053.}

%
\Title{}{\vbox{\centerline{M-Theory, Topological Strings and}
\vskip 0.2in
\centerline{Spinning Black Holes}}}
\vskip 0.2in

\centerline{Sheldon Katz\footnote{}{
email: katz@math.okstate.edu, klemm@ias.edu, vafa@string.harvard.edu }, 
Albrecht Klemm and Cumrun Vafa}
\vskip 0.1in
\centerline{\it Department of Mathematics, Oklahoma State University,
Stillwater, OK 74078, USA}
\centerline{\it Institute for Advanced Study, 
Princeton, NJ 08540, USA}
\centerline{\it Jefferson Laboratory of Physics, Harvard University, 
Cambridge, MA 02138, USA}

\vskip 0.3 in

\centerline{\bf Abstract}
We consider M-theory compactification on Calabi-Yau
threefolds.  The recently
discovered connection between the BPS states of wrapped M2 branes
and the topological string amplitudes on the threefold is used both as
a tool to compute topological string amplitudes at higher
genera as well as to unravel the degeneracies and quantum
numbers of BPS states.  Moduli spaces of $k$-fold symmetric
products of the wrapped M2 brane play a crucial role.  We
also show that the topological string partition function is the
Calabi-Yau version of the elliptic genus of the symmetric
product of $K3$'s and use the macroscopic
entropy of spinning black holes in 5 dimensions to obtain new
predictions for the
asymptotic growth of the topological string amplitudes at high genera.

\Date{HUTP-99/A056, IASSNS-HEP-98/107, OSU-M-99-9}

\vskip 0.1in

\centerline{Table of Contents}
$$\eqalign{
&\hbox{\bf 1.  Introduction}\cr
&\hbox{\bf 2.  Topological Strings (A-model)}\cr
&\qquad\hbox{\sl 2.1  The genus 0 contribution }\cr
&\qquad\hbox{\sl 2.2  The genus 1 contribution}\cr
&\qquad\hbox{\sl 2.3  The constant map contribution}\cr
&\hbox{\bf 3.  M-theory/Type IIA interpretation of the $F_\gg$}\cr
&\qquad\hbox{\sl 3.1  The new invariants}\cr
&\qquad\hbox{\sl 3.2  The higher genus contributions} \cr
&\hbox{\bf 4.  Computation of $n^\gg_d$} \cr
&\qquad\hbox{\sl 4.1 Computational scheme for $n^\gg_d$}\cr
&\hbox{\bf 5.  Considerations of Enumerative Geometry}\cr
&\qquad\hbox{\sl 5.1 Alternative interpretation of the $n^\gg_d$} \cr
&\hbox{\bf 6.  Application to counting $M2$ branes in $K3\times T^2$}\cr
&\qquad\hbox{\sl 6.1  Zero winding on the $T^2$} \cr
&\qquad\hbox{\sl 6.2  More general $H_2$ classes in $K3\times T^2$} \cr
&\hbox{\bf 7. Black Hole Entropy and Topological String}\cr
&\hbox{\bf 8. Computations in local Calabi-Yau geometries }\cr
&\qquad\hbox{\sl 8.1  Basic concepts}\cr
&\qquad\hbox{\sl 8.2  ${\cal O}(-1)\oplus {\cal O}(-1)\rightarrow {\bf 
P}^1$ }\cr
&\qquad\hbox{\sl 8.3  Local ${\bf P}^2$: ${\cal O}(-3)\rightarrow {\bf P}^2$ }\cr
&\qquad\hbox{\sl 8.4  Local ${\bf P}^1\times {\bf P}^1$: ${\cal O}(K)\rightarrow 
{\bf P}^1\times {\bf P}^1 $ }\cr
&\qquad\hbox{\sl 8.5  Other local Del Pezzo geometries $E_5$, $E_6$, $E_7$ and 
$E_8$}\cr
&\qquad\hbox{\sl 8.6  The topological string perspective}}$$
\vfill
\eject

\centerline{Table of Contents (continued)}
$$\eqalign{
&\hbox{\bf 9. Computations in compact Calabi-Yau geometries  }\cr
&\qquad\hbox{\sl 9.1  Compact one modulus cases}\cr
&\qquad\hbox{\sl 9.2  Higher genus results on the quintic}\cr
&\qquad\hbox{\sl 9.3  The sextic, the bicubic and four conics}\cr
&\hbox{\bf 10. Appendix A: Low degree classes on the del Pezzo Surfaces}\cr
&\hbox{\bf 11. Appendix B: B-model expression for $F_g$}}$$

\vfill
\eject

\newsec{Introduction}

The study of
Calabi-Yau threefolds has been a source of many new ideas
in string theory.  Not only they are useful as building
blocks of various string compactifications, but they
also provide interesting examples of
exactly computable quantities in string theory. In particular they correspond
to the ``critical dimension'' for the 
($N=2$) topological string.

Topological strings, roughly speeking, count the number
of holomorphic curves inside the Calabi-Yau.  As such,
one would expect that they should correspond to the partition
function of M2 (or D2)-branes wrapped around them.
The connection at first sight seems somewhat confusing:
The topological string amplitudes exist for each genus, whereas
the M2 brane (or D2 brane) degeneracies only care about the
charge and not the genus of the curve representing it.
It turns out, as discovered in \vafagop , that the genus
dependence of the topological string amplitudes captures the
$SU(2)_L$ representation content of
BPS states corresponding to wrapped M2 branes
upon compactifications of M-theory on Calabi-Yau threefolds.
Here
 $SU(2)_L$ denotes a subgroup of the $SO(4)$ rotation group in 5 dimensions.
 This identification
was based on the target space interpretation
of what the topological string computes \bcovII 
\ref\naret{I. Antoniadis, E. Gava, K.S. Narain, T.R. Taylor,
``Topological amplitudes in String Theory'' , Nucl. Phys. $\us{B413}$ (1994) 162, 
hep-th/9307158, I.\  Antoniadis, E.\ Gava, K.\ S.\  Narain, 
``Moduli Corrections to gravitational Couplings from Sring Loops'', 
Phys. Lett. $\us{B238}$ (1992) 209, hep-th/9203071
and ``Moduli Corrections to Gauge and gravitational Couplings in four-dimensional 
Superstrings'',  Nucl. Phys. $\us{B383}$ (1992) 93, hep-th/9204030. } 
and the contribution of BPS states to such terms
(using a Schwinger 1-loop computation) 
\ref\agnt{I. Antoniadis, E.\ Gava, K.\ S.\ Narain and 
T.\ R.\ Taylor, Nucl. Phys. $\us {B455}$ (1995) 109,  
hep-th/9507115.}.

Topological string amplitudes at genus 0 can be computed
using mirror symmetry \ck .
For higher genera, mirror symmetry is still a powerfull principle
and can be used to compute the amplitudes up to a finite number
of undetermined constants at each genus \bcovII . Fixing the constant
is called fixing the `holomorphic ambiguity', and for the certain
cases they were fixed for genus 1 and genus 2 in \bcovI \bcovII .
The number of unknown constants grows with the genus.  In certain cases
one can use direct $A$-model localization \kz\ to fix these constants
and in particular checking the integrality properties of
topological string partition functions, anticipated in \vafagop, 
at higher $g \le 5$.

In this paper we wish to use
the reformulation of topological string amplitudes as a computation
of BPS states in M-theory compactifications \vafagop\ to make progress
in explicit computations of topological strings at higher genera.
The reorganization this introduces into topological string amplitudes
is to fix the BPS charge and consider {\it all} allowed genera 
of the M2 brane at the same time.  For a given degree, there
typically is a highest genus curve embedded in the 3-fold which
realizes that class\foot{This highest genus is the arithmetic genus which we will often denote by $g$.
The topological string amplitude at genus $g$ usually denoted $F_g$ 
has contributions from curves of different arithmetic genera. If 
the distinction is important we use the label $r$ to refer to the 
worldsheet genus and write $F_r$ etc.} . 
One then studies  the moduli space of that curve, together with the flat bundle over it.  Understanding
of the cohomology of this moduli space and the $SU(2)_L$ action
on it, will in particular determine the BPS degeneracy and
its $SU(2)_L$ quantum numbers.  This in particular affects
the topological string amplitudes for all $F_\gg$ with $\gg\leq \ag$
in a well defined way.  The main aim of our paper is to 
develop techniques that at least in some cases allows us
to extract from the geometry of this moduli space
the $SU(2)_L$ action on its cohomology.  We relate
the degeneracies for a fixed $SU(2)_L$ spin, and in 
particular its contribution to $F_\gg$, to the Euler
characteristic of the $\delta =\ag-\gg$ fold symmetric product of the holomorphic
curves in the Calabi-Yau 3-fold and to higher $F_k$'s ($\gg < k\leq \ag$).
For $\delta$ sufficiently small this space is smooth and 
its Euler characteristic can be computed.  For $\delta$ too big, in general
this space is not smooth and the computation of its Euler characteristic
requires more care.  We will consider examples of both types.
We also use these results to fix the holomorphic ambiguities
for higher genera in some examples (and in particular
we push up the computation of topological
strings to higher genera).

We will also discuss the connection of topological string amplitudes
and the entropy of spinning black holes corresponding to $M2$ branes wrapped
over ``large'' cycles in the Calabi-Yau.  In particular we see how
in the case of $K3\times T^2$ the elliptic genus of the symmetric
product of $K3$'s predicts complete answers to
 the $SU(2)_L$ action on the moduli spaces that
we study.  For a general Calabi-Yau threefold, we see how the
black hole entropy predicts new growth properties for the topological
string amplitudes at higher genera that would be interesting
to verify.

The organization of this paper is as follows:  In section 2 we review
the definition and some results related to A-model topological strings.
In section 3 we review the definition of some of the new invariants
which allows one to rewrite topological string amplitudes using
integral data.  In section 4 we show how the
new invariants can be effectively
computed in certain cases. In section 5 we show
that the same invariants can also be computed
in a different way and be given a related geometric
interpretation.  In section 6 we show in the
case of $K3\times T^2$ how the elliptic genus of symmetric
products of $K3$
captures the BPS degeneracies of a wrapped M2 brane
and show how our methods can predict some of these results.
In section 7 we use predictions of macroscopic entropy
of black holes to estimate the growth of topological
string amplitudes for high genera.
In section 8 we give some examples involving non-compact
Calabi-Yau 3-folds and show how our methods work in those
cases.  In section 9 we do the same, but in the context
of compact CY 3-folds.  In appendix A we discuss some
aspects of del Pezzo surfaces and in appendix B we discuss
some aspects of B-model topological strings.

\newsec{Topological Strings (A-model)}

In topological string theory (A-model) one considers
maps from a Riemann surface $\Sigma_g$ of genus $g$ to a manifold
which in the case of interest in this paper we take
to be a Calabi-Yau threefold $X$.  The partition function
depends only on the complexified K\"ahler moduli of $X$
denoted by $(t_i,{\overline t}_i)$.  In the limit whereby one fixes
$t_i$ and takes the limit ${\overline t}_i\rightarrow \infty$,
a holomorphic anomaly decouples, and
the theory becomes purely topological.  In particular, in this limit the
$F_g(t_i)$ are obtained by considering {\it holomorphic} maps from the Riemann
surface to $X$.  Roughly speaking one has
$$F_g(t_i)=\sum_{hol. map f: \Sigma_g \rightarrow X}{\rm exp}
(-\int_{\Sigma_g}f^* (k))$$
where $k$ is the Kahler class on $X$ and $f^*(k)$ is its pullback to $\Sigma$.
The above formula is not quite general because often holomorphic
maps come in families.  In these cases the sum is replaced by
an integral over the moduli space of holomorphic maps 
representing some top characteristic class on the moduli space.
More precisely, in the special case of Calabi-Yau threefolds 
that we are considering the 
formal dimension of the moduli space of maps is zero and when
there is a moduli space of maps there is an equal dimensional space
corresponding to a cokernel of a bundle map.  Thus the cokernel
vector space forms a bundle on the moduli space of curves whose Euler
class enters the relevant topological computation which enters
in the above formula (for a more precise mathematical definition
and a review of the subject see \ck ).\foot{Strictly speaking, the obstruction
spaces need not form a bundle, and there can be a virtual fundamental class
in place of an Euler class.}
The result of such integrals for each
fixed topological class of the image curve in $X$ is known as
the Gromov-Witten invariants.
In other words one can write
$$F_g(t_i)=\sum_{d^i}f_{g,d} {\rm exp}(-d_i t^i)$$
where $d_i$ denotes the homology class of the image curve
in terms of some basis for $H_2(X,{\bf Z})$ and
$f_{g,d}$ are the Gromov-Witten invariants.  Since in most
cases of interest the computation of $f_{g,n}$ involves integrals
over moduli spaces, there is a priori no reason for them to be integers,
as they are not ``counting'' the number of holomorphic curves.
 However some surprising
integrality properties have already been observed for small genus
which we will review below.  From the viewpoint of the topological
string this integrality is very surprising and has not been explained.
An explanation of the observed integrality and its
generalization to all genera has
been found in \vafagop\ based on M-theory/type IIA duality which
recasts Gromov-Witten invariants in terms of some new integral
invariants $n_d^\gg$.  In this section we first review some aspects
of topological strings.  We then review the results of \vafagop\
in section 3.

\subsec{The genus $0$ contribution}

\def\li#1{{\rm Li}_{#1}}

Let $\li{r}(x)=\sum_{k=1}^\infty k^{-r} x^k$, i.e. 
$\li{-r}(x)=-\left(x {\dd\over \dd x } \right)^r \log(1-x)$ and  
$\li{r}(x)=-\left(\int {\dd x\over x} \right)^r \log(1-x)$ then 
\cdgp~gives a formal expansion 
\eqn\fnull{F_0 ={K_0 t^3\over 3!}+ t\int_X c_2 J+ {\chi \over 2} \zeta(3)+
\sum_{d=1}^\infty n^0_d \ \li{3}(q^d) \ .}
Here $K_0$ is the classical triple intersection number on $X$, which comes 
from the degree zero maps. 

The curve counting function in genus zero is $K_{ttt}=(\partial_t)^3 
F_0=K_0+\sum_{d=1}^\infty K_d q^d.$ 
By \fnull, $K_d$ is related to the $n^0_d$ by
\eqn\gendoub{K_d=\sum_{n|d} {n^0_{d/n}\over n^3}.}
It was observed that in this way of writing the Gromov-Witten
invariants the $n^0_d$ are integers \cdgp\ for the case
of the quintic in $\IP^4$ at least for degrees up to $300$. This was later 
extended to all $d$ which are not multiples of 5.\
\foot{Lian and Yau also proved integrality of the coefficients of the 
mirror map for 
the quintic, and in all the applications to toric hypersurfaces no non-integer 
$n^0_d$ 
ever appeared.}
An explanation of this integrality was suggested in \cdgp\ as
counting the ``number of rational holomorphic curves'' in  Calabi-Yau space.
This was further supported by the fact that it was shown in 
\earlymulticover\ that the $n$ fold covering of an isolated holomorphic
curve of degree $d$ gives a contribution of $1/n^3$ to the Gromov-Witten
invariant for degree $dn$ in perfect accord with \gendoub .  However
the interpretation of $n^0_d$ as counting holomorphic rational
curves in $X$ 
is in general {\it not} the right interpretation. In particular a 
counter-example occurs even for isolated curves in the quintic. In 
\ref\vain{ I.\ Vainsencher, {\sl Enumeration of
$n$-fold tangent hyperplanes to a surface\/}, J.\ Algebraic Geom.\
{\bf 4}, 1995, 503--526, alg-geom/9312012} a contribution 
of $n^0_{5, \ nod}=17,601,000$ plane curves with six nodes to total number
of curves at degree five $n^0_5=229,305,888,887,625$ was found. 
There are three contributions to $K^{10}$: degree 10 curves, double covers
of degree 5 curves, and additional integer contributions for double
covers of the 6-nodal curves corresponding to double covers with 2
components.
Correspondingly for each 
double covering of a nodal curve there is a higher dimensional stratum
{\sl and} six points in ${\cal M}_{0,0}(X,10)$ \ck . 
The contribution of the higher dimensional strata to the degree 10 
curves must 
therefore be calculated by a virtual fundamental class calculation,
which yields the usual ${1\over 2^3}$ and so the double covering contribution 
is ${1\over 2^3}\cdot n_{5\ smooth}^0$ for the smooth $d=5$ curves but  
$(6+{1\over 2^3})\cdot n_{5\ nod}^0$ for the nodal ones.  
That means that the number\foot{We assume that there are finitely many curves 
(Clemens' Conjecture). $N_{10}$ could still contain multiplicity factors for 
certain
curves.} of degree 10 curves on the quintic 
is not given by $n_{10}^0$, but rather by $N_{10}=K^{10}-{n_1^0\over 1000}
-{n_2^0\over 125}-
{n^0_{5}\over 8}-6 {n^0_{5, \ nod}}$, see Example 7.4.4.1 and Theorem  
9.2.6 in \ck .   
In particular $n_{10}^0$ as defined in \gendoub~has no interpretation 
as a ``number'' of curves, and there is currently no known mathematical 
reason to expect it to be integer! One aim of this paper is to outline a 
physically motivated geometrical definition of the $n^\gg_d$, 
which makes the integrality manifest.

More generally, let $C\subset X$ be a sufficiently general smooth curve of genus 
$g$ satisfying appropriate genericity hypotheses. Then $C_g(h,d)$ denotes the 
contribution to $F_{g+h}$ of maps whose image is $C$ whose image has class $d[C]$. 
 It is not yet clear if this notion is well-defined for $g\ge2$.

Extending \earlymulticover~Faber and Pandharipande 
\fp~prove that the multicover contribution $C_0(h,d)$ 
of a $\IP^1$ is described by
\eqn\buble{C_0(g,d)=\chi_g d^{2g-3} = {|B_{2 g}| d^{2g-3}\over 2 g (2g-2)!}\quad 
{\rm with }\quad 
\chi_0=1,\,\, \chi_1={1\over 12}.}
Here $\chi({\cal M}_g)={|B_{2 g}| \over 2 g (2g-2)!}$ is the Harer-Zagier formula 
for the orbifold Euler characteristic of ${\cal M}_g$ in complete accordance with 
the predictions
of M-theory \vafagop\ which will be discussed
later. 

\subsec{The genus $1$ contribution}

For $\gg=1$ the situation is more interesting. The localization \fp~
gives
\eqn\geomsubstraction{ C_1(0,d)={\sigma_1(d)\over d},\ C_1(h,d)=0\ (h>0).}
There is no bubbling contributions of genus 1 curves to higher genus 
curves, i.e.\ $C_1(h,d)=0$  for $h>0$ in accordance with the above and the 
zero-mode analysis in \bcovI . This is a feature one finds in the M-theory 
approach discussed in the next section.

However the form of $F_1$ discussed in the next 
section (which is most natural from the M-theory perspective) is 
(up to the $t$-terms)
\eqn\ansatzgIa{F_1=
{t \int c_2 J \over 24} + \sum_{d=1}^\infty \left( {1\over 12} n^0_d  + n^1_d  
\right)\li{1}(q^d)\ .}
Here, $n^1_d$ is an invariant of certain BPS states typically associated
to wrapping M2 branes around degree $d$ elliptic curves $E$.
This differs from the geometric substraction scheme 
\geomsubstraction, as it does 
not subtract all the multicovering maps from the torus to itself in the 
definition of 
the $n_d^1$, but instead subtracts $1/d$ for the class $d[E]$.
Substraction of these would yield\foot{The genus zero contribution follows
from  \buble~ in both cases.} \bcovI  
\eqn\ansatzgIb{F_1=
{t \int c_2 J \over 24} + \sum_{d=1}^\infty \left( {1\over 12} n^0_d \li{1}(q^d)  
+ n^{*1}_d 
\log\eta(q^d) \right)}
where $n^{*1}_d$ corresponds to elliptic curves rather than BPS states.
The reason for adding back in the multicover contributions is discussed
from the BPS perspective in the next section.
 
Comparing \ansatzgIa~with \ansatzgIb~
$$\sum_{d,n} {n_d^1\over n} q^{dn}=\sum_{d,n} n_d^{* 1}{\sigma_1(n)\over n} q^{nd} 
$$
and keeping in mind the definition ${\sigma_1(n)\over n}=\sum_{k|n}{1\over k}$, we 
see
that  the number of BPS states of charge $d[E]$ is $n^1_d=\sum_{m|d} n^{* 1}_m$ as 
expected from adding up all bound states.

\subsec{The constant map contribution}

We can compute for arbitrary genus in the simple case when
the holomorphic maps from the Riemann surface to $X$
are just the constant maps.  This is already a case where there is a moduli
space of such maps.
If the degree of the map $f$ is $0$ its moduli space  splits into 
\eqn\split{\MS{g}{n}(X,0)\sim \MS{g}{n}\times X}
where $\MS{g}{n}$ in this case corresponds to moduli space of
genus $g$ domain curves.  The relevant Gromov-Witten invariant in this
case is given by $\half e(X)\int_{{\cal M}_g} c_{g-1}^3(H)$ \bcovII\ 
where $e(X)$ denotes the Euler characteristic of $X$ and
 $H$ denotes the Hodge bundle (coming from the space of holomorphic
one forms on the Riemann surface) over the moduli space.
For the Hodge integrals 
involved in  the above formula
a closed formula was recently proven in \fp~
following the approach of \wittdgrav\kontdgrav\faber, which yields
\eqn\be{\langle 1 \rangle_{g,0}^X=(-1)^g {\chi\over 2} 
\int_{{\cal M}_g}\lambda^3_{g-1}=(-1)^g {\chi \over 2} {|B_{2g} B_{2g-1}|\over 
2g(2g-2)(2g-2)!}\ .} 
This is in perfect agreement with the prediction for
constant map contribution from the 
viewpoint of duality of type IIA and M-theory \vafagop .

\newsec{M-theory/Type IIA  interpretation of the $F_g$}

Recently  a series of integer invariants $n^\gg_d$ were defined
\vafagop~
for each Calabi-Yau threefold $X$, labeled
by a degree $d\in H_2(X,{\bf Z})$ and a positive
integer $\gg$.  Their definition was motivated by
 consideration of M-theory on Calabi-Yau threefolds.
The topological significance of these new invariants is that they can be used
to rewrite $F_\gg$ in terms of them.
In particular they explain
the integrality properties of $F_\gg$ for all $\gg$ and
reproduce the expected integrality properties for genus $\gg=0,1$.  

The definition of these invariants was motivated by consideration of
the spectrum of BPS states in M-theory compactification on Calabi-Yau
3-fold.  The spectrum in turn (in simple situations) can be computed
by considering a certain $SU(2)$ action on the cohomology of the
moduli space of holomorphic curves in a Calabi-Yau $X$ together with a
flat bundle.  We will now review the definition of these
integral invariants.

\subsec{The new invariants}

The invariants defined in \vafagop~ are given as follows:
Consider M-theory on a Calabi-Yau threefold $X$.  Thise gives
an $N=2$ theory in $d=5$. This theory has $b_2(X)$ gauge fields.
The states in $d=5$ are labeled in particular by the charge
under these $U(1)$'s, which thus correspond to an integral
element $d\in H_2(X,{\bf Z})$.  Fix the subsector
of the Hilbert space with charge $d$ and consider all
states in this sector which are BPS states.  These arise
in M-theory by considering M2 branes in $X$ wrapped around supersymmetric
cycles in the class given by $d$.  In particular their mass is fixed
by the Kahler class of $X$ and is given by $d_it_i$.
The BPS state in addition is labeled by how it transforms under
the spatial rotation group in $4+1$ dimensions which is
$SO(4)$, or more precisely $SU(2)_L\times SU(2)_R$.  In particular
we can write the degeneracy of the BPS states together with their
$SO(4)$ quantum numbers as
\eqn\spincontent{\left[\left({1\over 2},0\right)\oplus 2 
(0,0)\right]\otimes \bigoplus_{j_L,j_R} N^d_{j_L,j_R}\ [(j_L,j_R)].}
The numbers $N^d_{j_L,j_R}$ denote the number of BPS states with
charge represented by the class $d$ and with $SU(2)_L\times SU(2)_R$
representation given by the rerpresentation $(j_L,j_R)$, where
$j_L,j_R\in (1/2){\bf Z}$ and denote the spin of the representations.

The number of BPS states is not an invariant of the theory
and it can jump. Two (short) BPS multiplets can
join and become a (long) non BPS multiplet.  For example,
changing the complex structure of the Calabi-Yau $X$ will change
the numbers $N^d_{j_L,j_R}$.  However, the left index of the representation
does not change.  In other words, if we consider the degeneracies
with respect to $SU(2)_L$ and sum over all $SU(2)_R$ quantum
numbers multiplied by $(-1)^{2j_R}=(-1)^{F_{R}}$, 
then this weighted sum of left representations does
not change.
\foot{There are well known examples, e.g.\ \ref\kmp{S.~Katz,
D.R.~Morrison, and M.R.~Plesser, Nucl.\ Phys. $\us {B477}$ (1996) 105,
hep-th/9601108.}, where the
individual right spin content changes under complex structure
deformation. Consider a $\IP^1$ fibered over a genus $g$ curve
$C_g$. The right Leshetz decomposition of the base ${\cal M}=C_g$ is
$\left[1\over 2\right]+ 2 g [0]$. Vanishing complex volume of the
$\IP^1$ corresponds to a special value on the Coulomb branch $\phi=0$
with a $SU(2)$ gauge enhancement and $g$ hypermultiplets in the
adjoint representation. Higgsing w.r.t. to the diagonal components of
the hypers corresponds to a complex structure deformation and breaks
the gauge group to $U(1)$ and $2g-2$ charged hypers, which
geometrically corresponds to a splitting of the $\IP^1$ fibration into
$2g-2$ isolated $\IP^1$'s, whose right spin content is $(2g-2)[0]$.}
It is more useful for comparison with topological strings to choose
a different basis for the $SU(2)_L$ representations.  Let
$$I_r=[(\half )+2(0)]^{\otimes r}.$$
Using this basis, the procedure is
\eqn\indsu{
N^d_{j_L,j_R}[(j_L,j_R)]\rightarrow 
\sum N^d_{j_L,j_R}(-1)^{2 j_R}(2j_R+1)[(j_L)]=\sum_r n_{d}^{r} I_r}
The above equation defines the invariants $n_d^r$ which
appear in the partition function of the topological string.
According to \vafagop\ we have:
\eqn\allg{F=\sum_{\gg=0}^{\infty}\lambda^{2\gg-2} 
F_\gg=\sum_{\gg=0}^\infty\sum_{d=0 \atop 
m=1}n^\gg_d 
{1\over m} \left(2 \sin {m \lambda \over 2}\right)^{2\gg-2} q^{d m}.}
The argument leading to this identification is that the topological
string can be viewed as computing $\sum F_\gg R_+^2 F_+^{2\gg-2}\lambda^{2\gg-2}$
amplitudes in four dimensions upon considering type IIA compactification
on the Calabi-Yau. Here $R_+$ and $F_+$ denote self-dual parts
of the Riemann tensor and graviphoton field strength, respectively
\bcovII\naret . 
Then a 1-loop Schwinger computation as in \agnt\ with the BPS states running 
around
the loop relates the BPS content of states
in 5-dimensional M-theory to corrections to $R_+^2 F_+^{2\gg-2}$ amplitudes.
The appearance of the extra sum over $m$ in $\allg$ is related
to the momentum  a BPS state in 5 dimensions can have when
compactified on a circle down to 4 dimensions.  These appear
as `multi-cover' contributions in the topological string 
context, as first noticed for the case of genus 0 by
\ref\lawnek{A.\ Lawrence and  N.\ Nekrasov,``Instanton 
sums and five-dimensional Gauge Theories'', Nucl. Phys. $\us {B513}$ (1998) 239, 
hep-th/9706025.}.  The term $(2\, \sin {m\lambda \over 2})^{2\gg}$
in the above formula arises from
computation of ${\rm Tr}(-1)^{2j_L+2j_R} \exp(2i m \lambda J^3_L)$
in the $I_\gg$ representation, where $J^3_L$ is one of the generators
of $SU(2)_L$.

\subsec{The higher genus contributions}

Expanding \allg~gives back \fnull, i.e. the naive multicovering 
formula \gendoub~for the rational curves, which first was empirically 
observed in \cdgp . Of course the physical picture relates the integers  
$n^\gg_d$ naturally to the number of BPS states.  Note also that with 
the $\zeta$-function renormalization one gets the well known 
\cdgp ~subleading  contribution to the genus zero pre-potential 
$\chi(X) \zeta(3)/2$ from \split\allg .

Using genus $\gg=0,1$ as a model, we can try to recursively define $n_d^\gg$ by
$$K_d^\gg=\sum_{k\mid d\atop h\le \gg} n_k^hC_h(\gg-h,{d\over k}),$$
where $K_d^\gg$ are the Gromov-Witten invariants defined by $F_\gg=\sum K_d^\gg
q^d$.  This is the approach taken in \ck~for $\gg=0,1$; there the numbers
$n_d^\gg$ are called instanton numbers 
(see \ck\ for a precise version of the integrality conjecture for
$\gg=0$).  

For the  elliptic curve $T^2$ and any $n$, we
can consider $n$ D2 branes wrapped on $T^2$.  In
this case as discussed in \vafagop\ to count
the number of BPS states we should consider the moduli
space of stable rank $n$ bundles on $T^2$.
There are indecomposable
(semi)stable $U(n)$ bundles over 
$T^2$, which corresponds to a BPS bound state of $n$ D2 branes wrapping $T^2$
(the corresponding space for genus 0 is empty which is why we
do not have bound states of $n$ D2 branes on a genus 0 curve).
This explains the scheme used in the previous section in defining
the BPS numbers for genus 1.

For the genus 2,3 expansion we have
\eqn\ansatzgII{F_2= 
{\chi\over 5760} + \sum_{d=1}^\infty \left({1\over 240} 
n^0_d+n^2_{d}\right)\li{-1}(q^d)}
\eqn\ansatzgIII{F_3=-{\chi\over 1451520  } +  
\sum_{d=1}^\infty \left({1\over 6048} n_d^0 -{1 \over 12}n^2_d+ 
n_d^3\right)\li{-3}(q^d)}
and similarly for higher genus
\eqn\ansatzgIV{F_\gg\! =\! 
{(-1)^\gg\chi |B_{2\gg} B_{2\gg-2}|\over 4 \gg (2\gg-2)!(2\gg-2)}+\sum_{d=1}^\infty \! \!
\left(\! {|B_{2\gg}|n_d^0\over 2\gg (2\gg-2)!}  +
{2(-1)^\gg  n_d^2\over (2\gg-2)!}\pm\! \ldots \! - {{\gg-2}\over 12} n_d^{\gg-1}\! +n_{d}^\gg 
\! \right) 
\! \li{3\! -\! 2\gg}}

There is a subtlety for genus $r>2$ in that $n$ D2 brane bound states can deform
off the supporting genus $r$ curve.  This is briefly discussed
in \vafagop\ and
is a topic for further study.

\newsec{Computation of $n^\gg_d$}
The identity \allg\ reexpresses topological string amplitudes
in terms of integral quantities $n^\gg_d$ defined in terms of the 
BPS spectrum of M-theory on Calabi-Yau threefolds according to
\indsu .  If one finds a simple way to compute the new invariants
$n^\gg_d$ this
would translate to a practical method of computing topological
string amplitudes.

In \vafagop\ it was shown how one goes about computing
$n^\gg_d$ (at least in certain good cases).  The basic idea
is to consider the moduli space of $M2$ branes, which gets
translated using M-theory/IIA duality
to the study of certain aspects of moduli space of $D2$ branes.
One considers supersymmetric $D2$ branes whose class in $X$
is given by $[D2]=d$. The moduli space of such configurations
is given, in addition to the embedding of the $D2$ brane,
by the choice of a flat bundle on the brane.  In general
if we have $N$ coincident branes, we will have to consider
also the moduli of flat $U(N)$ bundles in addition to the
moduli of the embeddings of the $D2$ branes.  Let us consider
the simple case where we have a single $D2$ brane in class $d$ and
let us denote by ${\widehat {\cal M}}$ the moduli space of holomorphic
curves in $X$ in class $d$, together with the choice of the
flat bundle on the Riemann surface.  Let ${\cal M}$ denote
the moduli space of holomorphic curves in class $d$, without
the choice of the flat bundle.  Then we have a map
$${\widehat {\cal M}}\rightarrow {\cal M}$$ 
Let us assume that generically the Riemann surface has genus $g$.
Then the above map has generically a fiber which is $T^{2\ag}$, i.e.
the Jacobian of the Riemann surface.  However generally speaking
there are loci where the genus $g$ surface become singular.
For example it can develop nodes by having some pinched cycles.
Similarly the Jacobian torus becomes singular in this limit.\foot{More precisely, 
the compactified Jacobian becomes singular in this limit.}
Nevertheless,
one expects the total space $\widehat{{\cal M}}$ to be smooth (similar
to description of elliptic fibration of K3 where the fibration
becomes singular at 24 points, but the K3 is smooth).
Because of this smoothness, for many questions
it is possible to treat the above
fibration {\it as if} there are no degenerate fibers.  In particular,
consider the integral (1,1) form $k$ corresponding to the fiber $T^{2g}$
(usually denoted on each non-degenerate
fiber by $k|_{fiber}=\theta=dz({\rm Im} \Omega^{-1})dz^*$). 
We will assume, as is the case with smooth Jacobian varieties,
that  $k$ makes sense as an integral (1,1) class in ${\widehat {\cal M}}$.

Consider the cohomology of the manifold ${\widehat {\cal M}}$.
These will correspond to BPS states in M-theory compactification.
Moreover the $SU(2)_L$ quantum numbers get morally identified with the
$SL(2)$ Lefschetz decomposition in the fiber direction (i.e. using
$k$ as a raising operator) and the $SU(2)_R$ quantum numbers
get morally identified with the $SL(2)$ Lefschetz decomposition in the base
direction (i.e. using the K\"ahler form on the base).  In other words
we have \vafagop :
\eqn\como{H^*({\widehat {\cal M}})=\sum N^d_{j_1,j_2}[j_1^{fiber},j_2^{base}]}
from which we can read off $n^\gg_d$ according to 
\indsu.
There are
precise statements that can be made: the usual Lefschetz decomposition
of the cohomology
${\widehat{\cal M}}$ is identified with the diagonal $SL(2)\subset
SL(2)_{fiber}\times SL(2)_{base}$, and the $SU(2)_R$ content of the highest
left spin is identified with the Lefschetz decomposition of $H^*(\cM)$.

There are two particularly easy cases to compute from
the above definition, namely:  
\eqn\hilow{\eqalign{
n^\ag_d&=(-1)^{\dim\cM}e(\cM)\cr
n^0_d&=(-1)^{\dim\widehat{\cM}}e({\widehat{\cal M}})
}}
where $e(...)$ denotes the Euler characteristic of the space.
The relations follow from the definition of what the double
Lefschetz action is.  As we will demonstrate in the next section,
the other non-vanishing $n$'s, i.e.\ the $n^\gg_d$ for $0<\gg<\ag$ can
also be related to particular combinations of
Euler characteristics of certain subspaces in ${\widehat {\cal M}}$.
Sometimes we will write $r=g-\delta$ where $\delta $ is a
positive integer less than or equal to $g$.

The existence of such a double Lefschetz decomposition is 
expected from the M-theory
description of the BPS states in 5 dimensions and so it should 
be possible to rigorize the existence of the above double $SL(2)$ 
decomposition of the cohomology of ${\cal {\widehat M}}$.  However
here we would like to get the new invariants with the minimal
amount of assumptions about the properties of ${\widehat {\cal M}}$.
As we will discuss in the next section all we really need
for computation of $n^\gg_d$ is the existence of a smooth manifold
${\widehat {\cal M}}$ and an integral (1,1) class $k$ which on smooth fibers
is the canonical $(1,1)$ class on the Jacobian torus.  This will also
lead to a simple formulation for the computation of all $n^\gg_d$
in terms of Euler characteristics of relative Hilbert schemes, which are
frequently easy to compute.

\subsec{Computational Scheme for $n^\gg_d$}

As is clear from \como\ and \indsu\ all we need to compute
the $n^\gg_d$ is the Lefschetz action in the fiber direction.
We will see that
in fact we can compute $n^\gg_d$ without this assumption in a reasonably
general setting.

For each point on the base ${\cal M}$ let $C$ denote the 
 corresponding Riemann surface and $\cJ(C)$ its Jacobian.  
  The Riemann
surface together with the choice of $p$ points on it, is what
is called the Hilbert scheme of $p$ points on $C$, and denoted by
${\rm Hilb}^p(C)$.
We have the Abel-Jacobi mapping \ref\gh{P.~Griffiths and J.~Harris, {\sl 
Principles of Algebraic Geometry\/},
Wiley-Interscience, New York 1978.}:
\eqn\abjac{f_p:{\rm Hilb}^p(C)\to \cJ(C)}
whose image is denoted by $W_p$.  We can relate the cohomology of $W_p$
to the cohomologies of both ${\rm Hilb}^p(C)$ and $\cJ(C)$, thereby relating
these two cohomologies directly.

We have the map $H_*(W_p)\to H_*(\cJ(C))$, which by Poincar\'e duality
is identified with a map $i:H^*(W_p)\to H^*(\cJ(C))$ whose image we wish
to compute.  Let $\theta\in H^{1,1}(\cJ(C))$ be the cohomology class of
the zero locus of the theta function on $\cJ(C)$.  Since the image $W_p$
of $f_p$ is dual to $\theta^{g-p}/(g-p)!$\gh, the composition of the
restriction map $r:H^*(\cJ(C))\to H^*(W_p)$ with $i$ is (up to the
constant which we ignore) just the multiplication map
$$\theta^{g-p}:H^*(\cJ(C))\to H^*(\cJ(C)).$$
Since there is also a map
$f_p^*:H^*(W_p)\to H^*({\rm Hilb}^p(C))$, we expect to be able to relate
$H^*({\rm Hilb}^p(C))$ with the image of $\theta^{g-p}$.

Here is our strategy.  Once we understand this relation, we
consider varying the point on the base ${\cal M}$.  In this way
  the Abel-Jacobi map $f_p$ gets promoted to a map %
\eqn\famab{{\hat f}_p: {\cal C}^{(p)}\rightarrow {\widehat {\cal M}}.}
Here ${\cal C}^{(p)}$ denotes the moduli space of holomorphic curves
of degree $d$ together with the choice of $p$ points on the Riemann
surface.\foot{More precisely, we choose a length $p$ subscheme $Z$ of
the curve $C$, which means that $\dim\cO_Z=p$.  For smooth curves, a
length $p$ subscheme is the same thing as a subset of $p$ points of
$C$ (including multiplicity).  If the curve is singular, these notions
can differ.  In Section~5, we will see how the difference plays a
crucial role in relating our methods to geometry.}  Therefore we
relate $H^*({\cal C}^p)$ to the image of
multiplication by $k^{g-p}$ on $H^*(\widehat{\cM})$, where $k$ is
the $SU(2)$ raising operator in the fiber direction.
This can be used to compute
$n^{g-\delta}_d$ according to \como\ and \indsu.

Before we carry this out, 
it is first convenient to review some facts about the cohomology
of the Hilbert scheme of $p$ points on the Riemann surface $C$.
Let $C$ be a smooth curve of genus $g$.  
Then we have for its cohomology, as an $SU(2)$ Lefschetz representation

\eqn\csu{H^*(C)=({\bf{1\over2}})\oplus(2g)({\bf 0}).}
For a smooth curve, its Hilbert scheme is the same as its symmetric product.
Taking symmetric products, we have for the Lefschetz $SU(2)$ decomposition
\eqn\cksu{\eqalign{
H^*({\rm Hilb}^k(C))&=\bigoplus_r
{\rm Sym}^{r}\left({\bf{1\over2}}\right)\otimes\wedge^{k-r}(2g)({\bf 0})\cr
&=\bigoplus_{r=0}^k{2g\choose k-r}\left({\bf {r\over2}}\right)}}
Note that since the $2g({\bf 0})$ represent odd cohomology of $C$, we must
antisymmetrize.

For convenience, we explicitly list the first two cases of \cksu

\eqn\lowcksu{\eqalign{
H^*({\rm Hilb}^2(C))&=({\bf 1})\oplus (2g)({\bf{1\over2}})\oplus (2g^2-g)
({\bf 0})\cr
H^*({\rm Hilb}^3(C))&=({\bf{3\over2}})\oplus(2g)({\bf 1})\oplus
(2g^2-g)({\bf{1\over2}})\oplus ({4\over3}g^3-2g^2+{2\over3}g)({\bf 0}).
}}

The Jacobian $\cJ(C)$ is a principally polarized abelian variety \gh,
so as we have already mentioned has a canonical
K\"ahler class $\theta\in H^{1,1}(\cJ(C))$, whose corresponding divisor
is the zero locus of the theta-function on $\cJ(C)$.  It is
straightforward to check that the resulting Lefschetz $SU(2)$ representation 
content 
of  $H^*(\cJ(C))$ can be identified with $I_g$, which is
the representation we defined before.    Furthermore,
the class
$\theta$ in this context is identified with the $SU(2)$ raising
operator $k$.  So $\theta^{g-p}H^*(\cJ(C))$ is the same as $k^{g-p}I_g$,
and we are just dealing with a simple problem in the representation
theory of $SU(2)$.

One easily proves by induction that
\eqn\igsu{I_g=\bigoplus_{r=0}^g\left({2g\choose g-r}-{2g\choose g-r-2}\right)
\left({\bf {r\over 2}}\right)}
as an $SU(2)$ representation. It follows immediately that

\eqn\lefI{\eqalign{
\theta^{g-1}H^*(\cJ(C))&=\left[{\bf {1\over2}}\right]\oplus (2g)\left[{\bf 
0}\right],\cr
\theta^{g-2}H^*(\cJ(C))&=\left[{\bf 1}\right]\oplus (2g)\left[{\bf 
{1\over2}}\right]\oplus (2g^2-g-1)\left[{\bf 0}\right], 
\cr
\theta^{g-p}H^*(\cJ(C))&=\bigoplus_{r=1}^p  \left(\left(2g \atop p-r\right) -
\left(2 g \atop p-r-2\right)\right)\left[{\bf {r\over2}}\right]\qquad{\rm in\ 
general.}}}

In \lefI\ we are being a bit imprecise with notation, since the $k^{g-p}I_g$
are not representations of $SU(2)$.  What we mean by $[{\bf r/2}]$ is
a collection of $r+1$ classes of the form
$v,kv,k^2v,\ldots k^rv$.  We are not assuming that $v$ is killed by the
$SU(2)$ annihilation operator.  Here $v$ can have any $U(1)$ charge 
$m\ge -r$, so that $[{\bf r/2}]$ has $U(1)$ charges shifted to
$m,m+1,\ldots,m+r$.

Now we are ready to relate $H^*({\rm Hilb}^k(C))$ and the image of
multiplication by $\theta^{g-p}$.  
We can write the precise relationship by comparing \csu\ and
\cksu\ with \lefI :
\eqn\alldecomp{\eqalign{
H^*(C)&=k^{g-1}I_g\cr
H^*({\rm Hilb}^2(C))&=k^{g-2}I_g\oplus({\bf 0})\cr
H^*({\rm Hilb}^3(C))&=k^{g-3}I_g\oplus H^*(C) \cr
H^*({\rm Hilb}^p(C))&=k^{g-p}I_g \oplus H^*({\rm Hilb}^{p-2}(C))
\ \ {\rm in\ general.}}}
Note that $i$ takes $H^i(W_p)$ to $H^{2g-2p+i}(\cJ(C))$.
This shift by $2g-2p$ is precisely what is needed to match up the $U(1)$
charges in \alldecomp, which is understood as an identification of $U(1)$ 
charges.

Again by induction we note from \lefI\ that 
\eqn\projI{{\rm Tr}(-1)^F k^{g-p}I_g= 
{2\over p!} ( g - p) \prod_{i=1}^{p-1} (2 g-i)\equiv a(g,p)}

Now we vary over $\cM$.
We write the representation of the BPS states as
$R=\sum_\delta n^{g-\delta}I_{g-\delta}$.
Allowing the curve to vary over the moduli space of the curve $\cM$, 
we get from \alldecomp\

\eqn\relate{\eqalign{
H^*(\hc1)&=k^{g-1}R\cr
H^*(\hc2)&=k^{g-2}R\oplus H^*(\hc0)\cr
H^*(\hc3)&=k^{g-3}R\oplus H^*(\hc1)\cr
H^*(\hc{p})&=k^{g-p} R\oplus H^*(\hc{p-2})\ \ {\rm in\ general}}}
with the definitions $\cC^{(0)}\equiv\cM$ and $\cC^{(1)}\equiv \cC$. 
We now apply ${\rm Tr}(-1)^F$ to both sides of \relate, and get, using
\projI
\eqn\trequations{
(-1)^{{\rm dim} (\cM)+\delta}(
e(\hc{\delta})-e(\hc{\delta-2}))=\sum_{p=0}^\delta a(g-p,\delta-p)\,
n^{g-p}, \ \ {\rm for} \ \ \delta=1,\ldots} where we set
$e(\cC^{(-1)})\equiv 0$, $a(g,0)\equiv 1$. In particular, the first
two equations read

\eqn\almostdone{\eqalign{
(-1)^{\dim(\cM)+1}e(\hc1)&=(2g-2)n^g+n^{g-1}\cr
(-1)^{\dim(\cM)}\left(e(\hc2)-e(\hc0)\right)
&={1\over 2}(2 g-2) (2g-1)n^g+(2g-4)n^{g-1}+n^{g-2}
}}
If we solve \almostdone\ and $n^g=(-1)^{{\rm dim}( \cM )}e(\hc0)$ for $n^{g}$, 
$n^{g-1}$, $n^{g-2}$ we get
$$\eqalign{n^{g-1}&=(-1)^{\dim(\cM)+1}\left(e(\hc1)+(2g-2)e(\cM)\right)\cr 
           n^{g-2}&=(-1)^{\dim(\cM)}\left(e(\hc2) + (2g-4) e(\hc1)+{1\over 2} 
(2g-2) (2g-5) e(\hc0)\right).}
$$
In general one shows that the solution to \trequations\ yields 
\eqn\generalformng{n^\gg=n^{g-\delta}=(-1)^{\dim(\cM)+\delta}\sum_{p=0}^\delta 
b(g-p,\delta-p)\, e(\hc{p}),}
with 
$$ b(g,k)\equiv {2\over k!}(g-1)\prod_{i=1}^{k-1} (2 g-(k+2)+i), \quad 
b(g,0)\equiv 0 .$$ 

Note that we do not require the Lefschetz action on $\widehat{\cM}$
to apply these formulas, only the Lefschetz action on the spaces $\hc{k}$.
These exist whenever the spaces $\hc{k}$ are smooth.

\newsec{Considerations of Enumerative Geometry}
 
In this section, we put forward some natural geometric principles which allow
us to relate the invariants $n^\gg_d$ to computations on other, but related
geometrical objects.  Moreover, this reasoning points
us
to introduce correction terms for certain families of reducible curves.
We illustrate our formulas by a few examples which yield numbers which
can be checked by other methods.  More systematic checks are done
in the remaining sections of this paper.

In the previous section we have seen that the contribution
of a family of genus $g$ curves to $F_0$ comes from
the $e({\widehat {\cal M}})$.  Since ${\widehat {\cal M}}$ is
a Jacobian variety, one would expect, by torus action on the fibers, that
this can also be computed by a localization principle.  This is similar
to the situation considered in \yz.  There, the calculation of
$e(\widehat{\cM})$ can be
localized to a calculation on the set of nodal curves.
For a genus $g$ curve we would like to be able to compute $e(
\widehat{\cM})$ by localizing on curves with $g$ nodes.

  An additional
motivation comes from a glance at \hilow.  It is natural
to expect from \hilow\ that there would be
subspaces
\eqn\chain{\cM=\widehat{\cM}_0\subset \widehat{\cM}_1\subset\cdots
\subset\widehat{\cM}_\delta \subset ...\subset \widehat{\cM}_g=\widehat{\cM}}
such that $n_d^{g-\delta}=(-1)^{\dim\widehat{\cM}_\delta}
e(\widehat{\cM}_\delta)$ for some suitable spaces ${\widehat{\cM}_\delta}$.

Quite independently of the existence of such subspaces, we can still
ask for a localication type of computation for all these cases, as in \yz .
Consider for example the $\delta =g$ case.  In this case
we are degenerating the curve of genus $g$ to genus zero with $g$ nodes.
On the other hand, the genus 0 isolated curves have information
only about the $I_0$ content of BPS states. Since an isolated
genus $g-\delta$ curve has information only about the $I_{g-\delta}$
content of BPS states,
one would expect that $n_d^{g-\delta}$ which counts BPS states
in the representation $I_{g-\delta}$ is localized on curves
of genus $g-\delta$.    This reasoning would thus lead us to identify
\eqn\lowergenus{n_d^{g-\delta}=(-1)^{\dim \widetilde\cM_\delta} 
e(\widetilde\cM _\delta )}
where $\widetilde\cM_\delta \subset \cM$ denotes the moduli space
of irreducible curves with $\delta$ ordinary nodes, i.e.\
with genus $\gg=g-\delta$. Note that this proposal also
fits with the top genus contribution where $\delta =0$, namely
\eqn\topgenus{n^g=(-1)^{\dim \cM}e(\cM).}
where we have noted that $\cM$ parameterizes generic genus
$g$ curves.
 Regardless of the
existence or definition of $\widehat\cM_\delta$ and the
localization of its Euler characteristic to $e(\widetilde\cM_\delta)$
we would
like to explore the potential validity of \lowergenus\ and in particular
see if we get a match with the computations done in the previous section.\foot{
We could use the relation of invariants to classes considered
in the previous section to construct spaces formally related
to what we want, but that viewpoint does not appear useful in the present
context.}

\subsec{Alternative interpretation of the $n^\gg_d$}

In this section, we undertake the geometric calculation of the desired
formulas.  Recall that in \yz, a formula for $n^0$ was derived assuming
that the singular curves in the relevant family of curves had only
nodes as singularities.  In our situation, we will make similar
simplifying assumptions on the geometry of the singular curves in our
family.  Our viewpoint is that since we expect to derive formulas
which are generally valid, we are free to make extra assumptions
in order to derive them.  In fact, our assumptions rarely hold,
but we will be able to argue that the formulas obtained are sound.
In this way, we greatly enhance our ability to calculate geometrically.

Calculation of invariants of $\widetilde{\cM}_\delta$ can be tricky due to the
irreducibility requirement.  It is easier instead to study the spaces
$\cM_\delta$, the set of curves with
$\delta$ nodes, dropping the irreducibility hypothesis.  It is easier still
to consider
$\overline{\cM}_\delta$, the closure of the $\cM_\delta$ in $\cM$. The spaces
$\overline{\cM}_\delta$ parametrize curves with at least $\delta$ nodes,
and also curves with possibly more complicated singularities or higher 
multiplicities.  We will calculate
$e(\overline{\cM}_\delta)$ using a simple topological argument in
certain good situations.

Let us consider a special case where many of the difficulties are supressed.  
Suppose that all curves $C_i$ parametrized
by the points of
$\overline{\cM}_i-\overline{\cM}_{i+1}$ have exactly $i$ nodes for
$0\le i\le \delta$, where we have put $\cM_0=\cM$.  
Then the Euler
characteristic of $C_i$ is $2+i-2g$. 

If in addition $\cM_{\delta+1}$ 
is empty, then $\cM_\delta=\overline{\cM}_\delta=\overline{\cM}_\delta
-\overline{\cM}_{\delta+1}$.  We next set out to calculate $e(\cM_\delta)$.
Note that if the curves of $e(\cM_\delta)$ are irreducible, so
that $\cM_\delta=\widetilde{\cM}_\delta$, then we can
calculate $n^\gg=n^{g-\delta}$ from this Euler characteristic using 
\lowergenus. We are continuing to assume that $\cM_\delta$ is
smooth.

Recall that the Hilbert scheme ${\rm Hilb}^k(C)$ of degree $k$
subschemes of a single curve $C$ parametrizes subsets $S\subset C$ of
$k$ points.  The points of $S$ are allowed to occur with multiplicity.
There is a bit more structure placed on the higher multiplicity points
which are located at the singularities.  We will give examples later.
For the moment, we just observe that this Hilbert scheme has dimension
$k$.

Let $\pi:\cC\to\cM$ be the universal curve, so that if $m\in \cM$ corresponds
to a curve $C_m\subset X$, then $\cC\subset X\times\cM$ is such that
$\pi^{-1}(m)=C_m\times\{m\}$.
For each $k$, let 
\def\hc#1{\cC^{(#1)}}
$\pi_k:\hc{k}\to\cM$ be the relative Hilbert scheme of degree $k$
subschemes of the fibers of $\pi$.  In other words, we build $\hc{k}$
from the universal curve by taking the Hilbert scheme of the curves
$C_m=\pi^{-1}(m)$ for each $m\in\cM$, and $\hc{k}$ is constructed as the
union of these as $m$ varies in $\cM$.  Thus the fiber of $\pi_k$ over
$m$ is ${\rm Hilb}^k(C_m)$, and $\hc{k}$ has dimension $\dim\cM+k$.

Our assumptions imply that all fibers of $\hc{k}$ over points of
$\overline{\cM}_i-
\overline{\cM}_{i+1}$
have the same computable Euler characteristic, which will be an explicit
function $f(g,i,k)$ of $g,\ i$, and $k$.  We will calculate $f(g,i,k)$
explicitly soon for all $g$ and a few values of $i$ and $k$.

Then letting $\hc{k}_i$ be the preimage under $\pi_k$ of $\overline{\cM}_i$,
we get 
$$
e(\hc{k}_i)-e(\hc{k}_{i+1})=f(g,i,k)\left(
e(\overline{\cM}_i)-e(\overline{\cM}_{i+1})\right).
$$
Note that $\hc{k}_{\delta+1}$ is empty.  Summing these equations from
$i$ from 0 to $\delta$, we get an equation expressing $e(\hc{k})$ in
terms of the $e(\overline{\cM}_i),\ 0\le i\le \delta$.  If we generate
$\delta$ such equations by taking $k$ from 1 to $\delta$, then these
equations can be solved for the $\delta$ variables
$e(\overline{\cM}_i),\ 1\le i\le\delta$.  In particular, we can solve
for $e(\overline{\cM}_\delta)=e(\cM_\delta)$ in terms of these
$e(\hc{k})$ and $e(\cM)$.

As already stated, this gives the desired formula for $n^\gg$ 
when the irreducibility assumption holds.

We now carry out this procedure for small $\delta$.   The results are
\eqn\smallnodes{\eqalign{
e(\cM_1)&=e(\cC)+(2g-2)e(\cM)\cr
e(\cM_2)&=e(\hc2)+(2g-4)e(\cC)+{1\over 2}(2g-2)(2g-5)e(\cM)\cr
e(\cM_3)&=e(\hc3)+(2g-6)e(\hc2)+{1\over2}(2g-4)(2g-7)e(\cC)+\cr
        &\phantom{=}{1\over6}(2g-2)(2g-6)(2g-7)e(\cM)\cr
e(\cM_4)&=e(\hc4)+(2g-8)e(\hc3)+{1\over2}(2g-6)(2g-9)e(\hc2)+\cr
        &\phantom{=}{1\over6}(2g-4)(2g-8)(2g-9)e(\cC)+
        {1\over24}(2g-2)(2g-7)(2g-8)(2g-9)e(\cM)\cr
}}
These formulas suggest that in general, we have
\eqn\allnodes{
\eqalign{
e(\cM_\delta)&=e(\hc{\delta})+(2g-2\delta)e(\hc{\delta-1})+\cr
&\phantom{=}\sum_{i=2}^\delta{1\over i!}(2g-2\delta+2i-2)(2g-2\delta+i-3)
(2g-2\delta+i-4)\cdots(2g-2\delta-1)e(\hc{\delta-i})}
}
where we have put $\hc1=\cC$ and $\hc0=\cM$.  These are precisely
the formulas given in \generalformng, as we asserted at the beginning
of this section!

Consider for example the case when $X$ is a local $\IP^2$.  Since homogeneous
polynomials of degree $d$ in the three variables have
$(d+2)(d+1)/2$ coefficients and scalar multiplication of the equation
does not alter the curve, with get $\cM=\IP^{d(d+3)/2}$.  In particular,
if $d=4$, we get that $\cM=
\IP^{14}$, with Euler characteristic 15.  To understand $\cC$, we
consider the projection $\cC\to\IP^2$.  The fiber over $p\in\IP^2$ is
the set of plane quartic curves which contain $p$, and this is a
$\IP^{13}$ for all $p$, as the equation $f(p)=0$ imposes one linear
equation on the 15 coefficients of $f$.  Thus $\cC$ is a $\IP^{13}$
bundle over $\IP^2$, hence smooth, and $e(\cC)=(3)(14)=42$.  We
therefore get $e(\overline{\cM}_1)=42+4(15)=102$.  

Note that
$\widetilde{\cM}_1=\cM_1$ in this instance.  To see this, observe that 
a reducible curve of degree 4 would have to be the union of a line
and a degree 3 curve, a union of two degree 2 curves, or more 
degenerate configurations.  All such curves must have at least 3 nodes
or worse singularities, so are not contained in $\cM_1$.
But it is not true that
$\cM_2$ is empty in this case.  Nevertheless, we find from table 4
that $n_4^2=-102$, exactly as we would have found if $\cM_2$ were
empty!

This situation turns out to be quite common.  We derive formulas for
the $n^\gg$, and they turn out to have greater validity.

A few words are in order now about the assumption we made that
$\cM_{\delta+1}$ is empty.  Recall that we want to derive a
formula for $e(\overline{\cM}_\delta)$.  But this Euler characteristic
is only asserted to correctly calculate the appropriate $n^\gg$ if
$\overline{\cM}_\delta$ is smooth.  This is relevant because at a point of
$\cM_{\delta+1}\subset\overline{\cM}_\delta$, the space
$\overline{\cM_{\delta}}$ tends to be singular.  Here is the reason.
If $C$ has $\delta+1$ nodes, then choosing any subset of $\delta$ of
these nodes, we get a branch of $\overline{\cM_{\delta}}$ which
parameterizes curves for which the last node is allowed to smooth out
while the original subset of $\delta$ nodes remains nodal.  Since there
are $\delta+1$ choices of subsets of $\delta$ nodes, we see that
$\overline{\cM}_\delta$ has $\delta+1$ branches at a general point of
$\cM_{\delta+1}$. In particular, $\overline{\cM}_\delta$ is singular.

Said differently, once we assume that $\overline{\cM}$ is smooth,
then the assumption that $\cM_{\delta+1}$ is empty is quite natural.
Once we drop the smoothness assumption, then there is no reason that
the formula $n^{g-\delta}=e(\overline{\cM}_\delta)$ should be valid.
The pleasant surprise is that we have discovered that the formula we derive
for the $n^{g-\delta}$ are correct more generally.  In fact, these
formulas do not always compute the $e(\overline{\cM}_\delta)$, but that is
of no concern to us: the bottom line is that the formulas compute
the invariants that we are actually interested in.

We now have to say something about a common situation when
$\widetilde{\cM}_i\ne\cM_i$, namely when there are reducible curves in our
family with exactly $i$ nodes.  We expect a nice geometric
situation when all components parameterizing reducible families 
are irreducible components of $\overline{\cM}_\delta$.

Here is the problem.  If a curve $C$ has $\delta$ nodes and is
irreducible, then its desingularization $\tilde{C}$ has genus
$g-\delta$ and comes with a map $\tilde{C}\to C$ which gives an
explicit geometric contribution to the instanton sums.  However, if
$C$ has $\delta$ nodes but is reducible, then its desingularization at
$\delta$ nodes can split $C$ into disjoint components.  Since the
worldsheet must be connected, such a configuration does not contribute
to the instanton sums.  However, it does contribute to our
calculations which have ignored the issue of irreducibility.  We must
find a way to correct for this.

We consider such a component which parameterizes reducible curves of the
form $C=C_1\cup C_2\cup\ldots\cup C_k$.  Some of the curves $C_i$ may
also have a fixed number of nodes, hence a fixed geometric genus $r_i$.
Explicitly, if $C_i$ has degree $d_i$ and $\delta_i$ nodes,
then $\gg_i=(d_i-1)(d_i-2)/2-\delta_i$.  Since each degree $d_i$ is strictly
less than the degree $d$ of $C$, we can inductively compute the instanton
numbers $n^{\gg_i}_{d_i}$ for their respective degrees and geometric genus.

Suppose that these curves split into disjoint components after
desingularizing at the $\delta$ nodes.  We propose that this component
contributes $\prod_i n^{\gg_i}_{d_i}$ to the numbers naively computed by
multiplying formulas \smallnodes\ and \allnodes\ by the appropriate
sign.  In other words, we are proposing the following algorithm for
computing the instanton number of genus $\gg\le g$ associated to a
family of curves of arithmetic genus $g$.

\bigskip\noindent
Supposing that $\cC^{(k)}$ is smooth for $0\le k\le g$, we put $\delta=g-r$ 
and  calculate $(-1)^{\dim\cM_\delta}e(\overline{\cM}_\delta)$, where by
$e(\overline{\cM}_\delta)$ we mean the value calculated from
\smallnodes\ or \allnodes.  Then identify any components $M_1,\ldots
M_\gg$ of $\overline{\cM}_\delta$ which parameterize reducible curves.
Each component $M_j$ gives a contribution of the form $\prod_i
n^{\gg_{ij}}_{d_{ij}}$ as explained above.  We have introduced a second
subscript on $d$ and $g$ to emphasize that we may have to consider
several components.  Our proposal is then
\eqn\bestcorrection{
n^\gg=(-1)^{\dim\overline{\cM}_\delta}e(\overline{\cM}_\delta)-\sum_{j=1}^r
\prod_i n^{\gg_{ij}}_{d_{ij}}.
}

We illustrate again when $X$ is a local $\IP^2$ and again $d=4$.  This time, we
will calculate the genus~0 instanton number.  We have to impose
$\delta=3$ nodes to get the genus to 0.  We have already
computed that $e(\cM)=15$ and $e(\cC)=42$.  We consider the map 
$\rho_2:\hc2\to {\rm Hilb}^2\IP^2$ which takes a multiplicity~2 scheme in a
degree 4 curve and views it as a multiplicity~2 scheme in $\IP^2$.
We can compute the Euler characteristic of ${\rm Hilb}^2\IP^2$ either
from counting fixed points of a torus action or from the generating
function
\eqn\hilbgen{
\sum_{k=0}^\infty e\left({\rm Hilb}^k\IP^2\right)
q^k=\prod_{n=1}^\infty(1-q^n)^{-3}.
} Either way, we get 9 for this Euler characteristic.  It is not hard
to see that the fiber of $\rho_2$ over any point $Z\in{\rm
Hilb}^2\IP^2$ has codimension 2 in the space of all degree 4 curves,
as this fiber is just the space of all quartics containing $Z$.  Said
differently, the condition that $f|_Z=0$ places 2 independent linear 
conditions on the 15 coefficients of $f$. If $Z=\{p,q\}$, these two conditions
are just $f(p)=f(q)=0$. If $Z$ is concentrated at a single point, then after 
a change of coordinates we can write $Z$ locally as $y=x^2=0$.  The space
of $f$ which contain $Z$ is just the space of (not necessarily homogeneous) 
degree 4 polynomials in $x$ and $y$ whose constant terms and coefficient of
$x$ vanish, again a codimension 2 linear subspace.  After projectivizing,
This space is
therefore a $\IP^{12}$, with Euler characteristic 13.  
We therefore see that $\hc2$ is smooth, and we
compute that $e(\hc2)=9\cdot13 =117$.

Similarly, we get that $\hc3$ is smooth and
$e(\hc3)=22\cdot 12=264$, since ${\rm Hilb}^3\IP^2$
has Euler characteristic 22 by \hilbgen\ and the space of quartic curves
containing a fixed multiplicity~3 scheme is a $\IP^{11}$.

Now using these numbers and $g=3$ in \smallnodes, we get $e(\overline{\cM}_3)
=222$.

But this is not the entire story, since there are reducible quartics which
are unions of lines and cubic curves which have three nodes.  Lines and
cubics have respective instanton numbers $3$ and $-10$.  Since the space
of three nodal curves has dimension 11, we therefore
get the corrected number $n_4^0=(-1)^{11}222-(3)(-10)=-192$.  This is
in agreement with the value we will exhibit from the B-model in Table~4
of Section~8.3.

\medskip
We think of our calculational method as giving corrections to
\smallnodes\ and \allnodes.  Unfortunately, it does not apply in all
cases, since the $\hc{k}$ can be singular.  The simplest case we are
aware of is $n_6^2$ in local $\IP^2$.  This case has $\delta=8$, and
for $\delta<8$ our method applied successfully every time we are able
to check it by mirror symmetry or localization \kz.  Our proposal is therefore a very
powerful check of the M-theory integrality prediction.  We presume
that the eventual reconciliation with more general cases (including
$n_6^2$) will come from more subtle corrections.

As an interesting aside, we note that our method sometimes applies
nevertheless when $\hc{k}$ is singular.  The simplest case is if
$C\subset X$ is a single isolated curve of arithmetic genus 1 with a
single node.  The identity map $C\to C$ is a genus 1 stable map, and the
normalization map $\IP^1\to C$ is a genus 0 stable map.  It is clear that
these are the only degree 1 stable maps onto $C$ up to isomorphism.
We arrive at the conclusion that $n^1=1$ and $n^0=1$.\foot{The higher degree 
invariants have recently been computed in 
\ref\bkl{J.\ Bryan, N.C.\ Leung, S.\ Katz, in preparation.}.}  Since $\cM$ is
a point, we get from \topgenus\ that $n^1=1$.  As for the genus 0 
contribution, the first line of \smallnodes\ (or more simply, the
method of \yz ) gives $n^0=1$, since $\cC=C$ has Euler characteristic 1.

Continuing with this digression, note that this reconciles the count
of BPS states with stable maps of degree 1 for any isolated
irreducible curve of arbitrary genus and number of nodes.  Suppose
that an irreducible curve $D$ has arithmetic genus $g$ and $k$ nodes.
By \ref\beau{A.\ Beauville, Duke Math.\ J.\ $\us {97}$ (1999) 99--108.},
the compactified Jacobian of $D$ is isomorphic to a product of
factors.  One factor is the Jacobian $\cJ(\tilde{D})$ of the smooth
genus $g-k$ desingularization of $D$.  There are $k$ other factors,
one for each node, and each of these are isomorphic to the curve $C$
above with genus 1 and a single node.  By \vafagop, $\cJ(\tilde{D})$
contributes an $I_{g-k}$ representation, and we have just established
that each copy of $C$ contributes an $I_1+I_0$ representation.  So the
total representation is
\eqn\totalrep{
I_{g-k}\otimes(I_1+I_0)^k=\sum_{\delta=0}^k{k\choose\delta}I_{g-\delta}.}
The right hand side of \totalrep\ predicts $n^{g-\delta}={k\choose\delta}$.
This matches the stable maps perfectly, as we get a genus $g-k$ stable map
by picking $\delta$ of the $k$ nodes and partially normalizing $D$ only
at this subset.  Since there are ${k\choose\delta}$ ways to do this, we
have complete agreement.

\bigskip
We now derive \smallnodes, beginning with $\delta=1$.
Since all smooth curves of genus $g$ have Euler characteristic $2-2g$,
we get
$$e(\cC)-e(\cC_1)=(2-2g)\left(e(\cM)-e(\overline{\cM}_1)\right).$$ By our
assumption that all curves of $\overline{\cM}_1$ have exactly one node, we
have
$$e(\cC_1)=(3-2g)e(\overline{\cM}_1),$$
since one nodal curves have Euler characteristic $3-2g$.
Adding these two equations, we obtain 
\eqn\none{e(\overline{\cM}_1)=e(\cC)+(2g-2)e(\cM),}
which is the first equation in \smallnodes.

The case of $\delta=2$ requires additional explanation.

We have to calculate $e(\overline{\cM}_2)$.  
We have the equations
$$\eqalign{e(\cC_2)&=(4-2g)e(\overline{\cM}_2)\cr
e(\cC_1)-e(\cC_2)&=(3-2g)(e(\overline{\cM}_1)-e(\overline{\cM}_2))\cr
e(\cC)-e(\cC_1)&=(2-2g)(e(\cM)-e(\overline{\cM}_1))\cr}$$
obtained as before.  Adding these equations gives
\eqn\ntwohone{e(\cC)=(2-2g)e(\cM)+e(\overline{\cM}_1)+e(\overline{\cM}_2).}
We next derive another equation to eliminate $e(\overline{\cM}_1)$ by considering
$\hc2$.  As above, let $\hc2_i$ be the restriction of the map
$\hc2\to\cM$ to the part lying over
$\overline{\cM}_i\subset\cM$ for $i=1,2$.  
We calculate the Euler characteristics of the strata $\hc2_i-\hc2_{i+1}$, 
a new point needing explanation being the role of the 
nodes.  Writing the node locally as $xy=0$, we see that there is a $\IP^1$ 
moduli space for the schemes of multiplicity 2 at the origin.  Recall that
locally schemes are the same thing as ideals, so a scheme of multiplicity
2 at the origin is just an ideal $I$ of polynomials in $x,y$ such that
the origin has multiplicity 2.  It is easy to see that $I$ must be
generated by a linear and a quadratic polynomial in $x,y$, both vanishing
at the origin.  Given a 
linear polynomial $ax+by$, there is actually no need to specify a
choice of quadratic polynomial $q(x,y)$, since $q$, taken together with
the quadratic polynomials $x(ax+by)$ and $y(ax+by)$ spans a 3 dimensional
space, necessarily the entire space of quadratic polynomials vanishing at
the origin.  Explicitly,
these are the schemes $Z_{a,b}$ defined by the ideals
$I_{a,b}=(ax+by,x^2,xy,y^2)$, where $(a,b)\in\IP^1$.  Note that $xy\in 
I_{a,b}$, so that the $Z_{a,b}$ are indeed contained in the nodal curve.

This is the new ingredient that we need to calculate the Euler characteristics
of the strata.   To get ${\rm Hilb}^2(C)$, where
$C$ is a curve with $i$ nodes, 
we take its second symmetric product,
and replace the single point $2({\rm node})$ with a $\IP^1$, for each node.
This says that since $C$ has $i$ nodes, then
\eqn\nodeffect{\eqalign{
e({\rm Hilb}^2C)&=e({\rm Sym}^2C)+i\cr
&={i+3-2g\choose2}+i.\cr
}}

This leads immediately to the equations
$$\eqalign{
e(\hc2_2)&=\left({5-2g\choose 2}+2\right)e(\overline{\cM}_2)\cr
e(\hc2_1)-e(\hc2_2)&=\left({4-2g\choose 2}+1\right)
(e(\overline{\cM}_1)-e(\overline{\cM}_2))\cr
e(\hc2)-e(\hc2_1)&=\left({3-2g\choose 2}\right)
(e(\cM)-e(\overline{\cM}_1))\cr
}$$

We add these equations and obtain
\eqn\ntwohtwo{e(\hc2)={3-2g\choose 2}e(\cM)+(4-2g)e(\overline{\cM}_1)+
(5-2g)e(\overline{\cM}_2).}
We now can eliminate $e(\overline{\cM}_1)$ from \ntwohone\ and \ntwohtwo\ and get 
\eqn\ntwo{e(\overline{\cM}_2)=e({\rm 
Hilb}^2(\cC/\cM))+(2g-4)e(\cC)+(g-1)(2g-5)e(\cM),}
the second equation in \smallnodes.

We turn next to $\delta=3$.
The calculation begins as in the previous cases, and we get the
equations
\eqn\hthreeqs{\eqalign{
e(\cC)&=(2-2g)e(\cM)+e(\overline{\cM}_1)+e(\overline{\cM}_2)+e(\overline{\cM}_3)
\cr
e(\hc2)&={3-2g\choose2}e(\cM)+(4-2g)e(\overline{\cM}_1)+(5-2g)e(\overline{\cM}_2)
   +(6-2g)e(\overline{\cM}_3).}}
We now have to bring in $\hc3$ to derive one more equation for the
purpose of eliminating $e(\overline{\cM}_1)$ and $e(\overline{\cM}_2)$.
We have to explain how to calculate 
$\hc3$.  So we need to know how to calculate the Euler characteristic
of ${\rm Hilb}^3$ of a curve $C$ of arithmetic genus $g$ with $i$
nodes for $i=1,2,3$.  We study the map ${\rm Hilb}^3C\to {\rm Sym}^3C$
and see where it fails to be an isomorphism.  This is precisely over the
points $2p+q$ and $3p$ of ${\rm Sym}^3C$, where $p\in C$ is a node
and $q\ne p$ is arbitrary.  As in the discussion leading to
\nodeffect, we
replace $2p+q$ by $\IP^1\times {q}$, where $\IP^1$ is the $\IP^1$ of
tangent directions to $C$ at $p$.  So for each node $p_i$, we replace
a subset $\{p_i\}\times C-\{p_i\}$ by $\IP^1\times C-\{p_i\}$, adding
$i(i+1-2g)$ to the Euler characteristic.  As for $3p$, we write the
node locally as $xy=0$ and look for multiplicity 3 schemes contained in $xy=0$
and concentrated at $(0,0)$.  It suffices to compute the Euler 
characteristic, which is just the number of fixed points of a torus
action $(x,y)\mapsto(t^ax,t^by)$ where $a\ne b$ are arbitrary.  These are
just the points $(x,y^3),(y,x^3)$, and $(x^2,xy,y^2)$, so the
Euler characteristic is 3.\foot{It can be seen that the set of all
multiplicity 3 schemes is isomorphic to $\IP^2$, while those contained in the
nodal curve are $\IP^1\cup\IP^1$, the first $\IP^1$ being $\{(ax+by^2,
x^2,xy,y^3)\}$ for $(a,b)\in\IP^1$, the other $\IP^1$ being obtained
by interchanging $x$ and $y$.}  This gives 
$$
e\left({\rm Hilb}^3C\right)=
{4+i-2g\choose 3}+i(i+1-2g)+2i = {4+i-2g\choose 3}+i(i+3-2g).
$$
We now immediately get the equations
$$\eqalign{
e(\hc3_3)&={4-2g\choose3}e(\overline{\cM}_3)\cr
e(\hc3_2)-e(\hc3_3)&=\left({5-2g\choose3}+4-2g\right)
\left(e(\overline{\cM}_2)-e(\overline{\cM}_3)\right)\cr
e(\hc3_1)-e(\hc3_2)&=\left({6-2g\choose3}+2(5-2g)\right)
\left(e(\overline{\cM}_1)-e(\overline{\cM}_2)\right)\cr
e(\hc3)-e(\hc3_1)&=\left({7-2g\choose3}+3(6-2\right)g)
\left(e(\cM)-e(\overline{\cM}_1)\right)}
$$
Adding, we get the formula
\eqn\lasteqn{\eqalign{e(\cC^3)&={4-2g\choose3}e(\cM)+
\left({4-2g\choose2}+4-2g\right)e(\overline{\cM}_1)+\cr
&\phantom{=}\left({5-2g\choose2}+6-2g\right)e(\overline{\cM}_2)+
\left({6-2g\choose2}+8-2g\right)e(\overline{\cM}_3).}}
We can now solve our 3 equations in \hthreeqs\ and \lasteqn\ for 
$e(\overline{\cM}_3)$, obtaining
\eqn\nthree{\eqalign{e(\overline{\cM}_3)= & \ 
e(\hc3)+(2g-6)e(\hc2)+(g-2)(2g-7)e(\cC)+
\cr &
\ \ {2\over3}(g-1)(g-3)(2g-7)e(\cM)\ ,}}
the third equation in \smallnodes.

The same method applied to $\delta=4$ yields the fourth equation in
\smallnodes.  We use the general fact that the Hilbert scheme of
multiplicity $k$ points concentrated at a node has Euler characteristic $k$.
This fact can be easily verified using fixed points of torus actions.  We
have already seen this result explicitly for $k\le3$.

Note that it is clear that this method of calculation generalizes to
arbitrary genus.   We do not know how to carry this out in closed form,
but do presume that the answer is given by \allnodes\ since we have 
derived this in \generalformng\ and we will 
offer several checks of these formulas in later sections.

\newsec{Application to Counting M2 branes in $K3\times T^2$}

Before we come to the application of the formalism developed in
Sections 4 and 5 to the general case of Calabi-Yau threefolds, we consider
the simpler $K3\times T^2$ case\foot{A similar analysis can be done
for $T^4\times T^2$ with symmetric product of $K3$ playing the same
role as symmetric product of $K3$ plays here.}. 
In this case the topological string amplitudes are rather 
trivial (except for genus 1 which is 24 times the logarithm of the 
$\eta$ function).  The reason
for this triviality is also easy to explain from the view point
of BPS degeneracy of wrapped M2 branes in
M-theory compactification on $K3\times T^2$:  The BPS spectrum
of states which preserve exactly $1/4$ of the supersymmetry
(which has the same amount of supersymmetry preservation as for
the $M2$ branes wrapped around a generic Calabi-Yau) are longer:
They are in general of the form
\eqn\repco{R\otimes I_1^L\otimes I_1^R}
where $R$ is some representation of $SU(2)_L\times SU(2)_R$ and
$I_1^L=(1/2,0)+2(0,0)$ and $I_1^R=(0,1/2)+2(0,0)$. These states were recently 
considered in \ref\ls{ W.\ Lerche and S.\ Stieberger, ``1/4 BPS States and Non-Perturbative 
Couplings in N=4 String Theories'', hep-th/9907133 .} starting from a 
type II one-loop computation. 
We observe here that there is an extra factor of $I_1^R$ in the above representation.  
When we consider the relevant index contributing to topological string
amplitudes, by summing over the right representation with a
$(-1)^{F_R}$, the $I_1^R$ factor kills the contribution.  The
geometric explanation of this, in term of the moduli space we have
discussed is that the moduli space of M2 brane configurations is a
product space with a factor including a $T^2$.  This is because we can
use the $U(1)\times U(1)$ symmetry of the $T^2$ to obtain a new
holomorphic curve from any given one.  This introduces an extra factor
$I_1^R$ in the representation.  The only case where this action is
trivial, and the $I_1^R$ is absent from the above representation, is
if the $M2$ brane lies entirely in the $T^2$, i.e. it is a point in
$K3$ and wraps the $T^2$.  From the point of view of representation
theory the fact that the $I_1^R$ does not appear in that case is that
this BPS state preserves $1/2$ of the supersymmetry and so it is a
shorter multiplet.  The moduli space of the $M2$ brane wrapping $T^2$
and projecting to a point in $K3$ is simply $K3$, whose Euler
characteristic is $24$.  Taking into account the $N$ fold bound state
which always exists at genus 1, we reproduce the prediction of the
topological string amplitude at genus 1 and its vanishing at all other
genera.

However, clearly there is an enormous amount of information in precisely
which representations $R$ appear in \repco .  In particular,
for these BPS states, we can omit the factor of $I_1^R$, and
again concentrate on the $SU(2)_L$ action
and sum over the right states (with a $(-1)^{F_R}$) and define the degeneracy
number $n_d^\gg$, just as in the generic Calabi-Yau case.  Here
$d$ is an integral $H_2$ class of $K3\times T^2$ and $\gg$ labels
the $SU(2)_L$ representation content in terms of $I_\gg$.  We
can still ask how to compute $n_d^\gg$ numbers using the techniques
of this paper.  For simplicity 
of notation we consider a ``topological string amplitude''
$F_\gg$
using these numbers as input parameters, without worrying
whether or not they come from any 2d topological theories\foot{It
is likely that they do.  For example it is natural to expect
that by some insertion of operators at higher genera one can
effectively ``cancel'' the $I_1^R$ contribution above.  For example
an insertion of $\int J_LJ_R$ on the world sheet, where $J_{L,R}$ denote
the left- and right-moving $U(1)$ currents of $N=2$ algebra may do the job.}.

The class $d$ can be viewed as coming from a class $C\in H_2(K3)$ and
a class in $ H_2(T^2)$ defined by an integer $M$
times the basic class.  By diffeomorphism symmetry of $K3$ the number
of BPS states for $d=[C,M]$, as far as the $C$ dependence goes
should only depend on $C^2=2N-2$.  Thus we can recast the computation
in terms of finding the degeneracy associated to the choice of two integers
$(N,M)$.  

\subsec{Zero winding on the $T^2$}

Let us first consider the case where $M=0$, i.e. that the
BPS states correspond to wrapping only the $K3$ space, and to
a choice of a point on $T^2$.  The $T^2$ space enters in a rather trivial
way (simply giving the $I_1^R$ factor noted above) and essentially
drops out of further consideration for this case.
We are then just asking about the BPS spectrum of $M2$ brane
wrapped over M-theory compactifications on $K3$.
Since M-theory on $K3$ is dual to heterotic string on $T^3$
\ref\wittenstr{E.\ Witten,  ``String Theory Dynamics in Various Dimensions'',
Nucl. Phys. $\us {B443}$ (1995) 85,  hep-th/9503124.}, 
the heterotic string dual gives an immediate prediction for the number
of BPS states, as well as their $SU(2)_L\times SU(2)_R$ quantum numbers.
The answer for the dimension of the representation $R$ (summed over all
states weighted with $(-1)^{F_L+F_R}$) gives the degeneracy of the 
left oscillator
of the heterotic string at level $N$.  In other words
$$\sum n_{N,0}^{g=0}q^N =\prod_{n=1}^{\infty} {1\over (1-q^n)^{24}}.$$
This structure follows from the fact that there are
24 left oscillators $\alpha_{-n}^i$ where $i$ runs from 1 to 24 and $n$
runs over all positive integers.
 For example, if
$N=3$, the BPS states are specified by the 
symmetrization of the states 
$$\alpha^{i}_{-1}\alpha^{j}_{-1}\alpha^{k}_{-1}|0\rangle, \ \ 
\alpha^{i}_{-1}\alpha^{j}_{-2}|0\rangle, \ \ \alpha^{k}_{-3}|0\rangle ,$$ 
of the heterotic string.  In fact we can also easily read
off the $SU(2)_L\times SU(2)_R$ content of these states as well,
because the $SU(2)_L\times SU(2)_R=SO(4)$ is identified
with its canonical embedding in $SO(24)$.  In other words,
each oscillator $\alpha_{-n}^i$ corresponds to the representations\foot{The same 
reasoning applies to the $SU(2)_L\times SU(2)_R$ decomposition of 
rational elliptic surfaces. This makes it easy to calculate the
higher genus invariants in a fixed class $[B]+n[F]$ \hst .} 
$$[\alpha_{-n}^i]=20(0,0)+({\bf {1\over 2}},{\bf {1\over 2}})$$
We can thus decompose the BPS states above in terms of the
$SU(2)_L\times SU(2)_R$ quantum numbers inherited from each
oscillator.  For example,
 the contributions of the three types of states given above are
$$\eqalign{&\left(22 \atop 3\right)
({\bf 0},{\bf 0}) + \left(21\atop 2\right)\left({\bf {1\over 2}},{\bf {1\over 
2}}\right) +
20 (({\bf 1},{\bf 1})+({\bf 0},{\bf 0})) + 
\left({\bf {3\over 2}},{\bf {3\over 2}}\right)+\left({\bf {1\over 2}},{\bf {1\over 
2}}\right)+\cr
& 400 ({\bf 0},{\bf 0}) + 40\left({\bf {1\over 2}},{\bf {1\over 2}}\right) +
({\bf 1},{\bf 1})+({\bf 1},{\bf 0})+({\bf 0},{\bf 1})+({\bf 0},{\bf 0})+\cr
&20({\bf 0},{\bf 0})+\left({\bf {1\over 2}},{\bf {1\over 2}}\right)=
1984({\bf 0})-504\left({\bf {1\over 2}}\right)+64 ({\bf 1})- 4\left({\bf {3\over 
2}}\right)\ ,}$$
where we have summed in the last expression over the right representation with 
$(-1)^{F_R}$. Reexpressed in the $I_g$ the result reads 
\eqn\rowthree{3200 I_0-800 I_1+88 I_2-4 I_3 \ .} 
The above calculation can be easily systematized by writing the oscillator 
partition function for the oscillators in
the representation $20({\bf 0},{\bf 0})+\left({\bf {1\over 2}},
{\bf {1\over 2}}\right)$, one for each integer.  In particular one obtains
\eqn\oscillator{\prod_{n=1}^\infty {1\over { (1-q^n)^{20} (1- y q^n)^2 
(1-y^{-1}q^n)^2}}=
\sum_{\gg=0,d=0}^\infty (-1)^\gg 
 n_{N,0}^\gg (y^{1\over 2} -y^{-{1\over 2}})^{2 \gg} q^d\ . }
On the right-hand side $n_{N,0}^\gg$ is the number of BPS states
in the representation $I_\gg^L$ with charge whose square is $2N-2$.
The identification follows by noting that $I_g$ contains
$\left(2 g\atop g+i\right)$ states with $J^3_L$ eigenvalue $i/2$, or
alternatively since $I_1$ has one state with $J_L^3$ eigenvalue $\pm 1/2$,
while $I_g=I_1^{\otimes g}$. The expression
\oscillator\ contains information about all genus, and with \allg~one
can resum it to write the total free energy as 
$F=\sum_{m=1}^\infty {1\over m} F^{(m)}$, where the last sum is over the 
multicovering contributions with\foot{This decomposition into spins 
gives the higher genus result also for curves in the $K3$ classes of 
$K3$ fibered threefolds. If we take into account the multiplicity due to base 
$\IP^1$ integration $(-2)$ and the lattice sum, captured already at genus zero,
and 
multiply therefore by $(-2)E_4(q) E_6(q)$ we get the higher genus 
answer for the $K3$ fibration $X_{24}(1,1,2,8,12)$, 
which was obtained from a one loop computation in \mm .}  
\eqn\fullfreeenergy{F^{(m)}=\left(2 \sin {m \lambda\over 2}\right)^{-2} 
\prod_{n=1}^\infty {1\over { (1-q^{mn})^{20} (1- e^{i \lambda m} q^{mn})^2 
(1-e^{-i \lambda m} q^{mn})^2 }}\ . }       
Of course all information about the $n_d^\gg$ is already in $F^{(1)}$. 

Let us summarize for concreteness some of the low degree invariants.

\medskip
{\vbox{\ninepoint{
$$
\vbox{\offinterlineskip\tabskip=0pt
\halign{\strut
\vrule#
&
&\hfil ~$#$
&\hfil ~$#$
&\hfil ~$#$
&\hfil ~$#$ 
&\hfil ~$#$
&\hfil ~$#$ 
&\hfil ~$#$
&\hfil ~$#$
&\vrule#\cr
\noalign{\hrule}
&n^\gg_{N,0}
&\gg=0
&1
&  2
&  3
&  4
&  5
&  6
&\cr
\noalign{\hrule}
&N=0
&1
&
&
&
&
&
&
&\cr
&1
&24
&-2
&
&
&
&
&
&\cr
&2
&324
&-54
&3
&
&
&
&
&\cr
&3
&3200
&-800
&88
&-4
&
&
&
&\cr
&4
&25650
&-8550
&1401
&-126
&5
&
&
&\cr
&5
&176256
&-73440
&15960
&-2136
&168
&-6
&
&\cr
&6
&1073720
&-536860
&145214
&-25670
&3017
&-214
&7
&\cr
\noalign{\hrule}}\hrule}$$
\vskip -3 pt
\centerline{{\bf Table 1:} The weighted sum of BPS states $n^\gg_{N,0}$ for classes in the $K3$.}
\vskip7pt}}} 

These predictions for the spectrum of BPS states for
M-theory compactification on $K3$ is based on duality
with heterotic strings.  One can ask if one
can derive these spectra directly in M-theory
context.  In particular as discussed before, we would have to
study the moduli space of holomorphic curves in $K3$
together with a flat bundle, in the class
$C$ whose self-intersection is $2N-2$.  This has been considered
in \ref\bsv{M.\ Berschadsky, V.\ Sadov and C.  Vafa, ``D-Branes and Toplological 
Field Theories'', Nucl. Phys. $\us{B463}$ (1996) 420, hep-th/9511222 and
C.\ Vafa, ``Lectures on Strings and Dualities'', hep-th/9702201.},
and the above result from heterotic string was reproduced.
The basic idea is rather simple:   With some choice
of complex structure, we can assume $K3$ is an elliptic
surface over ${\bf P}^1$.  Moreover,
by global diffeomorphisms we can assume the cycle $C$ is represented
by the class $[B]+N[F]$ where $[B]$ denotes the class of
the base ${\bf P}^1 $ and $[F]$ denotes the class of the elliptic fiber.
The moduli space of curves in this class corresponds to degenerate
Riemann surfaces of genus $N$ and 
 is simply given by the choice of $N$
points on ${\bf P}^1$ to which the $N$ elliptic fibers are attached.
In the computation of the BPS states, we are instructed to consider
also flat bundles on the Riemann surface.  In this degenerate limit,
that choice is easy:  it simply corresponds to the choice of a flat
bundle on each elliptic fiber.  That in turn is equivalent to
a choice of a point on a dual elliptic fiber.  All said and done,
the choice of $N$ points on ${\bf P^1}$ and a point on the dual
elliptic fiber over each point, shows that the moduli
space of curves with the flat bundle is equivalent
to the choice of $N$ points on the $T$-dual $K3$.  Since
the ordering of the points are immaterial, this corresponds
to the $N$ fold symmetric product of $K3$, or more precisely, the Hilbert
scheme of $N$ points on $K3$\foot{It would be nice to make
this argument mathematically more rigorous.
What has to be checked is that this correspondence continues to hold
when several fibers are allowed to coincide.  The details will require
a mathematical study of sheaves on non-reduced curves.}.
 Thus the moduli
space is given by
$${\widehat {\cal M}}={\rm Hilb}^N(K3)$$
The cohomology of this space can be identified in the usual way
\ref\gotsche{L.\  G\"ottsche, Math. Ann.$\us{286}$ (1990) 193}\
with the Hilbert space of $24$ oscillators at level $N$, and exactly
reproduces the above results for the heterotic string. Moreover the 
$SU(2)_L\times SU(2)_R$ decomposition can also be deduced
from the corresponding decomposition for the cohomology
of a single copy of $K3$.  With the identification of $SU(2)_L$ with
the elliptic fiber direction and $SU(2)_R$ with the base directions,
we immediately get the decomposition
$$24\rightarrow 20(0,0)+({\bf {1\over 2}},{\bf {1\over 2}}),$$
as this is the unique representation whose diagonal $SU(2)$ content is
$({\bf 1})\oplus21({\bf 0})$, the Lefschetz representation of K3,
while the $SU(2)_R$ content of left-spin $1/2$ is $({\bf 1/2})$, the
Lefschetz representation on the base $\cM=\IP^1$.  This reproduces the
result based on duality with heterotic strings given above.

In \yz, the coefficients $c_N$ of $I_0$, which 
as discussed is the Euler
characteristic of ${\rm Hilb}^N(K3)$, were related to genus zero curves
coming from degenerate genus $N$ curves with exactly $N$ nodes.  As the $N$
continuous parameters of the moduli space $\IP^N$ of the genus $N$
curve are completely killed by the imposition of the $N$ nodes, this
eventually leads to the counting of points.  Here we consider the
intermediate cases, the genus $N-\delta$ curves, where we impose $0\le
\delta\le N$ nodes.  As this leaves a $\delta$ dimensional
moduli space, an appropriate virtual fundamental class on this
space is needed to reduce the dimension to 0\foot{A related problem
was considered in
\ref\bl{J.\ Bryan, N.C.\ Leung, ``The enumerative geometry of K3
surfaces and modular forms,'' alg-geom/971103 .}, where the dimensions
of the moduli space was reduced to 0 by forcing the curves to go
through $k-\delta$ points. }.  The formula \lowergenus\ is equivalent
to the assumption that the obstruction bundle in the case of a smooth
moduli space is the cotangent bundle, since the Euler class of the cotangent
bundle is the Euler characteristic of the moduli space up to sign.

For example, the coefficients in
\rowthree\ correspond to invariants $n_3^\gg$ associated to genus
$\gg=0,1,2,3$ curves obtained by putting nodes on the degree $d=3$ genus
$\ag=3$ curve. The moduli space of such curves has dimension $\gg=0,1,2,3$,
and the virtual fundamental class has the same codimension.
So the $n_3^\gg$ (and multiple cover/bubbling contributions)
can be thought of as computable by taking the virtual class and
performing an additional localization on the positive dimensional
moduli space $\bar{{\cal M}}_\delta$ of curves with $\delta$ nodes.  By the
discussion of Section 4 and 5, we can instead calculate these using the
invariants $e(\hc{k}_{[N]})$ for $k\le\delta$.
In other words, in this case we have two geometric models
for computing $n_{[N,0]}^\gg$: One is based on the Hilbert
scheme of $N$ points on $K3$, which we have already discussed,
and it agrees with the predicted answer from heterotic string.
Another way to compute these numbers is to follow the strategy
developed in previous sections and relate these numbers to
$e(\hc{k}_{[N]})$. This will be useful, as it will also tell
us how in some cases where these spaces are singular,
we may nevertheless define unambiguous answers.

Let's check a few cases of these numbers.  For any $N$, we have $\cM=\IP^N$.

For $N=0$, the moduli space is a
point, and $n^0_0=1$.

For $N=1$, the moduli space is $\cM=\IP^1$, giving $n^1_1=-2$ by \topgenus.
Choosing the complex structure so that the K3 is elliptically fibered and
our family of curves is the fiber class, we 
see that $\cC$ is just the K3.  So from \yz\ or from \smallnodes, we get
$n^0_1=e(\cC)=24$.

For $N=2$, we again get $n^2_2=3$ (and more generally,
$n^g_g=(-1)^{g}(g+1)$).  Let us choose the complex structure to be
that of $S=\IP(1,1,1,3)[6]$.  The projection $\pi:S\to\IP^2$ onto the
first 3 coordinates is a 2-1 cover.  The inverse images $C$ via $\pi$
of the lines in $\IP^2$ define the genus two curves.  To see this,
letting $H$ be the hyperplane class of $\IP^2$ we compute
$$C^2=(\pi^*(H))^2=\pi^*({\rm point})=2,$$ since 2 points of $S$ lie
over a point of $\IP^2$.  Since $C^2=2N-2=2$, this verifies that the
genus is $N=2$.  To calculate $\cC$, as usual we project $\cC$ onto
$S$ and note that the fiber is always $\IP^1$ as follows.  Given a
point $p$ of a curve $C$ (so that $(p,C)\in\cC$), the curves $C$
through $p$ are in 1-1 correspondence with the lines of $\IP^1$
through $\pi(p)$, and this is always a $\IP^1$.  This gives
$e(\cC)=e(S)e(\IP^1)=48$, and by
\smallnodes, we get $n^1_2=-(48+2\cdot 3) =-54$.  

But now something interesting happens.  The space $\cC^{(2)}$ is not a
projective bundle over ${\rm Hilb}^2(S)$.  To see this, let's pick a
point $Z$ of ${\rm Hilb}^2(S)$.  This usually projects via $\pi$ to a
point $Z'$ of ${\rm Hilb}^2(\IP^2)$.  When this happens, there is a
unique line $\ell$ connecting the two points of $Z'$ in $\IP^2$, hence
a unique curve $C=\pi^{-1}(\ell)$ in $S$ passing through the two
points of $Z$.  So we might have thought that $\cC^{(2)}$ is
isomorphic to ${\rm Hilb}^2(S)$, with Euler characteristic 324.  But
this is not true!  If the point $Z$ consisted of the 2 points of
$\pi^{-1}(p)$ for any point $p\in \IP^2$, then it maps via $\pi$ to
just the point $p$ of $\IP^2$.  There is a $\IP^1$ of lines through
$p$, whose inverse images via $\pi$ are a $\IP^1$ of curves in $S$
through $Z$.  In other words, there is an isomorphic copy of $\IP^2$
embedded in ${\rm Hilb}^2(S)$ via the map $p\mapsto\pi^{-1}(p)$, and
each of these points gets replaced by a $\IP^1$ in $\cC^{(2)}$.  It is not
difficult to see that $\cC^{(2)}$ is the blowup of ${\rm Hilb}^2(S)$
along $\IP^2$.  This tells us that $\cC^{(2)}$ is smooth, so that
\smallnodes\ applies.  It also tells us that $e(\cC^{(2)})=
e({\rm Hilb}^2(S))+e(\IP^2)=324+3=327$.  Now, \smallnodes\ gives
$n^0_2=324$.  We will give a physical way to calculate $e(\hc2)$ using
$K3\times T^2$ soon.

In general, if the map $\cC^{(k)}\to {\rm Hilb}^k(K3)$ is a projective
bundle, then we see that $\cC^{(k)}$ is smooth, and its Euler
characteristic is $(g-k+1)e({\rm Hilb}^k(K3))$, where $g$ is the genus
$N$. We can then calculate $n^{g-k}_g$ using \allnodes.  If it is not
a projective bundle, then there are two problems.  First, the Euler
characteristic is more difficult to calculate.  Second, and more
seriously, the space $\cC^{(k)}$ need not be smooth, so there may be a
virtual fundamental class to calculate instead of the Euler
characteristic and our formulas need not be valid.

Let's check $N=3$.  We have $\cM=\IP^3$ and $n^3_3=-4$.  It is not hard to
see (and we will check presently) 
that $\cC$ and $\hc2$ are projective bundles, so that 
$e(\cC)=3\cdot24=72$ and $e(\hc2)=2\cdot324=648$.  This gives, by \smallnodes,
$n^2_3=72+4\cdot 4=88$, and $n^1_3=-(648+2\cdot72+2\cdot4)=-800$.

To compute $e(\hc3)$, we choose our K3 to have the complex structure
of a degree 4 hypersurface $S\subset\IP^3$, and the genus 3 curves $C$ to be
the plane sections of $S$ (which are all degree~4 plane curves).  Note that
there is a $\IP^2$ of planes passing through any point of $S$, and a $\IP^1$
of planes containing any 2 points of $S$, verifying the projective bundle
structure on $\cC$ and $\hc2$ just mentioned.  Given three points of $S$,
there is typically a unique plane containing these points (the plane they
span), but this fails when the 3 points lie on a line $\ell$, in which case
we get a $\IP^1$ of planes.  In other words, the set of triples of collinear
points in $S$ forms a subset $T$ of ${\rm Hilb}^3(S)$, and we replace $T$
by a $\IP^1$ bundle over $T$ to get $\hc3$.  
Thus $e(\hc3)=e({\rm Hilb}^3(S))+e(T)$.   

There is an alternative way to describe $T$.  Let $p_0\in S$ be
arbitrary.  Let $\ell$ be any line containing $p_0$ (for fixed $p_0$,
there is a $\IP^2$ of such lines).  Then $\ell\cap S$ consists of 4
points, of which $p_0$ is one of them.  So there are 3 remaining
points $p_1,p_2,p_3$, which collectively give a point of $T$.  Thus the
data of $T$ is a point of $S$ and a point of $\IP^2$, and so $T$ is smooth and
$e(T)=e(S)e(\IP^2)=72$.  
Furthermore, it is again
straightforward to check that $\hc3$ is the blowup of ${\rm Hilb}^3(S)$
along $T$, so is smooth.  Thus $e(\hc3)=3272$ (which we will check soon
by a different method), and by \smallnodes, $n^0_3=3272+0-72+0=3200$.

In principle these calculations can be continued to higher $N$, at the
expense of having to use increasingly difficult projective geometry
to complete the calculation.  We will return to more of these calculations
shortly when we see that many of the moduli spaces $\cM$ for $K3\times T^2$ 
with $M\not= 0$ are
precisely the relative Hilbert schemes $\hc{k}$ for $K3$!

\subsec{More general $H_2$ classes in $K3\times T^2$}
We will now relax the assumption that $M=0$ and also consider
the case where the $M2$ brane wraps $M$ times around the $T^2$.
When $M=0$ we used three ways to compute it in the previous section:
One was the duality with heterotic strings.  The other was
the direct definition in M-theory, leading to Hilbert scheme
of $K3$ and the third one, was based on the methods we developed
in this paper.  The first two approaches gave a complete
answer, whereas the third approach was somewhat incomplete
(even though in principle one should be able to push that program
of computation sketched in some cases in the previous section). For
the case when $M\not=0$ it turns out we only know one method
to compute the exact answer and that is based on duality between
M-theory on $K3\times T^2$ and Type IIB string on $K3\times S^1$.  The
other two methods, are more difficult and we do not know how
to use the direct definition of BPS states of M2 brane to obtain
the results predicted by this duality.  We will discuss aspects of these
computations in special cases below.  But first
we show what the duality between M-theory and type IIB
predicts for all the numbers $n_{[N,M]}^\gg$ defined before.

M-theory on $T^2$ is dual to type IIB on $S^1$. Through this duality
the $M2$ brane in 9 dimensions, gets mapped
to a $D3$ brane wrapped around $S^1$.  Moreover the momentum
quantum number around $S^1$ of type IIB gets mapped to
the quantum number of the $M2$ brane wrapping number over $T^2$.
Now consider compactifying further on a $K3$.  Then an M2 brane
wrapped in some 2-cycle class of $K3\times T^2$ given by
$[C,M]$, gets mapped via this duality to $D3$ branes wrapped
around $C\times S^1$ carrying momentum $M$ around $S^1$.
In fact, in the type IIB setup, these are exactly the class
of black holes that were studied in \ref\stromva{A.~Strominger and C.~Vafa, 
``Microscopic  Origin of the Bekenstein-Hawking Entropy,''
Phys. Lett. $\us{B379}$ (1996) 99, hep-th/9601029. }.
Let us consider the limit where the $K3$ is small.  In this limit we have
an effective leftover $1+1$ dimensional worldvolume of the $D3$ brane
which is a supersymmetric
sigma model on ${\rm Sym}^N(K3)$. Thus the moduli
of D3 branes in the class $C$ of $K3$ with $C^2=2N-2$ is given
by ${\rm Sym}^N(K3)\times R^4$, where the extra $R^4$
comes from the position of the $D3$ brane in the rest of the space. 
Thus the low energy dynamics of the 1+1 dimensional
effective theory is a supersymmetric sigma model on this space.  
We are considering
the spatial direction to be wrapped over an $S^1$ and we are looking
for states which preserve $1/4$ supersymmetry in the full theory,
which correspond to states which preserve $1/2$ supersymmetry
in the $1+1$ dimensional sigma model. These
come from states with purely left-moving oscillator excitations,
and restricting to the ground states of the right-moving Hilbert space.
Moreover if we are interested in M2 brane BPS states
with wrapping number $M$ around $T^2$, we should look
at left-moving oscillator excitation $M$.  
The $SU(2)_L\times SU(2)_R$ symmetry is realized in the Hilbert
space of the sigma model by a left-moving current algebra
which realizes $SU(2)_L$ and a right-moving current algebra
realizing $SU(2)_R$ \ref\spinbl{J.C.\ Breckenridge, R.C.\ Myers, A.W.\ Peet, C. 
Vafa, ``D-Branes and  Spinning Black Holes'', Phys. Lett. $\us{B391}$ (1997) 93, 
hep-th/9602065.},
as far as the degrees of freedom is concerned on the symmetric
product of $K3$'s.  However for the $R^4$ factor, the $SU(2)_L$ current
algebra is a right-moving current and the $SU(2)_R$ current
is a left-mover\foot{The reason for this switch, relative
to the $K3$ degrees of freedom is that in the gauge theory
language in 6 dimensions (as in the dual $D1$--$D5$ brane dual systems), 
$K3$ degrees of freedom come from the
Higgs branch and the $R^4$ from the Coulomb branch, and the fermions
in these two multiplets have opposite 6-dimensional chiralities.  This
translates to the statement that the $SO(2)$ and $SO(4)$ chiralities
are oppositely correlated in the two cases.  This in particular
means that the left-moving fermions do not carry any $SU(2)_L$
quantum number for the $R^4$ sigma models.  Of course the bosonic
oscillators are vectors and do carry both quantum numbers.}.
Summing over the $R$ states with $(-1)^{F_R}$ is 
{\it exactly} the definition of the elliptic genus of
the sigma model on ${\rm Sym}^N(K3)\times R^4$.  It was in fact the elliptic genus
that was used to verify the predictions for black hole entropy
\stromva \spinbl !
So in this case we see that the left/right asymmetric treatment
of the $SU(2)_L\times SU(2)_R$ is exactly the same as the useful
elliptic genus index for supersymmetric sigma models, which has the
required stability under deformations which allows one to
predict at least a lower bound for the
black hole entropy.  The spinning quantum number of the black hole
gets mapped in this case to the $J^3_L$ quantum number.  We will
use this link with black holes in the next section to connect
the growth of $n_d^\gg$ with predictions based on black hole entropy
for M-theory compactification on a general Calabi-Yau 3-fold.

Due to interest from black hole physics the elliptic
genus of symmetric products of $K3$ has been computed, including
the modification due to the $J^3_L$ quantum number
\ref\dmvv{R.~Dijkgraaf, G.~Moore, E.~Verlinde and H.~Verlinde, ``Elliptic Genus 
of Symmetric 
Products and Second quantized Strings'', Commun. Math. Phys. $\us {185}$ 
(1997) 197-209, hep-th/9608096.}.
Let us denote the elliptic genus of a general manifold $M$   
\eqn\elldef{\chi(M;q,y)={\rm Tr}_{{\cal H}(M)}(-1)^F y^{F_L} q^H=
\sum_{M\ge 0,l\in Z} c(M,l) q^M y^l}
then the orbifold construction of \dmvv\  gives
for  the $N$-fold symmetric product ${\rm Sym}^N(M)$
\eqn\ellsymprod{
\sum_{N=0}^\infty p^N \chi(S^N K3;q,y)=\prod_{N>0,M\ge 0, l\in Z}{1\over (1-p^N 
q^M y^l)^{c(N\cdot M,l)}}\ . }

The elliptic genus  of the K3 can be easily made concrete from an 
orbifold representation of the $K3$ (e.g. $K3=T4/Z_2$) or a Landau-Ginzburg 
description as
$$ \chi(K3;q,y)= 24\left(\theta_3(q,y)\over \theta_3(q)\right)^2-2 
{\theta_4^4(q)-\theta_2^4(q)\over \eta^4}
\left(\theta_1(q,y)\over \eta \right)^2\ .$$
We note the first terms of this expansion
$$q^0:{\displaystyle \frac {2}{y}}+ \ 20 + 2\,y=2\left[1\over 2\right] +20 
\left[0\right], \quad q^1:\ 
{\displaystyle \frac {20}{y^{2}}} - {\displaystyle \frac {128}{y}}    + 216  - 
128\,y +  20\,y^{2} \ .  $$
Clearly the $q^0$ part of \ellsymprod\ gives \fullfreeenergy\ back. So far we 
ignored the 
fact that the momentum can be distributed in the space time directions as well. 
Taking this into account
(and noting as discussed before that the $SU(2)_L$ current algebra is
not active as far as the left-moving fermions from the $R^4$ part
are concerned) we get for the single cover contribution
\eqn\freenenergyII{\eqalign{
F^{(1)}&=\left(2 \sin {\lambda\over 2}\right)^{-2} \prod_{{{k>0 n>0} 
\atop {m\ge 0, l\in Z} }}{(1-q^k)^4 \over (1- e^{i \lambda} q^k)^2 
(1-e^{-i \lambda} q^k)^2} {1\over (1-p^n q^m e^{i l \lambda})^{c(n\cdot m,l)}}\cr 
&=:\left(2 \sin {\lambda\over 2}\right)^{-2}\sum_{N\ge 0,M\ge 0,L\in Z} 
{c_{N,M,L}} \ 
p^N q^M e^{iL \lambda}}\ . }
As with \fullfreeenergy, we can sum over all multicoverings $F^{(m)}$ 
to obtain the free energy. Moreover we may multiply by a lattice 
theta functions if we wish to exhibit the sum over classes 
in $H^2(K3,Z)$ .  

Let us recapitulate the dictionary which relates this to the counting
of $M2$ branes of M-theory in $K3\times T^2$. $N$ corresponds to the
genus $g$ of a smooth curve in the $K3$, which after a suitable
complex structure deformation can be always thought as a curve $C$
with degree $N$ with respect to an elliptic K3's fiber class which has
degree $1$ in the base class.  The powers $M$ correspond to the degree
with respect to the $T^2$. The powers $L$ (note the $y\rightarrow 1/y$
symmetry) correspond to the number of nodes $\delta$, and for a given
class $[M,N]$ we have $\delta=N+M-L$.

Let us consider the case $\delta=0$ first. Geometrically the moduli
space of these curves is modeled by ${\cal C}^{(M)}_{[N]}$, the degree
$M$ relative Hilbert scheme of the universal curve. To see this, note
that a general curve of this type consists of a sort of ``comb'', the
union of a single curve $C$ on the K3\foot{We have adjusted the
complex structure of the K3 away from the elliptically fibered
structure, so instead of $C$ being the base plus $g$ fibers, the curve
$C$ is typically irreducible.}  and $M$ copies of $T^2$. The only
moduli is where on $C$ to attach the copies of $T^2$.  This is given
by specifying the $M$ points on $C$, and the total moduli is
$\cC^{(M)}_{[N]}$, where $\cC_{[N]}$ is now the universal curve
of genus $N$ on the K3 (i.e.\ there is no $T^2$ being considered at
all).

By the general formula, $n^\ag_{[M,N]}=(-1)^{\dim {\cal M}}e({\cal
M})$, the arithmetic genus $\ag$ being $M+N$. 
The $c_{N,M,M+N}$ give the Euler number $e({\cal
C}^{(M)}_{[N]})$ at least for the case that ${\cal C}^{(M)}_{[N]}$ is
smooth and the relevant class for the obstruction theory is in the
cotangent bundle of ${\cal M}$. If ${\cal C}^{(M)}_{[N]}$ is singular
then the physical prediction $c_{N,M,M+N}$ should calculate an
integral over a suitable, but not yet understood, virtual fundamental
class, the result of which we may call in a slight abuse of notion
also $e({\cal C}^{(M)}_{[N]})$.  The important point here is that we
find that the relations between the $n_d^\gg$ and the so defined
$e({\cal C}^{(M)}_{[d]})$ are exactly reproduced by our general
formula
\generalformng !

Let us list the predictions for the values for the $e({\cal C}^{(M)}_{[N]})=
n^\gg_{[M,N]}=c_{N,M,M+N}$ and compare with the direct computation in several
cases.
       
\smallskip

{\vbox{\ninepoint{
$$
\vbox{\offinterlineskip\tabskip=0pt
\halign{\strut
\vrule#
&
&\hfil ~$#$
&\hfil ~$#$
&\hfil ~$#$
&\hfil ~$#$ 
&\hfil ~$#$
&\hfil ~$#$ 
&\hfil ~$#$
&\hfil ~$#$
&\vrule#\cr
\noalign{\hrule}
&n^g_{N,M}
&M=0
&  1
&  2
&  3
&  4
&  5
&  6
&\cr
\noalign{\hrule}
&N=0
&1
&-2
&3
&-4
&5
&-6
&7
&\cr
&1
&-2
&24
&-48
&72
&-96
&120
&-144
&\cr
&2
&3
&-48
&327
&-648
&972
&-1296
&1620
&\cr
&3
&-4
&72
&-648
&3272
&-6404
&9600
&-12800
&\cr
&4
&5
&-96
&972
&-6404
&26622
&-51396
&76955
&\cr
&5
&-6
&120
&-1296
&9600
&-51396
&185856
&-353808
&\cr
&6
&7
&-144
&1620
&-12800
&76955
&-353808
&1150675
&\cr
\noalign{\hrule}}\hrule}$$
\vskip -3 pt
\centerline{{\bf Table 2:} The weighted sum of BPS states $n^g_{N,M}$ 
for classes in $K3\times T^2$ with $\delta=0$.}
\vskip7pt}}}

These moduli spaces are the relative Hilbert schemes of
the K3.  The Euler characteristics of these relative Hilbert schemes
can be sometimes computed directly, and are in agreement when smooth.  
As usual, we have $\cM=\IP^g$. As for the computation of $\cC^{(M)}$, 
this is our standard method.  We
consider the map $\rho:\cC^{(M)}\to {\rm Hilb}^M(K3)$.  If this is a
projective bundle then we can immediately compute $e(\cC^{(M)})$.  If
not, then we analyze where it fails to be a projective bundle and
correct as appropriate.

Letting $\cC$ be the universal curve on a general K3 of genus $N$,
we have seen from the comb description that $\cM=\hc{M}$, leading
to $n^\ag_{[M,N]}=e(\hc{M})$ when $\hc{M}$ is smooth.  In particular,
this occurs if $N$ is large, with the asymptotic formula
$$n^\ag_{[M,N]}=(-1)^{N+M}(N+1-M)e({\rm Hilb}^M(K3)),\qquad N>>0 $$
which is consistent with the appearance of the columns of the above table.
The deviations from this formula occur when there is not a projective bundle
structure for $\hc{M}\to {\rm Hilb}^M(K3)$.  We have already explained this
for $M=N=2$ and for $M=N=3$.

Let's look at $N=4$ and $M=3$.  We choose the complex structure to be
that of $S=\IP^4[2,3]$, the intersection of a degree 2 hypersurface
$Q$ and a degree 3 hypersurface $T$.  The curves of genus 4 are the
hyperplane sections.  By dimension reasons, we might expect the fibers
of $\cC^{(3)}\to {\rm Hilb}^3(S)$ to be $\IP^1$, but if the 3 points
happen to lie on a straight line, there is a $\IP^2$ of hyperplanes
through the line, giving a fiber of $\IP^2$ rather than $\IP^1$.
Let's call this locus $B\subset {\rm Hilb}^3(S)$. Now, if 3 points of
$S$ lie on a line $\ell$, then the quadratic equation defining $Q$ has
at least 3 zeroes on the line $\ell$, hence vanishes identically.
Thus $\ell\subset Q$.  Conversely, any line contained in $Q$ will meet
$S$ in 3 points (the three points where the line meets $T$).  This
shows that $B$ is the set of lines contained in $Q$.  So $\hc3$ is a
$\IP^1$ bundle over ${\rm Hilb}^3(S)$, except over $B$ where it is a
$\IP^2$ bundle.  This gives $e(\hc3)=2\cdot e({\rm Hilb}^3(S))+e(B)$.
We can use standard techniques in algebraic geometry
\ref\altklei{A.~Altman and S.~Kleiman, ``Foundations of the theory of
Fano schemes'', Comp.\ Math.\ $\us{34}$ (1977) 3.} and the Schubert software
\ref\schubert{S.~Katz and S.A.~Str\o mme, Schubert, a Maple package
for intersection theory in algebraic geometry,
http://www.math.okstate.edu/\char 126katz/schubert.html\ .} to compute
$e(B)=4$.  This gives the invariant 6404, in agreement with the table.

We can also check the $N=1$ row.  We have by \topgenus\
$n^{M+1}_{[1,M]}= (-1)^{M+1}e(\hc{M})$, where $\cC$ is the universal
curve of the elliptic fibration.  If $C$ is a smooth fiber, then the
corresponding fiber of $\hc{M}$ is the $M^{\scriptstyle{\rm th}}$
symmetric product of $C$, which has Euler characteristic 0.  So it is
only the 24 singular fibers that contribute to $e(\hc{M})$.  Each of
these singular fibers $C$ has exactly one node, so $e({\rm Hilb}^M(C))=M$
by the discussion in the paragraph following \nthree.  This gives
$e(\hc{M})=24M$, or $n^{M+1}_{[1,M]}=(-1)^{M+1}24M$, in agreement with the
table.

Note that the symmetry of the table is immediate from duality and our 
calculations, but the geometric content of this symmetry is non-trivial,
even in identifying the $N=1$ row with the $M=1$ column.

\smallskip
We can also consider the more difficult situation where we put nodes 
$\delta=N+M-L>0$. Some invariants for $\delta=1$ are given in the table below.

{\vbox{\ninepoint{
$$
\vbox{\offinterlineskip\tabskip=0pt
\halign{\strut
\vrule#
&
&\hfil ~$#$
&\hfil ~$#$
&\hfil ~$#$
&\hfil ~$#$ 
&\hfil ~$#$
&\hfil ~$#$ 
&\hfil ~$#$
&\hfil ~$#$
&\vrule#\cr
\noalign{\hrule}
&n^{g-1}_{N,M}
&M=0
&  1
&  2
&  3
&  4
&  5
&  6
&\cr
\noalign{\hrule}
&N=0
&0
&0
&-6
&16
&-30
&48
&-70
&\cr
&1
&24
&0
&216
&-432
&768
&-1200
&1728
&\cr
&2
&-54
&720
&-162
&5712
&-9768
&15552
&-22680
&\cr
&3
&88
&-1488
&11832
&-4368
&83496
&-135456
&204872
&\cr
&4
&-126
&2376
&-23016
&138696
&-65112
&884184
&-1398486
&\cr
&5
&168
&-3360
&36000
&-258048
&1292712
&-701856
&7546536
&\cr
&6
&-214
&4440
&-50328
&396392
&-2324790
&10160160
&-6086258
&\cr
\noalign{\hrule}}\hrule}$$
\vskip -3 pt
\centerline{{\bf Table 3:} The weighted sum of BPS states $n^{g-1}_{N,M}$ 
for classes in $K3\times T^2$.}
\vskip7pt}}} 

Note that the $M=0$ column reproduces the K3 results found earlier.

\newsec{Black Hole Entropy and Topological Strings}
As discussed in the previous section in the context
of M-theory compactification on $K3\times T^2$, the
spectrum of M2 branes wrapped around various cycles
for large enough cycle classes would physically correspond
to black holes.    Moreover the 
$SU(2)_L$ content of  the BPS state corresponds to the spin
of the black hole. In such a case the growth of
wrapped M2 branes in a given $H_2$ class and with given spin
is anticipated by the Bekenstein-Hawking entropy
of macroscopic black holes, which has been verified.
Moreover, there are index-like quantities appearing in topological
string theories, computed as a
sum over the right-moving states with
$(-1)^{F_R}$,  which in the type IIB setup is perturbation invariant
and is related to the computation of the elliptic genus.
These are exactly the kind of computations which yield
black hole entropy in the regimes where string perturbation
breaks down.

It is thus natural to expect that this relation between BPS states in
M-theory compactification on a general Calabi-Yau threefold continue to hold,
namely the growth of the left BPS degeneracy
with fixed $SU(2)_L $ content but summed over right $SU(2)_R$ quantum
numbers with $(-1)^{F_R}$ will also yield the black hole entropy.
In what we shall write here we will not worry about
numerical constants in the formula for the growth of black
hole entropy.

If we consider the macroscopic prediction for black hole entropy
coming from a large charge $d>>1$ and for a given $J_L^3=m$ spin
one obtains \spinbl, in the regime
$d>>m$ and $d>>1$,
\eqn\macp{N_{d,m}\sim {\rm exp}\sqrt{d^3-m^2}}
This is meant to convey the exponential growth and its dependence
in uniform rescaling of $d$ and $m$, and is valid up to numerical
coefficients in the exponent. Also there may be a 
sub-leading power law correction
prefactor in front of the exponent.

One can try to compare this macroscopic prediction with the
microscopic prediction.  Since the $SU(2)_L$ content
of the black hole entropy is captured by the numbers $n_d^\gg$, we can compare
with the total number of BPS states with charge $d$ and with $J^3_L$ spin
$m$.  All we need to do is to recall that the full
representation of the BPS states with charge $d$
is given by
\eqn\reprs{R=[\sum_{\gg=0}^g n_d^\gg I_\gg]\otimes I_1}
 This sum is finite, as for a given degree there is a maximum
 genus curve which realizes it. Moreover if we consider the $I_g$ content
 of the state we have
$$Tr_{I_k} y^{J_L^3}=(y^{1\over 2}+y^{-1\over 2})^{2k}$$
This in particular means that the number of states in $I_k$
with $J^3_L=m$ spin is given by
$$\left(2 k\atop k+m\right)$$
Applying this to \reprs\ we see that the number $N_{d,m}$ is given by
$$N_{d,m}=\sum_\gg n_d^\gg \left(2 \gg+2\atop \gg+1+m\right)$$
Comparing this with the growth expected from
macroscopic considerations \macp , we get the prediction
in the limit $d>>1, d>>m$
\eqn\finmmp{\sum_\gg n_d^\gg \left(2 \gg+2\atop \gg+1+m\right)
\sim {\rm exp}\sqrt{d^3-m^2}}
This is a very delicate sum, in that $n_d^\gg$ are
typically very big numbers for $d$ large, which alternate
in sign when one changes $g$ by one unit, as we have seen
in many examples in this paper.  It is known that
$n_d^\gg$ for a fixed $\gg$ grows with large $d$ as ${\rm exp}(d)$.
This however, is not in contradiction with the prediction
\finmmp\ as one is summing over all non-vanishing $\gg$, and for large $d$'s
the allowed range in $\gg$ is also large.  In other words
one is considering a different region of parameters and the equation
\finmmp\ is a new prediction of the growth of these numbers in a different
direction.
It would be interesting to verify them (some special cases
of this formula for Calabi-Yau threefold has been verified in
\ref\blackcy{C.\ Vafa, Black Holes and Calabi-Yau Threefolds'', 
Adv. Theor. Math. Phys. $\us {2}$ (1998) 207, hep-th/9711067. }\ 
in the context of elliptic 3-folds).

\newsec{Computations in local Calabi-Yau geometries}

In this section we will apply the techniques developed in sections 4
and 5 to local Calabi-Yau geometries.  By a local Calabi-Yau model we
mean the total space ${\cal O} (K_B) \rightarrow B$ of the canonical
line bundle fibered over a (two) dimensional Fano\foot{The condition
for $B$ to be Fano can be relaxed, see \ckyz .}  variety $B$. For the
base $B$ we discuss here del Pezzo surfaces $\IP^2$, $\IP^1\times
\IP^1$ and $E_n$, the blow up of $\IP^2$ in $n=1,\ldots, 8$
points\foot{Quantum intersection rings on these surfaces were
discussed in \ref\dfi{M.~Kontsevich and Y.~Manin, ``Gromov-Witten
classes, Quantum Cohomology and enumerative Geometry'',
Commun. Math. Phys. $\us{164}$ (1994) 525--562; hep-th/9402147, 
P.\ Di Francesco and C.\ Itzykson, ``Quantum intersections rings'', 
hep-th/9412175, C.\ Itzykson, 
``Counting Rational Curves on Rational Surfaces'',
Int. J. Mod. Phys. $\us {B8} $ (1994) 3703--3724}. In these cases the
dimension of the moduli space of the curves is reduced to zero by
requiring the curves to go through fixed points. There is no obvious
relation between these invariants and the ones we calculate and relate
to M-theory BPS states
\ref\nekrasov{N.\ Nekrasov, ``In the Woods of M-theory'', hep-th/9810168.}.}. 
To calculate the invariants for a curve with $\delta$ nodes we have to 
calculate $e({\cal C}^{(k)})$ for $k\le\delta$, 
which is easy when there is a bundle 
structure $\rho_k:{\cal C}^{(k)}\rightarrow {\rm Hilb}^{(k)}(B)$. 
For a given degree $d$ we find in general a bound $\delta<\delta_{max}(d)$ for 
which this is true. For instance, on $\IP^2$ the first restriction comes from 
$\delta_{max}(6)=8$. In general one can show that in this case 
$\delta_{max}(d)=d+2$. Another complication arises when the curves are 
reducible. However we found a systematic procedure to account for the 
corrections due to reducible curves, which gives the expected answer 
whenever $e({\cal C}^{(k)})$ could be calculated using the 
bundle structure. The information from these calculations were sufficient 
to fix the $B$-model ambiguity up to genus $g=4,5$ in the examples. 
When there is no bundle structure for ${\cal C}^{(k)}$ the modeling
of these spaces becomes more complicated. While the bundle structure
guarantees smoothness, we find in examples\foot{$\hc{8}$ at $d=6$ and
$\hc{9}$ at $d=7$ on $\IP^2$ to be discussed below.} that some of the
spaces lacking the bundle structure are singular. Hence the Euler
number $e(\hc{k})$ has to be replaced by an integral
$\int_{[\hc{k}]_{vir}} c_{top}$ over a suitable virtual fundamental
class. We have not attempted to define this virtual fundamental
class. However in $K3\times T^2$ case the quantity
$\int_{[\hc{k}]_{vir}} c$ appeared in two places in our calculation
fitting exactly
our approximation of $H^*(\cJ(C))$ as expressed in $\hc{k}$ by \generalformng .
We are therefore optimistic that there is natural definition 
of $\int_{[\hc{k}]^{vir}} c_{top}$ for the non-smooth $\hc{k}$ in the 
general case, which calculates the $n^\gg_d$ invariants by the  
formulas in sections 4,5. 

\subsec{Basic concepts}   
First we need to calculate the Euler number of the moduli space $e({\cal M})$ 
for the non-degenerate curves $C$, which are embedded in the surface $B$. 
More precisely, these curves are characterized by the fact that their 
geometric genus and their arithmetic genus coincide. 
We obtain the dimension of the deformation space  $\IP H^0({\cal O}(C))$ from 
$\chi({\cal O}(C))=\sum_{i=0}^n (-1)^i h^i({\cal O}(C))$ under the assumption 
that $H^1({\cal O}(C))=H^2({\cal O}(C))=0$. This is true 
on del Pezzo surfaces with general moduli \ref\harb{B.\ Harbourne,
``Rational surfaces with $K^2>0$'', Proc.\ AMS $\us {124}$ (1996) 727--733.}. 
Then, using adjunction 
$$C^2+K C=(2 g-2)$$ 
and the Riemann-Roch formula 
$$\chi({\cal O}(C))={C^2-K C\over 2}+1,$$ 
we obtain $\chi({\cal O}(C))=g+d$, where $d$ is the degree of the curve with
respect to the anticanonical class $-K$ of $B$. 
Since the moduli space is obtained by projectivizing $H^0({\cal O}(C))$, 
we get as the result 
\eqn\em{e({\cal M})=e(\IP^{g+d-1})=(g+d).}

To be concrete, recall that the classes on a $E_n$ del Pezzo are 
the $\IP^2$ hyperplane class $H$ and the exceptional divisors of the blowups 
$e_i$ in $n$ points $p_1,\ldots, p_n$, with $H^2=1$ and $e_i
e_j=-\delta_{ij}$ (for $\IP^1\times \IP^1$ see the later discussion).
A curve of multidegree $(a;b_1,\ldots,b_n)$ refers to the class of
curves obtained from blowing up the plane curves of degree $a$ which
pass through the point $p_i$ with multiplicity $b_i$.  These curves have
class $aH-\sum b_iE_i$.  We will
typically rearrange the order of points to write the $b_i$ in
nonincreasing order.  We will also omit the $b_i$ which are 0.  And we
will use exponential notation $b^k$ to refer to a subsequence of $k$
copies of $b$.  The exceptional divisors $e_i$ do not fit into this
classification scheme and will be denoted separately. To follow the
subsequent discussion one needs an overview over the low degree
classes in the del Pezzos surfaces, which we provide in Appendix
A. They are ordered with respect to their arithmetic genus
\eqn\arithmeticgenus{g={(a-1)(a-2)\over 2}-\sum_{i=1}^n {b_i(b_i-1)\over 2}}
and their degree with respect to the anticanonical class of the del Pezzos 
$-K_{E_n}=3H-\sum_{i=1}^n e_i$
\eqn\degree{d=3 a-\sum_{i=1}^n b_i\ . }  

Let us consider as an easy example the $E_1$ del Pezzo surface. 
As it was observed by mirror symmetry in \ref\kkv{
S.\ Katz, A.\ Klemm and C.\ Vafa, ``Geometric Engineering of quantum Field Theories,'' 
Nucl.Phys. $\us{ B497}$ (1997) 173-195, hep-th/9609239.} 
there are invariants\foot{Which resum to the logarithm capturing the running of 
the gauge coupling in the N=2 gauge theory at the appropriate locus in the moduli 
space.} $n^0_{1,a}=2 a+1$ from curves wrapping $1$ times the base $H$ and $a$ 
times the fibre $H-e_1$ 
of the Hirzebruch surface $F_1=E_1$, i.e. since $a=H C$ and $1=(H-e_1) C$ these 
genus zero curves are in the class $(a;a-1)$ and by 
\degree\em\topgenus\ we immediately obtain the mirror symmetry prediction.     

To calculate the invariants for curves with one node $\delta=1$ we need next 
to determine the Euler number of the universal curve $e({\cal C})$. We first fix 
a class $[C]$ with a curve $C$ of arithmetic genus $g$ and degree $d$ in it. Then 
we calculate the contribution of a nodal curve in $[C]$ to the invariant 
$n_d^{g-1}$ at genus $g-1$ by \generalformng\ or \smallnodes\topgenus . 
In the simplest situation the nodal curve is irreducible. 
The space ${\cal C}$, comes with a fibration structure as follows. Fixing the 
location of a point on the curve gives a linear constraint in the moduli space 
$\IP^{g+d-1}$ of the genus $g$ curve. As the point is free to vary over $B$, 
$\cal C$ comes with a natural fibration $\cC\rightarrow B$ 
which, if it is smooth, gives rise to the Euler number 
\eqn\ec{e({\cal C})=e(\IP^{g+d-2}) e(B) = (g+d-1) e(B)\ . }      
In general if there is the bundle structure $\rho_k:{\cal C}^{(k)}
\rightarrow {\rm Hilb}^{(k)}(B)$ then
\eqn\eck{e({\cal C}^{(k)})= e(\IP^{g+d-1-k}) e({\rm Hilb}^{(k)}(B))  }
with $\sum_k e({\rm Hilb}^{(k)}(B))q^k=\prod_n {1\over (1-q^n)^{e(B)}}$.
The easiest example for the treatment of reducible curves is 
the calculation of  $n_4^0$ in $\IP^2$ in section 5.2 (after eq. 5.7). 
In the case of general del Pezzo surfaces we have to sum over 
various classes as will be explained in Section 8.5 for the case 
of the $E_5$ del Pezzo when we  calculate $n^0_4$.

\subsec{${\cal O}(-1)\oplus {\cal O}(-1)\rightarrow {\bf P}^1$ }

The conifold geometry does not support higher genus curves and the
only rational curve that exists is the ${\bf P}^1$. Therefore the
$F_g$ of the type IIA topological theory\foot{Not to be confused with
the IIB asymptotic behavior at the locus where an $S^3$ shrinks,
which we discuss in Sections~2,5,6.}  are completely governed by the
multicover and bubbling contribution of this one rational curve. In
this sense this simplest Calabi-Yau background to which 2-d gravity
can be coupled is completely solved by \buble~in full accordance to
\allg .  It is interesting that in this case the part of the $F_g$
which follows from the anomaly can be made vanishing in the
holomorphic limit by noting that there is a gauge choice in which all
propagators of Kodaira-Spencer become identically zero, compare \kz .
In this sense all holomorphic information here actually comes from the
ambiguity.

\subsec{  Local ${\bf P}^2$: ${\cal O}(-3)\rightarrow {\bf P}^2$ }

This is the next simplest case. This geometry supports an infinite 
number of curves of different arithmetic genera, making the
geometry much more interesting than in the conifold case.

We first present the result for the invariants. The numbers in the Table~4 
marked 
with a diamond were obtained with the technique discussed in sections 4 and
5, as explained in detail below. 
Once we can fix the holomorphic ambiguity completely for a genus $\gg$, which 
was possible up to $\gg=5$ in this case (see sect 8.6), 
the B-model gives an immediate answer for $F_\gg$ at all degree $d$. 
Using this $B$-model result integrality of the $n_d^\gg$ was checked up to $d=300$. 
For $\gg=6,7,8$  we could determine the ambiguity only up to $S=\{2,4,6\}$ 
constants, 
which can be parameterized by the $n^\gg_d$, $d=8,\ldots 8+S_i$. Assuming that they 
are integer we checked that the $B$-model gives rise to integer $n_d^\gg$ up to 
$d=300$.
As a further check we compared with the calculation in \kz , which uses the $c=1$  
KDV hierarchy and direct localization techniques. Numbers marked with a 
star were calculated this way. 

The $\gg=1$ numbers listed in Table 4 follow the geometric subtraction 
scheme 
\geomsubstraction . The differences in genus 1 between the geometric and the 
physical 
subtraction schemes $\Delta_d=(n^1_d-n^{*1}_d)$ are 
$\Delta_6=-10$, $\Delta_8=231$,
$\Delta_9=-10$ and
$\Delta_{10}=4452$ up to given order. 

\vskip 5 mm
{\vbox{\ninepoint{
$$
\vbox{\offinterlineskip\tabskip=0pt
\halign{\strut
\vrule#
&
&\hfil ~$#$
&\hfil ~$#$
&\hfil ~$#$
&\hfil ~$#$ 
&\hfil ~$#$
&\hfil ~$#$ 
&\hfil ~$#$
&\vrule#\cr
\noalign{\hrule}
&d
&\gg=0
&1
&2
&3
&4
&5
&\cr
\noalign{\hrule}
&1
&3_*^\diamond
&0
&0
&0
&0
&0
&\cr
&2
&-6_*^\diamond
&0
&0
&0
&0
&0
&\cr
&3
&27_*^\diamond
&-10_*^\diamond
&0
&0
&0
&0
&\cr
&4
&-192_*^\diamond
&231_*
&-102_*^\diamond
&15_*^\diamond
&0
&0
&\cr
&5
&1695_*^\diamond
&-4452_*^\diamond
&5430_*^\diamond
&-3672_*^\diamond
&1386_*^\diamond
&-270_*^\diamond
&\cr
&6
&-17064_*
&80948_*
&-194022_*
&290853_*^\diamond
&-290400_*^\diamond
&196857^\diamond
&\cr
&7
&188454
&-1438086
&5784837
&-15363990
&29056614
&-40492272
&\cr
&8
& -2228160
&25301295
&-155322234 
& 649358826
&-2003386626
&4741754985 
&\cr
&9
& 27748899
&-443384578
&3894455457
&-23769907110
&109496290149
&-396521732268 
&\cr
&10
&\! \! -360012150 \! \!
&\! \! 7760515332 \! \!
&\! \! -93050366010 \! \!
&\! \! 786400843911 \! \!
&\! \! -5094944994204 \! \!
&\! \!  26383404443193 \!\!
&\cr
\noalign{\hrule}}\hrule}$$}}}
\vskip - 8 mm
{\vbox{\ninepoint{
$$
\vbox{\offinterlineskip\tabskip=0pt
\halign{\strut
\vrule#&
&\hfil ~$#$
&\hfil ~$#$
&\hfil ~$#$
&\hfil ~$#$
&\hfil ~$#$
&\hfil ~$#$
&\hfil ~$#$
&\hfil ~$#$
&\hfil ~$#$
&\hfil ~$#$
&\hfil ~$#$
&\vrule#
\cr
\noalign{\hrule}
&d
&\gg=6
&7
&8
&9
&10
& \! 11 \!
& \! 12  \!
& \!13 \!
& \!14  \!
& \!15  \!
&\cr
\noalign{\hrule}
&1
&0
&0
&0
&0
&0
&0
&0
&0
&0
&0
&\cr
&\vdots
&
&\vdots
&
&\vdots
&
&\vdots
&
&\vdots
&
&\vdots
&\cr
&4
&0
&0
&0
&0
&0
&0
&0
&0
&0
&0
&\cr
&5
&21^\diamond
&0
&0
&0
&0
&0
&0
&0
&0
&0
&\cr
&6
&-90390^\diamond
&27538^\diamond
&-5310^\diamond
&585^\diamond
&-28^\diamond
&0
&0
&0
&0
&0
&\cr
&7
& \!\! 42297741^\diamond \!\!
& \!\!-33388020^\diamond \!\!
& \!\! 19956296^\diamond \!\!
& \!\!-9001908^\diamond \!\!
& \!\! 3035271^\diamond \!\!
& \!\!-751218^\diamond \!\!
& \!\! 132210^\diamond \!\!
& \!\!-15636^\diamond \!\!
& \!\! 1113^\diamond \!\!
& \!\!-36^\diamond \!\!
&\cr
\noalign{\hrule}}\hrule}$$
}
\vskip - 9 mm
\centerline{{\bf Table 4:} The weighted sum of BPS states $n^\gg_d$ for the local 
${\bf 
P}^2$ case. ($d$ is the degree w.r.t. $H$)}
\vskip7pt}

We now turn to a more detailed discussion of our methods outlined 
in Sections~4 and 5.  We will give the results in terms of the degree $d$ 
to the extent possible.

Since the degree $d$ plane curves have moduli space $\cM=\IP^{d(d+3)/2}$,
we have from \topgenus\ that
$$
n_d^{(d-1)(d-2)/2}=(-1)^{d(d+3)/2}(d+1)(d+2)/2.
$$
This is in complete agreement with Table 4.

Note that if $d\ge3$, then any reducible curves must have at least 2~nodes
or worse singularities.  This is seen
by considering the ways that a degree $d$ curve can split into curves
of degrees $d_i$ with $\sum d_i=d$, together with the
observation that general plane curves of degrees $d_i$ and $d_j$ meet in
$d_id_j$ points.  Therefore we only need the first term on the right hand side of 
\bestcorrection\ to compute instanton numbers with $\delta=1$.

Note that the fiber of
$\cC\to\IP^2$ is the set of degree $d$ plane curves containing $p$,
which is a $\IP^{d(d+3)-1}$.  Thus
$e(\cC)=e(\IP^2)e(\IP^{d(d+3)/2-1})$, or $3d(d+3)/2$.  Using the
first equation of \smallnodes\ and \bestcorrection\ together with
$g=(d-1)(d-2)/2$, we get a formula 
$$n_d^{d(d-3)/2}=(-1)^{d(d+3)/2}{d\choose2}(d^2+d-3),$$ 
which agrees with the results found for
$d=3,4,5,6$ in Table~4 above.

As our next check for $\IP^2$, we observe as before that for $d\ge4$, all 
reducible
curves have at least 3 nodes.  So as above, the calculation simplifies
for $\delta=2$ and $d\ge4$ by requiring only the first term on the
right hand side of \bestcorrection.

Consider the map $\rho_2:\hc2\to {\rm Hilb}^2
\IP^2$.  The fiber of $\rho_2$ over $Z\in{\rm Hilb}^2\IP^2$ is the set of plane 
curves
of degree $d$ containing $Z\subset\IP^2$.  Either $Z=\{p,q\}$ or $Z$
consists of a point $p\in \IP^2$ and a tangent direction at $p$.  Either
way, we see that it is two conditions on a plane curve to contain $Z$
(to say that $C$ contains $Z=(p,v)$ with $v\in T_p\IP^2$ means that
$p\in C$ and $v$ is tangent to $C$ at $p$).  Thus the fiber is a
$\IP^{d(d+3)/2-2}$.  This gives 
$$\eqalign{
e(\hc2)&=e(\IP^{d(d+3)/2-2})e({\rm Hilb}^2\IP^2)\cr
&=9\left({d(d+3)\over 2}-1\right).}$$
We get from the second equation of \smallnodes\ and \bestcorrection
$$\eqalign{n_d^{(d^2-3d-2)/2}&=
(-1)^{d(d+3)/2}\left(
9\left({d(d+3)\over2}-1\right)
+(d^2-3d-2)\cdot 3{d(d+3)\over2}\right.\cr
&\phantom{=}+\left.\left({d^2-3d\over2}\right)
(d^2-3d-3){d+2\choose2}\right).}$$
In particular, we verify that $n_4^1=231$, as noted earlier, 
as well as $n_5^4=1386$.

Continuing, we have that for $d\ge5$, all reducible curves have
at least 4~nodes, so the calculation for $\delta=3$ simplifies.  
We consider the map $\hc3\to{\rm Hilb}^3\IP^2$, and check that the
fiber is always a $\IP^{d(d+3)/2-3}$, yielding $e(\hc3)=22((d+4)(d-1)/2)$.
Using the third equation from \smallnodes\ and \bestcorrection, we
get
\eqn\tnp{\eqalign{n_d^{{d^2-3d-4\over 2}}&=
(-1)^{(d+1)(d+2)\over 2}\left(
22{(d+4)(d-1)\over2}+\left((d-1)(d-2)-6\right)
9\left({d(d+3)\over 2}-1\right)+\right.\cr
&\phantom{=}{\left((d-1)(d-2)-4\right)
\left((d-1)(d-2)-7\right)\over2}\left({3d(d+3)\over2}\right)+\cr
&\phantom{=}\left.{\left((d-1)(d-2)-2\right)\left((d-1)(d-2)-6\right)
\left((d-1)(d-2)-7\right)\over6}{d+2\choose2}
\right)}}
In particular, substituting $d=5,6$ into \tnp, we get
$n_5^3=-3672$ and $n_6^7=27538$, see Table~4.

We recall from Section~5 that 
we can also handle the situation where there are reducible 3~nodal curves,
namely $d=4$.  The problem is that we have the
locus of quartic curves which are unions of a cubic and a line.  
We saw how this leads to $n_4=-192$.

The case of $n_5$ is even more
interesting, in that there are two extra components
contained in $\overline{\cM}_6$.  The first component is the union of degree 2
and degree 3 curves, contributing $n_2^0n_3^1=60$, and the second component
is the union of
lines and degree 4 curves with 2 nodes, contributing $n_1^0n_4^1=693$.
According to \bestcorrection, we substitute $d=5$ in \allnodes\ with
$\delta=6$ and $g=6$ and subtract the contributions of 60 and 693, 
obtaining $n_5^0=1695$, as in Table~4.  The $e(\hc{k})$ can be computed as
above because in these cases the projection $\hc{k}\to{\rm Hilb}^k\IP^2$
is a bundle of projective spaces and $e({\rm Hilb}^k\IP^2)$ can be computed
by \hilbgen.

Using \allnodes\ and \bestcorrection, we can verify $n_6^\gg$ for $\gg\ge3$.
The projection $\hc{k}\to {\rm Hilb}^k\IP^2$ is again a bundle of
projective spaces for $k\le7$.  Recall that this bundle structure
implies that $\hc{k}$ is smooth, so that our method applies.
Thus we have a check of our method for
each $\delta\le7$.

This bundle structure does not occur in general.  In the computation
of $n_6^2$, we encounter the map $\hc8\to{\rm Hilb}^8\IP^2$.  Over a
general point of ${\rm Hilb}^8\IP^2$, the fiber is a
$\IP^{27-8}=\IP^{19}$, as a multiplicity 8 scheme usually imposes 8
conditions on curves of a given degree.  But now suppose the 8 points
all lie on a line $L$, and we identify the projective space of degree
6 curves containing these 8 points (some of which may coincide).  For
such a degree 6 curve, the restriction of its degree 6 equation to $L$
has 8 zeros, hence vanishes identically.  Thus the degree 6 equation
contains the equation of the line as a factor, and we can multiply it
by an arbitrary degree 5 polynomial and obtain a degree 6 polynomial
containing the desired 6 points.  Since the degree 5 polynomials form
a $\IP^{20}$, we get a fiber of $\IP^{20}$ rather than $\IP^{19}$ in
this case.

The lack of a bundle structure for $\rho_8:\hc8\to{\rm Hilb}^8\IP^2$
causes us to ask if our methods apply in this case.  We have to
see whether or not $\hc8$ is smooth.  Consider the projection
$\pi_8:\hc8\to\IP^{27}$ onto the other factor.  That is,
given an element $(C,Z)$ of $\hc8$, so that $C$ is a plane curve of
degree 6, and $Z$ is a multiplicity 8 scheme in $C$, i.e.\ an
element of ${\rm Hilb}^8C$, put $\pi_8(C,Z)=C$, where $C$ is now
identified with the corresponding element of $\cM=\IP^{27}$.
Since the fibers of $\pi_8$
are all 8 dimensional, we can see that $\hc8$ has dimension 
$27+8=35$.  To show that $\hc{8}$ is singular, we need only
exhibit a single element of $\hc8$ at which the tangent space of
$\hc8$ has dimension strictly greater than 35.  Since the tangent
space of $\hc8$ at $(C,Z)$ is naturally identified with the space of first order
deformations of the pair $(C,Z)$, we need only find a $(C,Z)$ for
which we can exhibit 36 independent first order deformations.

Towards this end, for $C$ we take a degree 6 curve $l(x_1,x_2,x_3)^2
f(x_1,x_2,x_3)=0$, where $l$ is linear and $f$ is homogeneous
of degree 4 in the homogeneous coordinates $(x_1,x_2,x_3)$ of
$\IP^2$.  For $Z$ we take any 8 points on the line
$l(x_1,x_2,x_3)=0$, which can possibly occur with 
multiplicity.  

Note first that we have 16 independent deformations obtained
by moving these 8 points arbitrarily in $\IP^2$ while
keeping $C$ fixed.  This is
because of the factor of $l^2$, which
ensures that all motions of the points stay within $C$ to
leading order.  Then we take the 20 deformations noted above,
where we fix $Z$, but now vary the degree 6 curve to an arbitrary
curve of the form $l(x,y,z)g(x,y,z)=0$, where $g$ is homogeneous
of degree 5.  These 20 deformations are in fact honest deformations, not
just first order deformations.  Combining these two classes of
first order deformations where we deform $C$ and $Z$ separately,
we have all together the needed $16+20=36$ deformations.

While our methods gave the correct results for singular moduli spaces
arising from isolated nodal curves in Section 4, we are not so
lucky this time.  The calculation based on \allnodes\ and \bestcorrection\
differs from the value of $n_6^2$ in Table 4 by 45.  We expect that
the singular locus of $\hc8$ provides a correction term to
\bestcorrection, much as the second term of \bestcorrection\ can be
viewed as providing a correction to \allnodes.  This is a topic for
future investigation.  If such a correction can be understood, then
we can find $F_g$ for all $g$ geometrically!

We close our discussion of the local $\IP^2$ with a tantalizing
observation
about the correction.  We see that the image of the singular locus
of $\hc8$ via the projection $\pi_8:\hc8\to\IP^{27}$ is the
set of curves which factor into a degree 1 factor with multiplicity
2 and degree 4 factor.
This locus is parameterized by $\IP^2\times\IP^{14}$, which has
Euler characteristic 45, exactly equal to the desired correction!  

But we can't get too excited yet about this observation.
The singular locus itself is a $\IP^8$
bundle over this space, since we must consider $Z$ itself, and
${\rm Hilb}^8$ of a line is just $\IP^8$.  So the singular
locus of $\hc8$ is parameterized by a space of Euler characteristic 
$9\cdot45$.

Nevertheless, we suspect that this is more than a coincidence.   For
$d=7$, we see that $\hc9$ can be singular when the degree 7 curve
factors into the square of a linear factor times a degree 5 factor.
This space is parameterized by $\IP^2\times\IP^{20}$, which has
Euler characteristic 63.  Once again, this is exactly the
discrepancy between the value of $n_7^6$ computed using the
B-model, and the value computed by \bestcorrection!  These
examples provide a big hint which needs to be better
understood.

\subsec{  Local ${\bf P}^1\times {\bf P}^1$: ${\cal O}(K)
\rightarrow {\bf P}^1\times {\bf P}^1 $ }

As expected from the Segre embedding of $P^1\times P^1$ into $P^3$ as
a degree 2 surface, we have the diagonal
perturbation of the local $P^1\times P^1$ case, i.e.\
$\sum_{i+j=r}n_{i,j}^{P^1\times P^1}=n_r^{X_{2}(1,1,1,1)}$, which we
have checked for genus $0,1$. Still saying it differently it
sums up the instantons in the compact elliptically fibered CY over
$F_0$ which survive the limit where the fibre volume becomes infinite.

{\vbox{\ninepoint{
$$
\vbox{\offinterlineskip\tabskip=0pt
\halign{\strut
\vrule#
&
&\hfil ~$#$
&\hfil ~$#$
&\hfil ~$#$
&\hfil ~$#$ 
&\hfil ~$#$
&\hfil ~$#$
&\vrule#\cr
\noalign{\hrule}
&d
&\gg=0
&1
&2
&3
&4
&
\cr
\noalign{\hrule}
&1
&-4^\diamond
&0
&0
&0
&0
&\cr
&2
&-4^\diamond
&0
&0
&0
&0
&\cr
&3
&-12^\diamond
&0
&0
&0
&0
&\cr
&4
&-48^\diamond
&9
&0
&0
&0
&\cr
&5
&-240^\diamond
&136^\diamond
&-24^\diamond
&0
&0
&\cr
&6
&-1356
&1616
&-812
&186^\diamond
&-16^\diamond
&\cr
&7
&-8428
&17560
&-17340
&9712
&-3156^\diamond
&\cr
&8
&-56000
&183452
& -302160
&307996
&-206776
&\cr
&9
&-392040
&1878664
&-4688912
&7590720
&-8583824
&\cr
&10
&-2859120
&19027840
&-67508988
&159995520
&-274149876 
&\cr
\noalign{\hrule}}\hrule}$$}
\vskip - 7 mm
\centerline{{\bf Table 5:} The 
weighted sum of BPS states $n^\gg_d$ for the local ${\bf P}^1\times {\bf P}^1$ 
case.}
\vskip7pt}
\noindent

We now compute the $^\diamond$ numbers in Table 5 geometrically.
We recall that a curve of bi-degree $(a,b)$ on $\IP^1\times\IP^1$ has
total degree $d=a+b$, genus $g=(a-1)(b-1)$, and $\cM=\IP^{(a+1)(b+1)-1}$.
By \topgenus, this gives a contribution of $(-1)^{(a+1)(b+1)-1}
(a+1)(b+1))$ to $n_{a+b}^{(a-1)(b-1)}$.  

If $d=2k$ is fixed, then relative to
all possibilities for $d=a+b$, the choice $(a,b)=(k,k)$ gives the maximum
possible genus $(k-1)^2$.  Thus 
$$n_{2k}^{(k-1)^2}=(-1)^{k^2+2k}(k+1)^2=(-1)^k(k+1)^2.$$
This verifies our results for $k=1,2$, and 3.

If $d=2k+1$, then the maximal genus $k(k-1)$ is attained for 
$(a,b)=(k+1,k)$ or $(k,k+1)$.  We may as well consider one of these cases
and multiply the result by 2.  By \topgenus\ we get
$$n_{2k+1}^{k(k-1)}=-2(k+1)(k+2),$$
where the definite sign follows since $(k+2)(k+1)$ is always even.  This
verifies our results for $k=0,1,2$.

We next turn to curves with 1 node, $\delta=1$.  If $d=2k+1$, then the only 
contributions
to $n_{2k+1}^{k(k-1)-1}$ come from $(a,b)=(k+1,k)$ or $(k,k+1)$.  
If $d\ge 5$, there are no reducible curves of these bi-degrees with only 1
node.  We restrict to the first case and will later multiply the contribution
by 2.  The map $\cC\to\IP^1\times
\IP^1$ has fiber over $p$ the space of curves of bi-degree $(k+1,k)$
which contain $p$, a $\IP^{(k+2)(k+1)-2}$.  This leads to 
$e(\cC)=4((k+2)(k+1)-1)=4(k^2+3k+1)$.  We get
$$n_{2k+1}^{k^2-k-1}=2\left(4(k^2+3k+1)+(2k(k-1)-2)(k+2)(k+1)\right),$$
which in particular verifies $n_5^1$=136.

If $d=2k$ and $k\ge2$, a new interesting possibility arises, which is
typical in the sequel.  We get contributions to $n_{2k}^{(k-1)^2-1}$
some of which have $\delta=0$, while others have $\delta=1$.  The
former type comes from $(a,b)=(k+1,k-1)$ or $(k-1,k+1)$.  This gives a
contribution of $2(-1)^{k(k+2)-1}k(k+2)$ by \topgenus .  The $\delta=1$
contribution comes from $(a,b)=(k,k)$, and we get
$(-1)^{(k+1)^2-2}(4((k+1)^2-1)+(2(k-1)^2-2)(k+1)^2)$ from the first
equation in \smallnodes\ together with \bestcorrection, as there are
no reducible curves of this bi-degree with only 1 node.  Combining
these two contributions gives
$$n_{2k}^{k^2-2k}=(-1)^{k(k+2)-1}\left(
2k(k+2)+4\left((k+1)^2-1\right)+\left(2(k-1)^2-2\right)(k+1)^2\right),$$
which agrees with Table 5 for $k=2,3$.

We next consider the case of $d=7,g=4$.  Here there are the possible
bi-degrees $(4,3)$ and $(5,2)$.  Since the latter case already has
$g=4$, we get a moduli space $\cM=\IP^{17}$, which gives a
contribution of $-18$ by \topgenus.  Curves of bi-degree $(4,3)$ have
genus $g=6$ and $\cM=\IP^{19}$, with $e(\cM)=20$.  We get as usual
$e(\cC)=19\cdot 4=76$ and $e(\hc2 )=18\cdot 14=252$ (we have used
$e({\rm Hilb}^2(\IP^1\times\IP^1)=14$ here).  So the second equation
of \smallnodes\ together with \bestcorrection\ gives a contribution of
$(-1)^{17}(252+8\cdot 76+5\cdot 7\cdot 20)=-1560$, as there are no
reducible curves of this bi-degree with 2 nodes.  Combining with the
$-18$ and multiplying by 2 to account for bi-degrees $(3,4)$ and
$(2,5)$, we get the consistent answer $-3156$.

More interesting is the geometric calculation of $n_5^0$.  We have
to consider bidegrees $(4,1)$ and $(3,2)$.  The first case is handled by
\topgenus, and we get $-10$.  In the second case, we have $g=\delta=2$,
$e(\cM)=12$, $e(\cC)=4\cdot11=44$, $e(\hc2)=14\cdot10=140$.  There are
also reducible curves of type $(1,0)\cup(2,2)$ with 2 nodes.  By \topgenus ,
we have $n_{1,0}^0=-2$ and $n_{2,2}^1=9$.  So by the second equation in
\smallnodes\ and \bestcorrection, we get
$n_{3,2}^0=-(140-12)-(-2)\cdot9=-110$.  Thus $n_5^0=2(n_{3,0}^0+n_{4,1}^0)=
-240$, in agreement with Table 5.  

\subsec{ Other local Del Pezzo geometries: $E_5$, $E_6$, $E_7$, and $E_8$}

\noindent
The $E_5$ del Pezzo:

{\vbox{\ninepoint{
$$
\vbox{\offinterlineskip\tabskip=0pt
\halign{\strut
\vrule#
&
&\hfil ~$#$
&\hfil ~$#$
&\hfil ~$#$
&\hfil ~$#$ 
&\hfil ~$#$
&\hfil ~$#$
&\vrule#\cr
\noalign{\hrule}
&d
&\gg=0
&1
&2
&3
&4
&
\cr
\noalign{\hrule}
&1
&16^\diamond
&0
&0
&0
&0
&\cr
&2
&-20^\diamond
&0
&0
&0
&0
&\cr
&3
&48^\diamond
&0
&0
&0
&0
&\cr
&4
&-192^\diamond
&5^\diamond
&0
&0
&0
&\cr
&5
&960^\diamond
&-96^\diamond
&0
&0
&0
&\cr
&6
&-5436
&1280^\diamond
&-80^\diamond
&0
&0
&\cr
&7
&33712
&-14816 
&  2512^\diamond
&-160^\diamond
&0 
&\cr
&8
&-224000
&160784
&  -51928
& 8710^\diamond
&-680^\diamond
&\cr
&9
&1568160
&-1688800
& 886400
&-274240
& 51040
&\cr
&10
& -11436720
&17416488
& -13552940
&6643472
& -2167656
&\cr
\noalign{\hrule}}\hrule}$$}
\vskip - 4 mm
\centerline{{\bf Table 6:} The weighted sum of BPS states $n^\gg_d$ for the local 
$E_5$ del Pezzo.}
\vskip7pt}

As in $\IP^1\times\IP^1$, to verify our calculations we need to break
up our degrees into subclasses.  Verification of the $E_n$ geometries
is complicated mainly by the fact that there are many possibilities
contributing to a given degree, and we must consider all the
possibilities for any fixed $d$ that we want to understand.

As typical examples, let us verify $n_4^1$ and $n_4^0$.  
We look at the $d=4,g=1$ data first.  We have the class $(3;1^5)$
with $\cM=\IP^4$ (by Riemann-Roch, one
expects $\cM=\IP^{d+g-1}$ in general).  This verifies $n_4^1=5$ by
\topgenus .  But there is also a $\delta=1$ contribution to
$n_4^0$.  We have $\cC\to\cM$ with fiber $\IP^3$, hence 
$$e(\cC)=e(\IP^3)e(E_5)=4\cdot8=32.$$  This gives a contribution of
$-32$ to $n_4^0$ by the first equation of \smallnodes\ and
\bestcorrection.

We next look at the $d=4,g=0$ data.  Each of these classes has
$\cM=\IP^3$.  Each $\IP^3$ gives a contribution of $-4$.  So we have
to count numbers of such families.  For $(2;1^2)$ we have to choose 2
out of the 5 points at which to put the two ones; there are 
${5\choose2}=10$ ways
of doing this.  For $(3;2,1^3)$ we choose 1 point for the two, and 3
of the remaining points for the ones, and there are $5{4\choose2}=20$
ways of doing this.  Finally, for $(4;2^3,1^2)$ there are 10 choices.
All together, there are $10+20+10=40$ distinct $d=4,g=0$ families.
The total contribution is $40(-4)=-160$.  Combining with the $-32$, we
have $-160-32=-192$, which agrees with the calculated value of
$n_4^0$.

Now let's calculate $n_5^0$.
We have for $d=5,g=1$ the families $(3;1^4),
(4;2^2,1^3),(5,2^5)$ and we want $\delta=1$.  Including permutations,
there are 16 such families.  In each case, we have reducible curves
with 1 node:
$$\eqalign{
(3,1^4)&=(3,1^5)\cup E_5\cr
(4,2^2,1^3)&=(3,1^5)\cup(1,1^2)\cr
(5,2^5)&=(3,1^5)\cup(2,1^5)}$$
Since the curves of type $(3,1^4)$ form a $\IP^4$, and the respective
curves $E_5,\ (1,1^2)$, and $(2,1^5)$ are all isolated, it follows that
the term to be subtracted on the right hand side of \bestcorrection\ is
$-5$.  Since $e(\cC)=5\cdot8=40$ in each case, we get for each the 
contribution $-40-(-5)=-35$ by the first equation of \smallnodes\ and
\bestcorrection.
Combining this calculation with the more standard calculations for 
$\delta=0$ based on \topgenus, we get $n_5=80(5)+16(35)=960$, in agreement
with Table 6.

We can similarly verify all the other cases indicated with a diamond using just
these techniques.  We can also check some cases requiring curves with 
$\delta>1$ nodes, but these become increasingly tedious.

\bigskip\noindent
The $E_6$ del Pezzo:

{\vbox{\ninepoint{
$$
\vbox{\offinterlineskip\tabskip=0pt
\halign{\strut
\vrule#
&
&\hfil ~$#$
&\hfil ~$#$
&\hfil ~$#$
&\hfil ~$#$ 
&\hfil ~$#$
&\hfil ~$#$
&\vrule#\cr
\noalign{\hrule}
&d
&\gg=0
&1
&2
&3
&4
&
\cr
\noalign{\hrule}
&1
&27^\diamond
&0
&0
&0
&0
&\cr
&2
&-54^\diamond
&0
&0
&0
&0
&\cr
&3
&243^\diamond
&-4^\diamond
&0
&0
&0
&\cr
&4
&-1728^\diamond
&135^\diamond
&0
&0
&0
&\cr
&5
&15255
&-3132^\diamond
&189^\diamond
&0
&0
&\cr
&6
&-153576
&62976
&-10782^\diamond
&789^\diamond
&-10^\diamond
&\cr
&7
&1696086
& -1187892 
&  397899
&-75114
&7641
&\cr
&8
& -20053440
&21731112
&  -12055770
& 4188726
&-948186
&\cr
&9
& 249740091
&-391298442
& 326385279
&-179998572
& 69918830
&\cr
&10
& -3240109350
& 6985791864
& -8218296072
&6602867631
& -3896482536
&\cr
\noalign{\hrule}}\hrule}$$}
\vskip - 6 mm
\centerline{{\bf Table 7:} The weighted sum of BPS states $n^\gg_d$ for the local 
$E_6$ 
del Pezzo.}
\vskip7pt}

Note that the second equation in \smallnodes\ is needed for $n_6^2$, 
and we get from several different families of curves, \topgenus, 
the first two equations in \smallnodes, and \bestcorrection, the result
$n_6^2=270(-8)+108(-72)-846 =-10782$, again in agreement with Table 7.

\bigskip\noindent
The $E_7$ del Pezzo:

{\vbox{\ninepoint{
$$
\vbox{\offinterlineskip\tabskip=0pt
\halign{\strut
\vrule#
&
&\hfil ~$#$
&\hfil ~$#$
&\hfil ~$#$
&\hfil ~$#$ 
&\hfil ~$#$
&\vrule#\cr
\noalign{\hrule}
&d
&\gg=0
&1
&2
&3
&
\cr
\noalign{\hrule}
&1
&56^\diamond
&0
&0
&
&\cr
&2
& -272^\diamond
&3
&0
&
&\cr
&3
&3240^\diamond
&-224^\diamond
&0
&
&\cr
&4
& -58432
&12042
& -844^\diamond
&7^\diamond
&\cr
&5
&1303840
&-574896
&116880
&-12112^\diamond
&\cr
&6
&-33255216
&26127574
&-10554800
&2654912
&\cr
&7
& 930431208
&-1163157616 
& 787322120
&-358173504
&\cr
&8
&-27855628544
&51336812456
&-52707074296
&37805931096
&\cr
&9
&878169863088
& -2258519658288
&3292406219040
&-3430645737616
&\cr
&10
&-28835628521920
&99301270680473
&-196037258631040
&280764828128124
&\cr
\noalign{\hrule}}\hrule}$$}
\vskip - 6 mm
\centerline{{\bf Table 8:} The
weighted sum of BPS states $n^\gg_d$ for the local $E_7$ 
del Pezzo.}
\vskip7pt}

Again the numbers marked with $\diamond$ have been checked. Here we determine 
the number $n_{5}^3$. 
First let us calculate the contribution of the smooth genus $3$  
curves with $d=5$. From Appendix A with simple combinatorics follows that 
there are
$1+35+105+7+140+7+140+105+35+1=576$ curves all with moduli space $\IP^7$ by 
\em\topgenus , hence  
contributing $(-1)^7 8\cdot 576=-4608$ to $n_5^3$. Also from the table we read off 
that there are 
$7+21+21+7=56$ genus\foot{Note the classical fact that for fixed $d$ these curves 
are in the 
representation of the Weyl-group of $E_7$.} $4$ curves with $d=5$. In fact we have 
$n_5^4=56(-1)^8\cdot 9=504$ from \em\topgenus\ as there are no $g=5$ curves 
in classes with $d=5$, which could degenerate to $g=4$. 
The universal curve for each of the of the $56$ $\delta=1$ curves 
has by \ec\ $e({\cal C})=-8\cdot 10$. Application of \generalformng\ gives 
hence a contribution of $-56\cdot 80-(2\cdot 4-2) 504$ of the nodal curves to 
give a total of $n_5^3=-4608- 7504=-12112$. 

\bigskip\noindent

For the  $E_8$ del Pezzo surface we obtained 
  
{\vbox{\ninepoint{
$$
\vbox{\offinterlineskip\tabskip=0pt
\halign{\strut
\vrule#
&
&\hfil ~$#$
&\hfil ~$#$
&\hfil ~$#$
&\hfil ~$#$ 
&\hfil ~$#$
&\hfil ~$#$
&\hfil ~$#$ 
&\hfil ~$#$
&\hfil ~$#$
&\vrule#\cr
\noalign{\hrule}
&d
&\gg=0
&1
&2
&3
&4
&5
&6
&7
&
\cr
\noalign{\hrule}
&1
&252^\diamond
& -2^\diamond
&
&
&
&
&
&
&
\cr
&2
& -9252^\diamond
&762^\diamond
&-4^\diamond
&
&
&
&
&
&
\cr
&3
&848628
&-246788
&30464
&-1548^\diamond
&7^\diamond
&
&
&
&\cr
&4
& -114265008
& 76413073
&-26631112
& 5889840
&-835236
&69587
&-2642^\diamond
&11^\diamond
&\cr
\noalign{\hrule}}\hrule}$$
\vskip -3 pt
\centerline{{\bf Table 9:} The weighted sum of BPS states $n^\gg_d$ for the local 
$E_8$ 
del Pezzo.}
\vskip7pt}}

The case of the $E_8$ del Pezzo is interesting because 
it can be easily can be related to the ${1\over 2} K_3$, which gives 
an additional check. Extending \mnw~ it was observed \hst\ that there is a modular 
anomaly in this case for higher genus. 

More precisely let $F=K_{{1\over 2}K_3}$ the fibre (also the canonical class) 
and $B=e_9$ the base class of the elliptic half $K_3$ surface. Following \hst\ we 
can solve the modular anomaly for this two parameter subspace and write the 
contribution to 
$F_\gg$ for a fixed base class $n$ as a quasimodular form $P_{2g+2n-2}(E_2,E_4,E_6) 
{q^{m\over 2}\over \eta^{12 n}}$. In particular we can compare the 
diagonal class $K_{E_8}=F+e_9$ and test the modular anomaly  
against the holomorphic anomaly calculation. We observed 
complete accordance up to genus 8 degree 10. 
 
\subsec{The topological string perspective}  

In the local case one has for all genera an explicit virtual 
fundamental class on the moduli spaces of maps and therefore a direct 
A-model localization calculation of the topological string amplitudes 
at higher genera is in principle possible \kz . 
At genus zero there is a virtual fundamental class also for the global 
case the and equivalence of the $A$-model and the $B$-model calculation 
was proven \mirrorproofs . This  sort of proof was adapted to the local 
case \ckyz .  Hence as an immediate check we can perform in certain 
cases $A$-model computations to compare with. However using the $c=1$ 
KDV hierarchy and localization on the toric ambient space becomes 
extremely difficult for high $g$, where the $M$-theory calculation
is still very easy, provided that $\delta< \delta_{max}(d)$.   

Furthermore we can consider the $B$-model and use the $M$-theory calculation to 
resolve
the holomorphic ambiguity. The $B$-model, which has some additional 
simplifications 
in the local case \kz, has the virtue that it comes naturally with the analytic 
techniques on the complex moduli space of the mirror manifold, which allows us
to relate the answers that we get in the infinite volume limit to physical 
systems at other degenerations by analytic continuation. This yields 
further consistency checks.
   
To describe the analytic structure we note that the vectors $\vec a=(a_1,a_2)$ 
$$\eqalign{
& 
\IP^2:\ \ \vec a =\left({1\over 2},{2\over 3}\right), 
\qquad
\IP^1\times \IP^1 :\ \ \vec a =\left({1\over 2},{1\over 2}\right), 
\qquad
E_5: \ \ \vec a=\left({1\over 2},{1\over 2}\right),\cr
& E_6:\ \ \vec a=\left({1\over 3},{2\over 3}\right), 
\qquad
E_7:\ \ \vec a=\left({1\over 4},{3\over 4}\right), \qquad 
E_8:\ \ \vec a=\left({1\over 6},{5\over 6}\right),}$$ 
determine the Picard-Fuchs differential equation, 
which governs the complex  geometry of the mirror for the local cases 
\eqn\genrallocalpf{\left(\theta^3-z\prod_{i=1}^3 
(\theta-a_i+1)\right)\int_{\gamma_i} 
\Omega =0\ . }
They are solved by Meier's G-function $G(a_1,a_2,1;x)$, 
compare with \ref\bate{A.\ Erd\'elyi, W.\ Magnus, F.\ Oberhettinger, 
F.\ Tricomi, ``The Bateman Project: Higher Transcendental Functions,´´ 
$\us {Vol. 1}$.}. In particular their Riemann symbol is 
\eqn\riemann{{\bf P}\left\{
\matrix{
0& \infty& 1 &\cr
0& 0  & 0 &\cr
0& a_1& 1 &\cr
0& a_2& 1 &}\!\!,z\right\} }
and shows that $z=0$ is the maximally unipotent point, $z=1$ is the 
conifold point. We can also read from the Riemann symbol that for the 
$\IP^1\times \IP^1$ and the $E_5$,
$z=\infty$ is not an orbifold point, but has also logarithmic 
solutions\foot{Further properties of the solutions have been made 
explicit in \kmv\lmw\kz . In particular the monodromy is related to 
$\Gamma_0(n)$ for $E_{9-n}$ \kz . }. 

As in the quintic case of \bcovII, it is convenient
for the higher genus calculation to work in the $\psi$ 
variable as it avoids fractional exponents 
in the B-model propagators. In this variables the Yukawa are 
\foot{The sign is minus for the local $\IP^1\times \IP^1$ case and 
minus for all other cases. For the somewhat exceptional $\IP^2$ and 
$\IP^1\times \IP^1$ case we have $\alpha(X)=1/3,\beta=3$ and 
$\alpha(X)=1$,$\beta=-2$.}
\eqn\yukawa{\psi^{-\beta}\equiv \pm z, 
\qquad Y_{\psi\psi\psi}={\alpha(X) \beta^3 
\psi^{\beta-3}\over (1\pm \psi^\beta)}, 
\quad {\rm with}\  \beta=\sum_{i=1}^l d_i}
The mirror map is normalized so that $\alpha(X)=\prod_{i=1}^k 
d_i/\prod_{j=1}^l w_l$ is just the classical triple intersection number expressed
in terms of the degrees $d_i$ of the complete intersections and the weights of the
ambient space. The $E_n$ $n=5,\ldots,8$ del Pezzo surfaces can be represented as
complete intersections of degree $(2,2)$ in $\IP^4$, a degree $3$ hypersurface in 
$\IP^3$ , 
a degree $4$ hypersurface in weighted projective space $\IP^3(1,1,1,2)$ and a 
degree $6$ hypersurface in $\IP(1,1,2,3)$.

The $B$-model multi-loop contributions to the free energy 
are determined from the holomorphic anomaly equations \bcovII .
In solving these differential equations one is left at the end 
with a holomorphic ambiguity, which is a holomorphic section of 
${\cal L}^{2-2r}$ over the complex moduli  space\foot{Here we denote the
worldsheet genus of the $B$-model $r$, to distinguish it from the arithmetic genus $g$.}. 
With the right choice of gauge there is no singularity of the holomorphic 
ambiguity at $z=0$, but from the singularity 
structure is clear that we have to generalize the ansatz for the ambiguity 
\bcovII\kz~to allow beside the singularities at the conifold also for 
residue terms at $z=\infty$. Hence we make in general the ansatz 
\eqn\ansatzII{f_\gg(\psi)=\sum_{k=0}^{2\gg-2} {A_k^\gg\over \mu^k} + 
\sum_{k=1}^{\gg-1} {B^\gg_k\over \rho^{2 k}}\ , }
where $\mu= c_1 (1\pm \psi^\beta)$ and $\rho= c_2 \psi^\beta$. 

The finite number of constants $A^\gg_k,B^\gg_k$ can be fixed by the direct 
calculation of the $n_d^\gg$, but at least part of it is also encoded in the 
universal 
behavior of $F_\gg$ at singular loci in the moduli space. 
In particular near a conifold singularity one expects \ghovaf\ from the 
duality with the $c=1$ critical string theory at the selfdual radius an expansion
\eqn\conifold{
F_{S^3}(\mu)={1\over2} \mu^2 \log(\mu) - {1\over 12} \log(\mu)+\sum_{\gg=2}^\infty 
{B_{2\gg}\over (2\gg -2) (2\gg)}\mu^{2-2\gg}\, }
where ${B_{2\gg}/(2\gg-2) 2 \gg}=-{1\over 240},{1\over 1008},-{1\over 1440},\ldots $ 
for genus $\gg=2,3,4,\ldots$. Relative to the  genus expansion of the type II free
energy $F(t)=\sum_{\gg=0}^\infty \lambda^{2\gg-2} F_\gg(t)$ \conifold~is a double 
scaling
limit in which the distance $t$ to the conifold in the moduli space and the 
string 
coupling go to zero while their ratio $\mu$ is kept fixed. This selects the 
corresponding leading term $\mu^{2\gg-2}$ form $F_\gg$ in \conifold. 
For the local propagators we use the same gauge as in \kz . In this 
gauge the singular behavior at the conifold is captured entirely 
by the ambiguity.  The relation between $t=(1\pm\psi^\beta)$ and $\mu$ 
can be fixed from the genus zero result by comparing the asymptotic 
of the Yukawa coupling with \conifold. We found the simple systematics 
$\mu= {(1\pm \psi^\beta)\over \sqrt{\alpha(X)}}$ for the local 
and $\mu={ i (1\pm \psi^\beta)\over \sqrt{\alpha(X)}}$ for the global cases.

It has been speculated that the next to leading order might be related 
to correlations functions of the $c=1$ string at the selfdual radius 
involving discrete states whose $SU(2)_L\times SU(2)_R$ charges could be read
off from next to leading order monomial in the perturbations of the local
equations \ref\gvtgtg{R. Gopakumar and C. Vafa,  
``Topological Gravity as Large N Topological Gauge Theory'', 
Adv. Theor. Math. Phys. $\us{2}$ (1998) 413-442, hep-th/9802016.}. 
It would be very interesting to make this more concrete.
At any rate we will report this behavior at the conifold and other
singularities along with our previous results at the large complex
structure limit. Beside the verification of the claim in \ghovaf\ we
find also some regularities for the next to leading order residue. For
instance, we observed up to $\gg=4$ that the $A^\gg_{2\gg-3}$ of the
quintic Calabi-Yau threefold are exactly $(4/5)$ times smaller as the
$A^\gg_{2\gg-3}$ for the local $\IP^1\times \IP^1$ case, which in turn
are identical to the analogous quantities on the $E_6$ del Pezzo.

For the $\IP^2$ case we fixed the constants in the ambiguity \ansatzgII~as follows 

{\vbox{\ninepoint{
$$
\vbox{\offinterlineskip\tabskip=0pt
\halign{\strut
\vrule#
&
&\hfil ~$#$
&\hfil ~$#$
&\hfil ~$#$ 
&\hfil ~$#$
&\hfil ~$#$
&\hfil ~$#$ 
&\hfil ~$#$
&\hfil ~$#$
&\hfil ~$#$
&\hfil ~$#$
&\vrule#\cr
\noalign{\hrule}
&\gg
&A^\gg_0
&A^\gg_1
&A^\gg_2
&A^\gg_3
&A^\gg_4
&A^\gg_5
&A^\gg_6
&A^\gg_7
&\! \! A^\gg_8 \! 
&
\cr
\noalign{\hrule}
&2
&{3 e-220\over 17280}
&{107\over 4320 \sqrt{3}}
&-{1\over 240}
&
&
&
&
&
&
&\cr
&3
&{9940-3 e\over 4354560 }
& -{2519\over 181440\sqrt{3}}
&{\frac{6479} {653184}}
&-\frac{13} {1440\,{\sqrt{3}}}
& \frac{1}{1008}
&
&
&
&
&\cr
&4
&{\frac{-1864975 + 27\,e}{2351462400}}
&{\frac{2117}{230400\,{\sqrt{3}}}}
&{\frac{-32949197}{2351462400}} 
&{\frac{14229281}{440899200\,{\sqrt{3}}}}
&{\frac{-143987}{10886400}} 
&{\frac{15041}{1814400\,{\sqrt{3}}}}
&  {\frac{-1}{1440}}
&
&
&\cr
&5
&\!\! \frac{252764050-243\,e}{587865600s_5 } \!\!
&\!\! \frac{-2607911 }{340200 \sqrt{3}s_5  } \!\!
&\!\! \frac{164095927 }{8398080s} \!\!
&\!\! \frac{-35952843613}{35952843613   \sqrt{3}s_5 } \!\!
&\!\! \frac{88814851411 }{1322697600 s_5} \!\!
&\!\! \frac{-616239319}{ 6123600\sqrt{3}s_5} \!\!
&\!\! \frac{348989 }{11664s_5 } \!\!
&\!\! \frac{-5519 }{378 \sqrt{3}s_5 } \!\!
&\!\! \frac{1}{s_5} \!\!
&\cr
\noalign{\hrule}}\hrule}$$}}
\noindent
with $\mu=\sqrt{3}(1+\psi^3)$ and $s_g$ as in \conifold . This is in perfect
agreement with expected behaviour at the conifold.

Because of the additional singularity at $\psi=0$ the holomorphic ambiguity 
is more interesting in the $\IP^1\times \IP^1$ case.
Using the Landau-Ginzburg orbifold description the
mirror manifold can be represented as a hypersurface \ghovaf\ 
\eqn\lgcone{W=x_1^2+x_2^2+x_3^2+x_4^2 -2 \left( \psi x x_1 x_2 x_3 x_4 +{\psi\over 
x} \right)}
in a non-compact  $\IP^4(-2,1,1,1,1)$ .
Note that at $x=1$, $x_i=1$  for $\psi=1+\epsilon$ one has the usual conifold 
singularity, 
but  $x=1$, $x_i=0$ for small $\psi$ does not correspond to a conifold 
singularity. 
Because of the $Z_2$ $(x\rightarrow x; x_i\rightarrow -x_i)$ identification 
symmetry in the 
$\IP^4(-2,1,1,1,1)$ it rather is a $S^3/Z_2$ lens space, called $L(2,1)$, which 
is 
shrinking. In the IIB picture this shrinking lens space creates a 
logarithmic singularity (with a logarithmic shift by $2$ at 
$\psi=0$ due to the bound state of D-3 branes wrapping $L(2,1)$). 
It explains why  we need  here the poles at $\psi=0$ in the 
generalized ansatz \ansatzgII . In fact using the relation
$\rho=2^4 \psi^4$, fixed from the asymptotic behaviour of the Yukawa 
coupling, we find 

{\vbox{\ninepoint{
$$
\vbox{\offinterlineskip\tabskip=0pt
\halign{\strut
\vrule#
&
&\hfil ~$#$
&\hfil ~$#$
&\hfil ~$#$ 
&\hfil ~$#$
&\hfil ~$#$
&\hfil ~$#$ 
&\hfil ~$#$
&\hfil ~$#$
&\hfil ~$#$ 
&\hfil ~$#$
&\hfil ~$#$
&\vrule#\cr
\noalign{\hrule}
&\gg
&A^\gg_0
&A^\gg_1
&A^\gg_2
&A^\gg_3
&A^\gg_4
&A^\gg_5
&A^\gg_6
&B^\gg_1
&B^\gg_2
&B^\gg_3
&
\cr
\noalign{\hrule}
&2
&{e-100\over 5760}
&{83\over 5760}
&-{1\over 240}
&
&
&
&
&-{2^3\over {240}}
&
&
&\cr
&3
&{10045-2 e\over 2903040 }
& -{413\over 51840   }
&{\frac{ 5783 } { 645120}}
&-\frac{193 } {40320 }
& \frac{1}{1008}
&
&
& -{23\over 48384}
&{2^5 \over {1008}}
&
&\cr
&4
&{8 e-935025\over 696729600 }
& {1001683 \over  2^{16}5 3^4 7  }
&{\frac{- 16034329} { 2^{15} 3^5 5^2 7   }}
&\frac{9924889 } {2^{14} 3^5 5^2 7   }
&\frac{-2871}{27648}
&\frac{10001}{2^{9} 3^3 5^2 7}
&\frac{1}{1440}
& {150671\over 2^{16}3^4 5^2 7}
&{-3641\over {2^{13} 3^3 5^2 7   }}
&{2^7 \over {1440}}
&\cr
\noalign{\hrule}}\hrule}$$}}
\noindent
Thus the leading behavior at $\psi=0$, captured in the $B^\gg_i$, is exactly 
as expected\gvtgtg~for two particles with half the mass $\rho$ leading 
to ${\cal F}_{L(2,1)}(\rho) = 2 {\cal F}_{S^3}({\rho\over 2})$ 
with  ${\cal F}_{S^3}$ as in \conifold .

For the $E_n$ $n=5,\ldots,8$ cases one expect a complicated singularity 
structure  at $\psi=0$, due to the simultaneous occurence of light electric 
and magnetic states. For the $E_5$ case we got

{\vbox{\ninepoint{
$$
\vbox{\offinterlineskip\tabskip=0pt
\halign{\strut
\vrule#
&
&\hfil ~$#$
&\hfil ~$#$
&\hfil ~$#$ 
&\hfil ~$#$
&\hfil ~$#$
&\hfil ~$#$ 
&\hfil ~$#$
&\hfil ~$#$
&\hfil ~$#$ 
&\hfil ~$#$
&\hfil ~$#$
&\vrule#\cr
\noalign{\hrule}
&\gg
&A^\gg_0
&A^\gg_1
&A^\gg_2
&A^\gg_3
&A^\gg_4
&A^\gg_5
&A^\gg_6
&B^\gg_1
&B^\gg_2
&B^\gg_3
&
\cr
\noalign{\hrule}
&2
&{2 e-155\over 11520}
&{83\over 11520}
&-{1\over 240}
&
&
&
&
&-{38 \over {240}}
&
&
&\cr
&3
& {26425-32 e\over 46448640}
& {-3997\over  3317760 }
& {6119 \over 2580480 }
& {-193\over 80640}
& \frac{1}{1008}
&
&
& {467\over 30240 }
&{158 \over {1008}}
&
&\cr
&4
& { {2^{11} e-9710925} \over {2^{22}  3^5 5^2  7 }}
&  {1405589 \over {2^{21} 5 3^4  7 }    }
& {-19054729 \over  {2^{19} 3^5 5^2 7 } }
& {10715449 \over {2^{17} 3^5 5^2 7 } }
& \frac{-2057}{2^{12} 3^3 7}
& {10001\over {2^{10} 3^3 5^2 7}}
& {-1\over 1440}
& {-625063\over {2^{15} 3^4 5^2 7} }
& {-137563 \over  {2^{10} 3^3 5^2 7 } }
& { -638\over 1440}
&\cr\noalign{\hrule}}\hrule}$$}}
\noindent
where we normalized  $(1+\psi^2)=\mu/2$ and  $\psi^2= \rho/2^{5}$.  Here 
we observe the same next to leading order behavior at the conifold 
as in the $\IP^1\times \IP^1$ examples. 
However the residue at $\psi=0$ has yet 
to be interpreted.

Let shortly summarize the remaining cases. For the $E_6$ we got

{\vbox{\ninepoint{
$$
\vbox{\offinterlineskip\tabskip=0pt
\halign{\strut
\vrule#
&
&\hfil ~$#$
&\hfil ~$#$
&\hfil ~$#$ 
&\hfil ~$#$
&\hfil ~$#$
&\hfil ~$#$
&\hfil ~$#$
&\hfil ~$#$ 
&\hfil ~$#$
&\hfil ~$#$
&\vrule#\cr
\noalign{\hrule}
&\gg
&A^\gg_0
&A^\gg_1
&A^\gg_2
&A^\gg_3
&A^\gg_4
&A^\gg_5
&A^\gg_6
&B^\gg_1
&B^\gg_2
&
\cr
\noalign{\hrule}
&2
&\frac{-1660 + 27\,e}{155520}
&\frac{107}{12960\,{\sqrt{3}}}
&-\frac{1}{240}
&
&
&
&
& 
&
&\cr
&3
&\frac{20300 - 81\,e}{117573120}
&-\frac{10301}{14696640\,{\sqrt{3}}}
&\frac{35419}{29393280}
&-\frac{13}{4320\,{\sqrt{3}}}
&\frac{1}{1008}
&
&
&1
&
&\cr
&4
&\!\!\! \frac{19683 e-11499775}{1714216089600}
&\!\!\!  \frac{6364177 }{ 95234227200\sqrt{3}}
&\!\!\! -\frac{43056877  }{190468454400}
&\!\!\!\!  \frac{ 15901121 }{11904278400 {\sqrt{3}}}
&\!\!\!\! - \frac{148787 }{97977600 }
&\!\!\!\! \frac{15041 }{5443200  {\sqrt{3}}}
&\!\!\!\!  -{1\over 1440} 
&\!\!\!\!  \frac{1}{54}
&\!\!\!\!  486
&\cr
\noalign{\hrule}}\hrule}$$}}
\noindent
with $(1+\psi^3)=\mu/\sqrt{3}$ and  
$\psi^3= \rho/3^{9}$. Here we note that the next to leading
order at the conifold is $1\over 3$-times the one of the ${\bf P}^2$ 
theory.

The $E_7$ case: 

{\vbox{\ninepoint{
$$
\vbox{\offinterlineskip\tabskip=0pt
\halign{\strut
\vrule#
&
&\hfil ~$#$
&\hfil ~$#$
&\hfil ~$#$ 
&\hfil ~$#$
&\hfil ~$#$
&\hfil ~$#$ 
&\hfil ~$#$
&\vrule#\cr
\noalign{\hrule}
&\gg
&A^\gg_0
&A^\gg_1
&A^\gg_2
&A^\gg_3
&A^\gg_4
&B^\gg_1
&
\cr
\noalign{\hrule}
&2
&{8 e -635 \over  46080}
&{143 \over 23040\sqrt{2}}
&-{1\over 240}
&
& 
&
&\cr
&3
& \frac{110215-512 e }{743178240 }
& -\frac{749 }{1658880\sqrt{2} }
& \frac{68857  }{61931520 } 
&-\frac{131  }{53760\sqrt{2} }
& \frac{1}{1008}
& 1
&\cr
\noalign{\hrule}}\hrule}$$}}

with $(1+\psi^4)=\mu/\sqrt{2}$ and  $\phi^4=\rho/2^{15}$.

Finally for the $E_8$: 

{\vbox{\ninepoint{
$$
\vbox{\offinterlineskip\tabskip=0pt
\halign{\strut
\vrule#
&
&\hfil ~$#$
&\hfil ~$#$
&\hfil ~$#$ 
&\hfil ~$#$
&\hfil ~$#$
&\hfil ~$#$ 
&\hfil ~$#$
&\hfil ~$#$
&\hfil ~$#$ 
&\vrule#\cr
\noalign{\hrule}
&\gg
&A^\gg_0
&A^\gg_1
&A^\gg_2
&A^\gg_3
&A^\gg_4
&A^\gg_5
&A^\gg_6
&B^\gg_1
&\cr
\noalign{\hrule}
&2
&{9 e-1100\over 51840}
&{251\over 51840}
&-{1\over 240}
&
&
&
&
&
&\cr
&3
&{12775-54 e\over  78382080 }
& -{\frac{517}{1399680}}
&{\frac{225613}{156764160}}
&-{\frac{251}{120960}}
&{\frac{1}{1008}}
&
&
&
&\cr
&4
&\! \! {\frac{-1547875 + 5832\,e}{507915878400}}
& {\frac{2590747}{96745881600}}
& -{\frac{19158899}{67722117120}}
&{\frac{122105573}{101583175680}}
&-{\frac{43501}{18662400}}
&{\frac{45281}{21772800}}
&-{\frac{1}{1440}}
&\! \! \!  {\frac{2525}{3869835264}}\!
&\cr
\noalign{\hrule}}\hrule}$$}}
with $\mu=(1+\psi^6)$ and $\rho=\psi^6$.

\bigskip

\newsec{Computations in compact Calabi-Yau geometries}

\subsec{Compact one modulus cases}
 
Here we  analyze the higher genus contribution for compact 
one modulus Calabi-Yau spaces. In this case there is no virtual
fundamental class for the higher genus topological string 
calculation known. In absence of this approach we combine the 
topological $B$ model calculation and the $M$-theory description 
of the invariants to determine the higher genus $F_\gg$.  We can carry
out the M-theory computation of certain $n^{g-\delta}_d$ for small $\delta$.
We expect to be able to make much further progress even in the compact case
once we understand how to systematically correct for singularities
in the relative Hilbert schemes $\hc{k}$ for $k\le\delta$.

Examples are hypersurfaces or complete intersections 
in weighted projective 
space  which avoid the singularities of the ambient space. Denoting a complete 
intersection of 
degree $(d_1,\ldots, d_k)$ 
in ${\bf P}^n(w_1,\ldots,w_l)$ by $X_{d_1,\ldots 
,d_k}(w_1,\ldots,w_l)$\foot{$n$-times repeating
weights will be denoted by $w_i^n$.} we have the following complete list of such 
examples (compare 
the the second ref. in \hkty~for the $\gg=0,\gg=1$ results)

$$
\eqalign{
X_5(1^5):\! \vec a &= ({1\over 5},{2\over 5},{3\over 5}, {4\over 5}),\ 
X_6(1^4,2):\! \vec a = ({1\over 6},{2\over 6},{4\over 6}, {5\over 6}),\ 
X_8(1^4,4):\! \vec a = ({1\over 8},{3\over 8},{5\over 8}, {7\over 8}),\cr
X_{10}(1^3\! ,2,5):\! \vec a &= (\! {1\over 10}\! ,\! {3\over 10}\!,{7\over 10}\!, 
\! {9\over 10}\!),\ 
X_{3,3}(1^6):\!\vec a = ({1\over 3},{2\over 3},{1\over 3}, {2\over 3}),\ 
X_{4,2}(1^6):\!\vec a = ({1\over 4},{1\over 2},{3\over 4}, {1\over 2}),\cr
X_{3,2,2}(1^7): \!\vec a &= ({1\over 3},{2\over 3},{1\over 2}, {1\over 2}),\ 
X_{2,2,2,2}(1^8):\! \vec a = ({1\over 2},{1\over 2},{1\over 2}, {1\over 2})\ 
X_{4,3}(1^5, 2): \!\vec a = (\!{1\over 3}\!,\!{2\over 3}\!,{1\over 4}\!, {3\over 
4}\!),\cr
X_{4,4}(1^4,2^2):\! \vec a &= ({1\over 4},{3\over 4},{1\over 4}, {3\over 4}),\
X_{6,2}(1^5,3): \!\vec a = ({1\over 6},{1\over 2},{5\over 6}, {1\over 2}),\cr
X_{6,4}(1^3,2^2,3): \!\vec a &= ({1\over 6},{5\over 6},{1\over 4}, {3\over 4}),\ 
X_{6,6}(1^2,2^2,3^2): \!\vec a = ({1\over 6},{5\over 6},{1\over 6}, {5\over 6})}$$

The components of the vector $\vec a$ specify the Picard-Fuchs operator for the 
mirror manifold as follows 
\eqn\generalpf{\left(\theta^4-z\prod_{i=1}^4 (\theta+a_i)\right) \int_{\gamma_i} 
\Omega =0\ .} 
{}From the  Riemann Symbol
\eqn\riemann{{\bf P}\left\{\matrix{
0& \infty& 1 &\cr
0& a_1& 0 &\cr
0& a_2& 1 &\cr
0& a_3& 2 &\cr
0& a_4& 1 &}\!\!,z\right\} }
we conclude that $z=0$ is the maximally unipotent point with three logarithmic solutions and
$z=1$ is the conifold point with one logarithmic solution. 
{}From the Riemann symbol it also clear that the hypersurfaces have a cyclic 
monodromy of order\foot{For the hypersurfaces this 
was just an orbifold singularity with non-vanishing $B$-field. 
In fact the triple intersection corresponds 
to canonical normalized kinetic terms, precisely to the one 
calculated in the associated Gepner model \cdgp.}
$\beta$ (with $\beta$ as in \yukawa ) at $z=\infty$, 
while the complete intersections have degenerate indices and therefore 
logarithmic degeneration of the periods with shift monodromy and vanishing 
cycles at $z=\infty$, which justifies the general ansatz for the ambiguity 
\ansatzII .
We choose the gauge so that the propagators of the $B$ are regular at the conifold 
and at
$z=\infty$. This amounts, in the notation of sections 7.2 of \bcovII, to a choice 
of 
$f(\psi)=\psi$ for quintic and sextic, $f(\psi)=\psi^2$ for the bicubic and 
$f(\psi)=\psi^4$ for the four conics, while $v(\psi)=1$ for all cases.      
The normalization of the Yukawa couplings is as in \yukawa .

\subsec{Higher genus results on the quintic }

We come now to the simplest compact Calabi-Yau, the zero locus of the quintic 
\eqn\quinticpf{\sum_{i=1}^5 x_i^5- 5\psi \prod_{i=1}^5x_i=0.}
in ${\bf P}^4$.
The unique analytic solution at $z=0$ is 
$w_0=\sum_{n=0}^\infty {5 n!\over (n!)^5}\left(z\over 5^5\right)^n$, the three-point 
coupling is as in \yukawa. The rational and elliptic curves 
have been computed in \cdgp\bcovI .
\vskip 3 mm
\noindent
{\vbox{\ninepoint{
$$
\vbox{\offinterlineskip\tabskip=0pt
\halign{\strut
\vrule#&
&\hfil ~$#$
&\hfil ~$#$
&\hfil ~$#$
&\vrule#\cr
\noalign{\hrule}
&d
&\gg=0
&1
&
\cr
\noalign{\hrule}
&1
&2875
&0
&\cr
&2
&609250
&0
&\cr
&3
&317206375
&609250
&\cr
&4
&242467530000
&3721431625
&\cr
&5
&229305888887625
&12129909700200
&\cr
&6
&248249742118022000
&31147299733286500
&\cr
&7
&295091050570845659250
& 71578406022880761750
&\cr
&8
&  375632160937476603550000
&154990541752961568418125
&\cr
\noalign{\hrule}}\hrule}$$}
\vskip - 9 mm
\centerline{{\bf Table 10.a:} The weighted sum of BPS states $n^\gg_d$ for the 
quintic.}}
\noindent
We agree with those results. Note however that the BPS numbers on the genus 1 
cases 
differ from the invariants defined in \bcovI, where the maps from the 
torus to itself are subtracted as explained in sec 3.3.  

Unlike the non-compact cases, the reduction of the holomorphic
anomaly involves a global property of the model, the Euler number. 
The general form of $F_2$ has been given in \bcovII. In appendix B we give the 
complete result of the reduction of $F_3$\foot{The $F_4,F_5$
expressions can be made available on request.}. In view of the fast growing 
number of terms in $F_\gg$ with the worldsheet genus $\gg$ one may hope that the ring of modular 
functions on the moduli space of the concrete Calabi-Yau -- here the quintic-- 
transforming in ${\cal L}^{2-2\gg}$ has a much lower dimension, so that 
there are many relations between the terms in $F_\gg$. 
Restricting the expression given in appendix B for $F_3$ to the 
one-moduli case one has $50$ terms of different functional form. 
Somewhat surprisingly we find only one relation between the terms, 
which is reported in the Appendix B.

With these formulas we obtain the following genus $\gg=2,3$ 
results.
\vskip + 3 mm
\noindent
{\vbox{\ninepoint{
$$
\vbox{\offinterlineskip\tabskip=0pt
\halign{\strut
\vrule#&
&\hfil ~$#$
&\hfil ~$#$
&\hfil ~$#$
&\hfil ~$#$
&\vrule#\cr
\noalign{\hrule}
&d
&\gg=2
&3
&4
&
\cr
\noalign{\hrule}
&1
&0
&0
&0
&\cr
&2
&0
&0
&0
&\cr
&3
&0
&0
&0
&\cr
&4
& 534750
&8625
&0
&\cr
&5
& 75478987900
&-15663750
&15520
&\cr
&6
&871708139638250
&3156446162875
&-7845381850
&\cr
&7
& 5185462556617269625
&111468926053022750
&243680873841500
&\cr
&8
&22516841063105917766750 
&1303464598408583455000
&25509502355913526750
&\cr
\noalign{\hrule}}\hrule}$$}
\vskip -9 mm
\centerline{{\bf Table 10.b:} The weighted sum of BPS states $n^\gg_d$ for the 
quintic.}}
\noindent
Note that for $d>7$ the numbers of genus 2 invariants 
are expected to differ from the ones given in \bcovII , because 
of the different definition of the integer invariants
related to $F_2$.

Using  the vanishing of $n^2_{1}=n^2_{2}=n^2_{3}=0$, as explained below, 
we can fix the genus 2 ambiguity. 
Comparing \yukawa~with~\conifold\ we learn that the distance to the
singularity $t=(1-z)=i \mu/\sqrt{\alpha}$.
Written in terms of
$\mu$ it reads
$$f_2=-{571\over 36000}+{83 i\over 7200 \sqrt{5}} \mu^{-1}- {1\over 240} 
\mu^{-2}\ .$$
Note that the leading singularity confirms \conifold. 
Further the constant map contribution is in 
accordance with \bcovII\fp.

To fix the genus 3 anomaly we used the vanishing 
of $n^2_{1}=n^2_{2}=n^2_{3}=0$ and the fact that $n_4^2=(-1)^2\cdot 3\cdot 
2875=8625$, as 
explained below.
The number $n_5^3$ turns out to be 
negative, which is plausible as these curves 
come in  singular families, but the actual value 
has not been checked.   
Together with \fp~for the constant 
map piece this determines 
$$f_3=
\frac{26857}{126000000} 
+ \frac{356921i}{567000000  \sqrt{5}}\mu^{-1}
- \frac{5393}{5040000} \mu^{-2}
- \frac{193i}{50400 \sqrt{5}}\mu^{-3} 
+ \frac{1}{1008}\mu^{-4} \ ,
$$
i.e.\ the leading behavior is in perfect agreement with the
$c=1$ interpretation at the conifold singularity. 
We note that the old  definition of the invariant
yielding $n^2_8=22516841063105918836250$ seems
incompatible with the BPS interpretation as it 
destroys the integrality of the expansion of $F_3$. The leading
behavior of the ambiguity at genus $4$ is
$$f_4=\ldots +{10001i\over 3024000\sqrt{5}}\mu^{-5}-{1\over 1440}\mu^{-6}$$
Most notably in order to get an integer expansion we had also to assume
that there is also a  residue at $\psi^5=0$ more precisely $f_4\sim 
{11853023518768\over 8859375}\psi^{-5}$.

\bigskip
We turn our attention to the application of \bestcorrection\ to verify
some of the instanton numbers that we have calculated for the quintic.
Before we can do this,  we have to understand the restrictions on
the degree and arithmetic genus of a curve in projective space.
This is part of the subject of Castelnuovo theory 
\ref\hart{R. Hartshorne, {\sl Algebraic Geometry\/}, Springer-Verlag, Berlin
1977.}
\ref\har{J.\ Harris, {\sl Curves in Projective Space\/}, University of
Montreal Press, 1982.}.  Castelnuovo theory gives the maximum
arithmetic genus of a nondegenerate irreducible curve $C$ of degree
$d$ in $\IP^r$.  Here, nondegenerate means that $C$ is not contained
in any hyperplane.  This is not a restriction, since any curve is
nondegenerate inside the linear subspace that it spans.

We give a somewhat detailed description for $\IP^2$, $\IP^3$, and a
general formula for $\IP^r$.  

In $\IP^2$,
Castelnuovo theory is trivial, since a degree $d$ curve necessarily has 
arithmetic genus $g=(d-1)(d-2)/2$.

In $\IP^3$, the result of \ref\castp{L.~Gruson and C.~Peskine,
``Genre des courbes de l'espace projectif, II'', Ann.\ Sci.\ ENS $\us{15}$
(1982) 401--418.}  says that a nondegenerate
curve of degree $d$ either lies on a quadric surface or else there is
a number $g(d)$ (to be described presently) such that any curve in $\IP^3$
has genus $g\le g(d)$.

Note that a quadric surface is just $\IP^1\times\IP^1$ so all
possibilities for $(d,g)$ can be computed from the bidegrees $(a,b)$
and $d=a+b$, $g=(a-1)(b-1)$.  Some easy algebra shows that a curve
with degree and genus $(d,g)$ can be found on a quadric precisely when
$(d-2)^2-4g$ is a perfect square.

For $g(d)$, we get
\eqn\castthree{\eqalign{
g(d)&={d^2-3d+6\over 6}\quad d\equiv 0 ({\rm mod}\ 3)\cr
&={d^2-3d+2\over 6}\quad d\not\equiv 0 ({\rm mod}\ 3)\cr
}}

For $\IP^r$, we give a less complete answer, and just note that the maximum
genus possible is found by writing
$$d-1=m(r-1)+\epsilon$$
with $0\le\epsilon < m$. The formula is
\eqn\castr{g\le {m\choose 2}(r-1)+m\epsilon.}
If $r=3$, it can be checked that the right hand side of \castr\ is the
genus of a degree $d$ curve on a quadric, compare with Section 8.4.

Let's apply these formulas to the quintic in low degree.

For $d=1,2$, it is clear that the only genus possible is 0.  For $d=3$, we
can get $g=1$ in the plane, and $g=0$ is possible, even in $\IP^3$ (e.g.\
bidegree $(2,1)$ in $\IP^1\times\IP^1$).  Thus $n_1^g=n_2^g=0$ for $g>0$
and $n_2^g=0$ for $g>1$, consistent with Tables 10.a and 10.b.

Now $d=4$ is more interesting.  This is possible in $\IP^2$ only for
$g=3$.  In $\IP^3$ it is either on a quadric or $g\le 1$ by
Castelnuovo theory.  But we check immediately that $g=0,1$ are the
only possibilities on a quadric as well.  Applying Castelnuovo theory
to $\IP^4$, we again get $g\le 1$.  So $g=2$ is impossible on a
quintic, and $n_4^g=0$ for $g\ge4$.

But this discussion does not mean that $n_4^2$ vanishes.  Quite the
contrary, we can apply \smallnodes\ and \bestcorrection\ to the family
of $g=3$ curves, which are plane quartics.  The plane intersects the
quintic in a quintic plane curve containing the quartic, leaving a
residual line.  So we get the moduli space of these $d=4,g=3$ curves
by taking a line on the quintic, and passing all possible 2-planes
through it, leaving a quartic over by reversing the above reasoning.
So the moduli space is 2875 copies of $\cM=\IP^2$.  In passing, we
note that $n_4^3=2875(3)=8625$ by \topgenus.  The
universal curve $\cC$ is a bit subtle.  The projection $\cC\to X$ ($X$
is the quintic) is 1-1 except over the line (since a point of $X$ not on the
line determines a unique 2-plane containing the line).  But the fiber
of this projection over the line is a $\IP^1$, so we must add 2 to the
Euler characteristic of the CY ($-200$) to get the Euler
characteristic ($-198$) of $\cC$.  Then the first equation in \smallnodes\
together with \bestcorrection\ gives for $n_4^2$
the quantity $2875(-1)(-198+(4)(3))=534750$, as there are no reducible curves
of degree 4 in the plane.  This is in agreement with Table~10.b.

We can similarly apply Castelnuovo theory to degree 5.  Here the result is
that plane quintics have genus 6, and otherwise the genus is at most 2.
The moduli space of plane quintic curves is the same as the moduli space
$G=G(2,4)$ of $\IP^2$s in $\IP^4$.
By \topgenus, we get $n_5^6=(-1)^6e(G)=10$.  We can also in principle
get $n_5^g$ for $3\le g\le 5$ from \smallnodes\ and \bestcorrection.
Unfortunately, we don't know how to compute the Euler characteristic of
$\hc3$, so we can only compute $n_5^g$ for $g=4,5$ at present.

We can calculate $e(\cC)$ as usual by considering the projection $\cC\to X$.
The fiber over a point $p\in X$ is identified with the set of 2 planes
in $\IP^4$ which contain $p$.  This in turn is identified with the 
Grassmannian $G(1,3)$ of lines in $\IP^3$.  So $e(\cC)=e(X)e(G(1,4))=
-200\cdot10=-2000$.  Then the first equation in \smallnodes\ together with
\bestcorrection\ gives $n_5^5=(-1)^5(-2000+(2\cdot3-2)10)=1960$.

The map $\hc2\to {\rm Hilb}^2X$ has fiber over $Z\in{\rm Hilb}^2X$ the
set of all 2 planes in $\IP^4$ containing $Z$.  Since $Z$ is either a
pair of distinct points or a single point of multiplicity 2 with a 
distinct tangent direction, we see that this fiber is isomorphic to 
$\IP^3$ in either case.  Also, ${\rm Hilb}^2X$ is obtained from 
the symmetric product ${\rm Sym}^2X$ by blowing up the diagonal,
replacing each point of the diagonal with a $\IP^2$.
Thus 
$$\eqalign{
e({\rm Hilb}^2)&=e({\rm Sym}^2X)+e(X)(e(\IP^2)-1)\cr
&={-199\choose2}+(-200)\cdot2=19500}$$
from which it follows from the second equation in \smallnodes\ together
with \bestcorrection\ that $n_5^4=15520$.

For $0\le g\le 2$, there are both smooth curves of degree 5 and genus $g$
in the quintic as well as a contribution from singular plane quintics.
In the case of $g=0$, there are actually finitely many singular curves,
which were enumerated in \vain.

\subsec{The sextic, the bicubic and four conics}
  
A further typical hypersurface example
is the sextic in the weighted projective space ${\bf P}^4(1,1,1,1,2)$.    
We find the following all integer\foot{In \bcovII~it was claimed 
that this example has a half-integral invariant for $n_4^{(2)}$, which would be 
in contradiction with the M-theory interpretation of the $R^2 
F^{2g-2}$ amplitude. Luckily we find that the problem relied on a computational 
error.}  $n_d^\gg$ invariants for this  case
\vskip + 3 mm
\noindent
{\vbox{\ninepoint{
$$
\vbox{\offinterlineskip\tabskip=0pt
\halign{\strut
\vrule#&
&\hfil ~$#$
&\hfil ~$#$
&\hfil ~$#$
&\vrule#\cr
\noalign{\hrule}
&d
&\gg=0
&1
&
\cr
\noalign{\hrule}
&1
&7884
&0
&\cr
&2
& 6028452
& 7884
&\cr
&3
&11900417220
&145114704
&\cr
&4
&34600752005688
&1773044322885
&\cr
&5
&124595034333130080
&17144900584158168
&\cr
&6
& 513797193321737210316
& 147664736456952923604
&\cr
&7
& 2326721904320912944749252
&1197243574587406496495592
&\cr
&8
&11284058913384803271372834984  
& 9381487423491392389034886369
&\cr
\noalign{\hrule}}\hrule}$$}}
\vskip - 3 mm
\noindent
{\vbox{\ninepoint{
$$
\vbox{\offinterlineskip\tabskip=0pt
\halign{\strut
\vrule#&
&\hfil ~$#$
&\hfil ~$#$
&\hfil ~$#$
&\vrule#\cr
\noalign{\hrule}
&d
&\gg=2
&3
&
\cr
\noalign{\hrule}
&1
&0
&0
&\cr
&2
&0 
&0
&\cr
&3
&17496
&576
&\cr
&4
&10801446444
&-14966100 
&\cr
&5
&571861298748384  
&1412012838168 
&\cr
&6
& 13753100019804005556
&403369763928730938
&\cr
&7
&233127389355701229349884
&552961951281452536352
&\cr
&8
&\ \,  3246006977306701566424657380
&\ \,  560485610266924061005490676 
&\cr
\noalign{\hrule}}\hrule}$$}
\vskip -9 mm
\centerline{{\bf Table 11:} The weighted sum of BPS states $n^\gg_d$ for the 
sextic.}}
\noindent
The ambiguity can be fixed from the vanishing of $n_1^{2}=n_2^{2}=0$ 
plus the general form of the constant map contribution. Again we find the 
leading term of $f_2$ in accordance with 
\conifold~$f_2=-{473\over 25920}-{463 \over 51840 \sqrt{3}} \mu^{-1}-
{1\over 240} \mu^{-2}$. 

For the genus 3 contribution we can fix a combination of 
$n_3^{3},n_4^{3}$ by demanding the expected behavior of the $\mu^{-4}$ 
term in
$$f_3=
\frac{3917}{ 26127360} 
+ \frac{61447i}{117573120 \sqrt{3}}\mu^{-1}
- \frac{107945}{94058496} \mu^{-2}
- \frac{25 i}{ 8064 \sqrt{3}}\mu^{-3} 
+ \frac{1}{1008}\mu^{-4} \ .
$$  
This yields $n_4^{3}=36 (-865581+781 n_3^{3})$. The fact the $\mu^{-3}$ term 
is a relatively simple fraction for other cases leads to the conjecture 
that the correct value is $n_3^{3}=576$.

This can be checked by geometry as follows.  Note that by projection to the
first four coordinates, $X$ admits a $3-1$ cover of $\IP^3$.  Writing the
equation of the CY in the form 
$$x_5^3+f_4(x_1,x_2,x_3,x_4)x_5+f_6(x_1,x_2,x_3,x_4)=0$$ 
We see that the branch locus $4f_4^3+27f_6^2$ has degree $12$.  A curve of
degree $3$ must either map isomorphically to a degree 3 curve in $\IP^3$
or be a triple cover of a line.  Since degree 3 curves have genus at most
1, and we are interested in genus 3, we must have a triple cover of a line
branched at 4 points, which has genus 4.  Thus $\cM=G(1,3)$, the
Grassmannian of $\IP^1$s in $\IP^3$.  This has dimension 4 and Euler
characteristic 6.
So we already see that $n_3^4=6$.  

We project the universal curve to $X$ as usual;
the fiber over $p\in X$ is the set of triple covers of lines which contain
$p$, which is in 1-1 correspondence with the set of lines in $\IP^3$ which
contain the image of $p$ in $\IP^3$.  This is isomorphic to $\IP^2$.  We
get $e(\cC)=e(\IP^2)e(X)=3(-204)=-612$.  
Now an application of the first equation in \smallnodes\ and \bestcorrection\
gives $(-1)^3(-612+6\cdot 6)=576$.

We next consider as simplest complete intersection cases two cubics in ${\bf 
P}^5$, i.e.
$X_{3,3}(1^6)$ and four conics in ${\bf P}^7$, denoted $X_{2,2,2,2}(1^8)$. 
{}

{\vbox{\ninepoint{
$$
\vbox{\offinterlineskip\tabskip=0pt
\halign{\strut
\vrule#
&
&\hfil ~$#$
&\hfil ~$#$
&\hfil ~$#$
&\hfil ~$#$ 
&\hfil ~$#$
&\vrule#\cr
\noalign{\hrule}
&d
&\gg=0
&1
&2
&3
&
\cr
\noalign{\hrule}
&1
&1053
&0
&0
&0
&\cr
&2
& 52812
&0
&0
&0
&\cr
&3
& 6424326
& 3402 
&0
&0
&\cr
&4
&1139448384
&5520393
&0
&0
&\cr
&5
& 249787892583 
&4820744484
& 5520393
&0
&\cr
&6
& 62660964509532
&3163476678678
&23395810338
&6852978
&\cr
&7
&17256453900822009
&1798399482469092 
&42200615912499 
& 174007524240
&\cr
&8
& 5088842568426162960
&944929890847710108
&50349477671013600
&785786604262830
&\cr
\noalign{\hrule}}\hrule}$$}
\vskip - 9 mm
\centerline{{\bf Table 12:} The weighted sum of BPS states $n^\gg_d$ for the compl. 
intersection $X_{3,3}(1^6)$.
}\vskip7pt}}

{\vbox{\ninepoint{
$$
\vbox{\offinterlineskip\tabskip=0pt
\halign{\strut
\vrule#
&
&\hfil ~$#$
&\hfil ~$#$
&\hfil ~$#$ 
&\hfil ~$#$
&\hfil ~$#$
&\hfil ~$#$ 
&\hfil ~$#$
&\hfil ~$#$
&\vrule#\cr
\noalign{\hrule}
&\gg
&A^\gg_0
&A^\gg_1
&A^\gg_2
&A^\gg_3
&A^\gg_4
&B^\gg_1
&B^\gg_2
&
\cr
\noalign{\hrule}
&2
&-{2861 \over 116640}
&{103 i\over 15552   }
&-{1\over 240}
&
&
& -{71 \over 64} 
&
&\cr
&3
& \frac{1921447 }{6348948480 }
& \frac{1358671 \,i}{2116316160 }
&- \frac{118421 }{ 70543872} 
&-\frac{ 377 \,i}{181440 }
& \frac{1}{1008}
& \frac{ 5}{27648 }
&- \frac{240501}{35840 }
&\cr\noalign{\hrule}}\hrule}$$}}

with $\mu  = 3 (1-\psi^6)/i$ $\rho=3^6 \psi^6$.

{\vbox{\ninepoint{
$$
\vbox{\offinterlineskip\tabskip=0pt
\halign{\strut
\vrule#
&
&\hfil ~$#$
&\hfil ~$#$
&\hfil ~$#$
&\hfil ~$#$ 
&\hfil ~$#$
&\vrule#\cr
\noalign{\hrule}
&d
&\gg=0
&1
&2
&3
&
\cr
\noalign{\hrule}
&1
& 512
&0
&0
&0
&\cr
&2
&9728
&0
&0
&0
&\cr
&3
&416256 
&0
&0
&0
&\cr
&4
&25703936
&14752
&0
&0
&\cr
&5
&1957983744 
&8782848
&0
&0
&\cr
&6
&170535923200
&2672004608
&1427968
&0
&\cr
&7
&16300354777600
&615920502784
&2440504320
&86016
&\cr
&8
&1668063096387072
&123699143078400
&1628589698304 
&2403984384
&\cr
\noalign{\hrule}}\hrule}$$}
\vskip - 9 mm
\centerline{{\bf Table 13:} The weighted sum of BPS states $n^\gg_d$ for the compl. 
intersection 
$X_{2,2,2,2}(1^8)$.}}
The indices of the Picard-Fuchs  system is 4 fold degenerate at $\psi^8=0$ and we 
find the leading behavior from the ambiguity

\vskip 1 mm
{\vbox{\ninepoint{
$$
\vbox{\offinterlineskip\tabskip=0pt
\halign{\strut
\vrule#
&
&\hfil ~$#$
&\hfil ~$#$
&\hfil ~$#$ 
&\hfil ~$#$
&\hfil ~$#$
&\hfil ~$#$ 
&\hfil ~$#$
&\hfil ~$#$
&\vrule#\cr
\noalign{\hrule}
&\gg
&A^\gg_0
&A^\gg_1
&A^\gg_2
&A^\gg_3
&A^\gg_4
&B^\gg_1
&B^\gg_2
&
\cr
\noalign{\hrule}
&2
&-{133741 \over 2949120 }
&{349  i\over 46080  }
&-{1\over 240}
&
&
& -{5377 \over 15} 
&
&\cr
&3
& \frac{110365853 }{1014686023680 }
& \frac{ 1859479 \,i}{1698693120 }
&- \frac{30277}{13762560  } 
&- \frac{149\,i}{64512}
& \frac{1}{1008}
& -\frac{2181905 }{221184 }
& \frac{23115884 }{2835  }
&\cr\noalign{\hrule}}\hrule}$$}}
\vskip - 3 mm 
\noindent
with $\mu =  4 (1-\psi^8)/i$ and  $\rho= \psi^8 2^{16}$.  It would be very 
interesting to find the analog of the $c=1$ model at the $\rho=0$ singularity.

\bigskip
\noindent {\bf Acknowledgement:} 
It is a pleasure to thank Tom Graber, Rajesh Gopakumar, 
Brian Harbourne, Shinobu Hosono,
J. Maldacena, Rick Miranda, S.-T. Yau and Eric Zaslow 
for valuable discussions and comments.  C.V. would also like to thank
Rutgers Physics Department for hospitality while this work
was in progress.
 The research of Sheldon Katz was 
supported in part by NSA grant
MDA904-98-1-0009.  The research of Albrecht Klemm was supported in part
by a DFG Heisenberg Fellowship and
NSF Math/Phys DMS-9627351.  The research of Cumrun Vafa was supported in part
by NSF grant PHY-92-18167.
This project received additional support
from the American Institute of Mathematics.

\vfill
\eject

\newsec{Appendix A: Low degree classes on the del Pezzo Surfaces}

{\ninepoint $$
\vbox{\offinterlineskip\tabskip=0pt
\halign{\strut
\vrule#&
~$#$~&
\vrule~$#$~
\hfil\vrule\cr
\noalign{\hrule}
& d & \phantom{xxxxxxxxxxxxxxxxxxxxxxxxxxxx} g=0 \ \ {\rm  classes} \cr
\noalign{\hrule}
&1&e_i \ ^1|  (1;1^2)\ ^2| \ 
 (2;1^5)\ ^5| \
 (3;2,1^6)\ ^7| \ 
(6;3,2^7)\ (5;2^6,1^2)\ (4;2^3,1^5)\ ^8| \cr
\noalign{\hrule}
&2&(1;1)\ ^1| 
 (2;1^4)\ ^4| 
 (3;2,1^5)\ ^6|  
 (5;2^6,1)\ (4;2^3,1^4)\ ^7| \cr && 
 (11;4^7,3)\ (10;4^4,3^4)\ (9;4^2,3^5,2)\ (8;4,3^4,2^3)\ (8;3^7,1)\ (7;4,3,2^6)\ 
(7;3^4,2^3,1)\ 
\cr &&(6;3^2,2^4,1^2)\ (5;3,2^3,1^4)\ (4;3,1^7)\ ^8| \ 
 \cr
\noalign{\hrule}
&3&(1;0)\ ^0| \ (2;1^3)\ ^3| 
 (3;2,1^4)\ ^5|  
 (5;2^6)\ (4;2^3,1^3)\ ^6| \cr && 
 (8;3^7)\ (7;3^4,2^3)\ (6;3^2,2^4,1)\ (5;3,2^3,1^3)\ (4;3,1^6)\ ^7|  \cr
\noalign{\hrule}
&4&(2;1^2)\ ^2|  
 (3;2,1^3)\ ^4|  
 (4;2^3,1^2)\ ^5| 
 (6;3^2,2^4)\ (5;3,2^3,1^2)\ (4;3,1^5)\ ^6| \cr
\noalign{\hrule}
&5&(2;1)\ ^1| 
 (3;2,1^2)\ ^3| 
 (4;2^3,1)\ ^4| 
 (5;3,2^3,1)\ (4;3,1^4)\ ^5| 
\cr && 
 (8;4,3^5)\ (7;4,3^2,2^3)\ (7;3^5,1)\ (6;3^3,2,1^2)\ (6;4,2^4,1)\ (5;3^2,1^4)\ ^6| 
\cr 
\noalign{\hrule}
&6& (2;0) ^0| \ (3;2,1)\ ^2|\ 
 (4;2^3)\ ^3|\  
 (5;3,2^3)\ (4;3,1^3)\ ^4|  
 (7;3^5)\ (6;3^3,2,1)\ (6;4,2^4)\ (5;3^2,1^3)\ ^5| \cr 
\noalign{\hrule}
&7&(3;2)\ ^1| 
 (4;3,1^2)\ ^3|  
 (6;3^3,2)\ (5;3^2,1^2)\ ^4|  \cr && 
 (8;4^2,3^3)\ (7;4,3^3,1)\ (7;4^2,2^3)\ (6;4,3,2,1^2)\ (5;4,1^4)\ ^5| 
\cr 
\noalign{\hrule}
&8&(4;3,1)\ ^2| 
 (5;3^2,1)\ ^3|  
 (7;4,3^3)\ (6;4,3,2,1)\ (5;4,1^3)\ ^4| \cr
\noalign{\hrule}  
\noalign{\hrule}
&  & \phantom{xxxxxxxxxxxxxxxxxxxxxxxxxxxx} g=1 \ \ {\rm  classes} \cr
\noalign{\hrule}
&1&(3;1^8)\ ^8| \ \cr
\noalign{\hrule}
&2&(3;1^7)\ ^7| 
 (9;4,3^7)\ (8;3^6,2^2)\ (7;3^3,2^5)\ (6;3,2^6,1)\ (5;2^5,1^3)\ (4;2^2,1^6)\ ^8| \ 
\cr
\noalign{\hrule}
&3&(3;1^6)\ ^6|  (6;3,2^6)\ (5;2^5,1^2)\ (4;2^2,1^5)\ ^7| \cr
\noalign{\hrule}
&4&(3;1^5)\ ^5| \
 (5;2^5,1)\ (4;2^2,1^4)\ ^6| \cr &&
 (9;4^2,3^5)\ (8;4,3^4,2^2)\ (7;3^4,2^2,1)\ (7;4,3,2^5)\ (6;3^2,2^3,1^2)\ 
(5;3,2^2,1^4)\ ^7| \cr
\noalign{\hrule}
&5&(3;1^4)\ ^4|
 (5;2^5)\ (4;2^2,1^3)\ ^5| \
 (7;3^4,2^2)\ (6;3^2,2^3,1)\ (5;3,2^2,1^3)\ ^6|  \cr
\noalign{\hrule}
&6&(3;1^3)\ ^3| 
 (4;2^2,1^2)\ ^4| 
 (6;3^2,2^3)\ (5;3,2^2,1^2)\ ^5| \cr && 
 (9;4^3,3^3)\ (8;4^2,3^2,2^2)\ (7;4,3^2,2^2,1)\ (6;4,2^3,1^2)\ (6;3^3,1^3)\ ^6| 
\cr
\noalign{\hrule}
&7&(3;1^2)\ ^2|
 (4;2^2,1)\ ^3|  
 (5;3,2^2,1)\ ^4| 
 (7;4,3^2,2^2)\ (6;4,2^3,1)\ (6;3^3,1^2)\ ^5| \cr
\noalign{\hrule}
&8&(3;1)\ ^1|   
 (4;2^2)\ ^2|   
 (5;3,2^2)\ ^3| 
 (6;4,2^3)\ (6;3^3,1)\ ^4| \cr 
\noalign{\hrule}
\noalign{\hrule}
&  & \phantom{xxxxxxxxxxxxxxxxxxxxxxxxxxxx} g=2 \ \ {\rm  classes} \cr
\noalign{\hrule}
&2&(6;2^8)\ ^8| \ 
 \cr
\noalign{\hrule}
&3&(14;5^7,4)\ (13;5^4,4^4)\ (12;5^2,4^5,3)\ (11;4^7,2)\ (11;5,4^4,3^3)\ 
(10;4^4,3^3,2)\ (10;5,4,3^6)\ 
\cr &&(9;4^2,3^4,2^2)\ (8;4,3^3,2^4)\ (8;3^6,2,1)\ (7;4,2^7)\ (7;3^3,2^4,1)\ 
(6;3,2^5,1^2)\ (5;2^4,1^4)\ 
\cr &&(4;2,1^7)\ ^8| \ 
 \cr
\noalign{\hrule}
&4&(8;3^6,2)\ (7;3^3,2^4)\ (6;3,2^5,1)\ (5;2^4,1^3)\ (4;2,1^6)\ ^7|  \cr
\noalign{\hrule}
&5&(6;3,2^5)\ (5;2^4,1^2)\ (4;2,1^5)\ ^6| \cr 
\noalign{\hrule}
&6&(5;2^4,1)\ (4;2,1^4)\ ^5| \ (8;4,3^4,2)\ (7;3^4,2,1)\ (7;4,3,2^4)\ 
(6;3^2,2^2,1^2)\ (5;3,2,1^4)\ ^6|  \cr
\noalign{\hrule}
&7&(5;2^4)\ (4;2,1^3)\ ^4| 
 (7;3^4,2)\ (6;3^2,2^2,1)\ (5;3,2,1^3)\ ^5|\cr
\noalign{\hrule}
&8&(4;2,1^2)\ ^3| 
 (6;3^2,2^2)\ (5;3,2,1^2)\ ^4| 
 (8;4^2,3^2,2)\ (7;4,3^2,2,1)\ (6;4,2^2,1^2)\ ^5| \cr
\noalign{\hrule}}}$$}

{\ninepoint $$
\vbox{\offinterlineskip\tabskip=0pt
\halign{\strut
\vrule#&
~$#$~&
\vrule~$#$~
\hfil\vrule\cr
\noalign{\hrule}
& d & \phantom{xxxxxxxxxxxxxxxxxxxxxxxxxxxx} g=3 \ \ {\rm  classes} \cr
\noalign{\hrule}
&3&(12;5,4^7)\ (11;4^6,3^2)\ (10;4^3,3^5)\ (9;4,3^6,2)\ (8;3^5,2^3)\ (7;3^2,2^6)\ 
(6;2^7,1)\ ^8| \  \cr
\noalign{\hrule}
&4&(6;2^7)\ ^7| \cr 
\noalign{\hrule}
&5&(11;4^7)\ (10;4^4,3^3)\ (9;4^2,3^4,2)\ (8;3^6,1)\ (8;4,3^3,2^3)\ (7;4,2^6)\ 
(7;3^3,2^3,1)\ 
\cr &&(6;3,2^4,1^2)\ (5;2^3,1^4)\ (4;1^7)\ ^7| \cr 
\noalign{\hrule}
&6&(8;3^6)\ (7;3^3,2^3)\ (6;3,2^4,1)\ (5;2^3,1^3)\ (4;1^6)\ ^6| \cr
\noalign{\hrule}
&7&(6;3,2^4)\ (5;2^3,1^2)\ (4;1^5)\ ^5| \cr  
&8&(5;2^3,1)\ (4;1^4)\ ^4| \ (8;4,3^4)\ (7;3^4,1)\ (7;4,3,2^3)\ (6;3^2,2,1^2)\ 
(5;3,1^4)\ ^5| \cr
\noalign{\hrule}
\noalign{\hrule}
&  & \phantom{xxxxxxxxxxxxxxxxxxxxxxxxxxxx} g=4 \ \ {\rm  classes} \cr
\noalign{\hrule}
&3&(9;3^8)\ ^8| \ \cr
\noalign{\hrule}
&4&(18;7^2,6^6)\ (17;7,6^5,5^2)\ldots ^8| \cr
\noalign{\hrule}
&5&(9;4,3^6)\ (8;3^5,2^2)\ (7;3^2,2^5)\ (6;2^6,1)\ ^7| \cr
\noalign{\hrule}
&6&(6;2^6)\ ^6| \cr && 
 (13;5^5,4^2)\ (12;6,4^6)\ (12;5^3,4^3,3)\ (11;5,4^5,2)\ (11;5^2,4^2,3^3)\ 
(10;4^5,2^2)\ (10;5,4^2,3^3,2)\ 
\cr &&(9;4^3,3,2^3)\ (9;4^2,3^4,1)\ (9;5,3^4,2^2)\ (8;4,3^3,2^2,1)\ (8;4^2,2^5)\ 
(7;4,2^5,1)\ (7;3^3,2^2,1^2)\ 
\cr &&(6;3,2^3,1^3)\ (5;2^2,1^5)\ ^7| \cr 
\noalign{\hrule}
&7&(9;4^2,3^4)\ (8;4,3^3,2^2)\ (7;4,2^5)\ (7;3^3,2^2,1)\ (6;3,2^3,1^2)\ 
(5;2^2,1^4)\ ^6| \cr
\noalign{\hrule}
&8&(7;3^3,2^2)\ (6;3,2^3,1)\ (5;2^2,1^3)\ ^5| \cr
\noalign{\hrule}
\noalign{\hrule}
&  & \phantom{xxxxxxxxxxxxxxxxxxxxxxxxxxxx} g=5 \ \ {\rm  classes} \cr
\noalign{\hrule}
&4&(17;6^7,5)\ (16;6^4,5^4)\ (15;6^2,5^5,4)\ (14;6,5^4,4^3)\ (14;5^7,3)\ 
(13;5^4,4^3,3)\ (13;6,5,4^6)\ 
\cr &&(12;5^2,4^4,3^2)\ (11;4^6,3,2)\ (11;5,4^3,3^4)\ (10;5,3^7)\ (10;4^3,3^4,2)\ 
(9;4,3^5,2^2)\ (8;3^4,2^4)\ 
\cr &&(7;3,2^7)\ ^8| \ 
 \cr
\noalign{\hrule}
&5&(18;8,7^3,5^4)\ldots ^8| \ 
 \cr
\noalign{\hrule}
&6&(12;5^2,4^5)\ (11;5,4^4,3^2)\ (10;4^4,3^2,2)\ (10;5,4,3^5)\ (9;4^2,3^3,2^2)\ 
(8;3^5,2,1)\ (8;4,3^2,2^4)\ 
\cr &&(7;3^2,2^4,1)\ (6;2^5,1^2)\ ^7| \cr
\noalign{\hrule}
&7&(8;3^5,2)\ (7;3^2,2^4)\ (6;2^5,1)\ ^6| \cr 
\noalign{\hrule}
&8&(6;2^5)\ ^5| \cr
\noalign{\hrule}
\noalign{\hrule}
&  & \phantom{xxxxxxxxxxxxxxxxxxxxxxxxxxxx} g=6 \ \ {\rm  classes} \cr
\noalign{\hrule}
&4&(15;6,5^7)\ (14;5^6,4^2)\ (13;5^3,4^5)\ (12;5,4^6,3)\ (11;4^5,3^3)\ 
(10;4^2,3^6)\ (9;3^7,2)\ ^8| \ \cr
\noalign{\hrule}
&5&(18;8,7,6^5,4)\ldots ^8| \cr
\noalign{\hrule}
&6&(11;4^6,3)\ (10;4^3,3^4)\ (9;4,3^5,2)\ (8;3^4,2^3)\ (7;3,2^6)\ ^7| \cr
\noalign{\hrule}
&7&(15;6^3,5^4)\ldots ^7|  \cr
\noalign{\hrule}
&8&(10;4^4,3^2)\ (9;4^2,3^3,2)\ (8;3^5,1)\ (8;4,3^2,2^3)\ (7;3^2,2^3,1)\ 
(6;2^4,1^2)\ ^6|
\cr
\noalign{\hrule}}}$$}

\newsec{Appendix B: B-model expression for $F_g$}
The $F_g$ can be determined by recursively solving the 
B-model anomaly equation \bcovII. As each contribution comes from the boundary 
of the moduli space of Riemann surfaces the result has a graph interpretation
in which each graph corresponds to a possible degeneration of the genus 
$g$ curve into components with lower genera. In the local case  the descendent of 
the dilaton
decouples and only one sort of propagator $S^{i,j}$ occurs. They are in 
one to one correspondence with the tubes connecting the irreducible components, 
in this way each contribution of $F_g$ corresponds precisely to one boundary 
stratum \kz . 
In the global case one has three sorts of propagators shown below
{
\goodbreak\midinsert
\centerline{\epsfxsize 3 truein\epsfbox{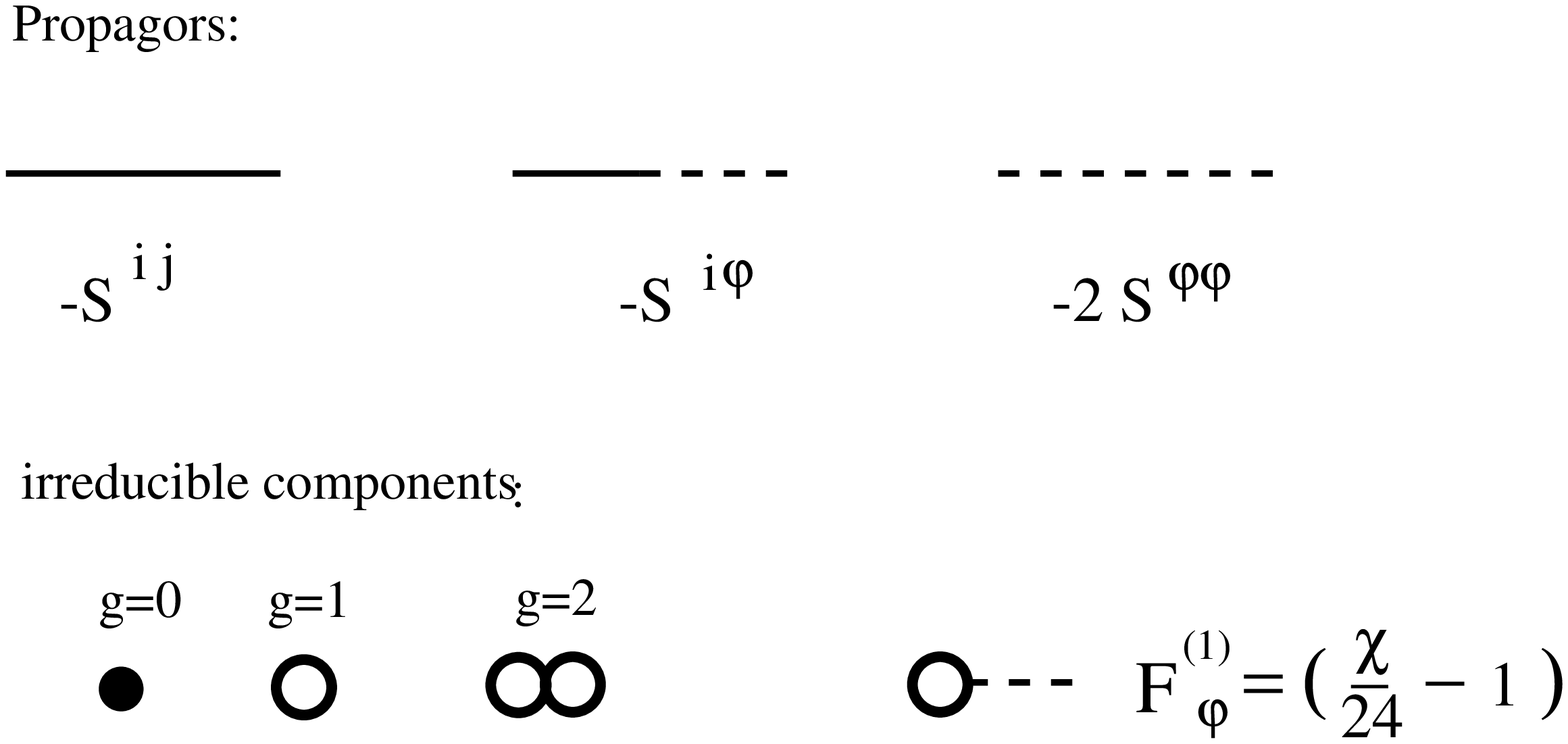}}
\leftskip 1pc\rightskip 1pc \vskip0.3cm
\endinsert}
The following Feynman rules hold for $n,m=0,1,2\ldots$ 
$$F^{(0)}_{\varphi^n}=0,\ F^{(0)}_{i\varphi^n}=0,\ F^{(0)}_{i j\varphi^n}=0,\ 
F^{(g)}_{i_1,\ldots i_m,\varphi^{n+1}}=(2g-2+n+m)=F^{(g)}_{i_1,\ldots 
i_m,\varphi^{n}}$$
The graph contribution is divided by the following symmetry factors: 
$k!$ for $k$ equal (self)links joining the same vertices, $2$ for each selflink 
$S^{\phi,\phi}$, $S^{i,i}$ times the order of the graph automorphism 
obtained by permuting the vertices. The generation of graphs proceeds along the 
line described for the the A-model in \kz .

Twelve graphs contribute to $F_2$, which were derived in \bcovII . 
Further we find that $193$ graphs contribute to the free energy at genus $3$. 
As it depends in an universal combinatorial way on the lower genus boundary 
components, which  also applies more generally to non-topological 
string $3$ loop  calculations, and we have a powerfull check via integrality of 
the BPS states on the quintic on its expression, 
we will give it below, despite its complicated
nature. Let $\bar \chi ={\chi \over 24}$ with $\chi$ the Euler number of the 
target space, $F$ the genus 0 prepotential, $G=F_1$, $H=F_2$ and $S^{ij}$, 
$S^{i}$, $S$ the propagators of Kodaira-Spencer gravity \bcovII~, then we found

\vbox{\ninepoint{
$$\eqalign{F_3&= 
2 (1+ 2{\bar \chi}) HS +  
{\bar \chi} ( 1+ {\bar \chi}- 2 {\bar \chi}^2) S^2 + 
2G_{i}HS^{i} + 
(2+{\bar \chi}) H_{i}S^{i}  - 
2 {\bar \chi} (1+ 2{\bar \chi})G_{i}S\, S^{i}-
\cr &
{1\over 2}(2+ 3{\bar \chi}) G_{i}G_{j}S^{i}S^{j} -  
{1\over 2}(2+3{\bar \chi}-{\bar \chi}^2) G_{i j}S^{i}S^{j} + 
{1\over 6} {\bar \chi} (2+3{\bar \chi} + {\bar \chi}^2) 
F_{i j k}S^{i}S^{j}S^{k}  + 
G_{i}H_{j}S^{i j} + 
\cr &
{1\over 2} H_{i j}S^{i j} - 
(1+2{\bar \chi} )G_{i}G_{j}S\,S^{i j} - 
(1+2 {\bar \chi}) G_{i j}S\, S^{i j} - 
G_{i}G_{j}G_{k}S^{k}S^{i j} - 
(2+{\bar \chi} )G_{i}G_{j k}S^{k}S^{i j} - 
\cr &
(1+{\bar \chi})G_{i j k}S^{k}S^{i j}  + 
{\bar \chi} (1+ 2 {\bar \chi})  F_{i j k}S\ S^{k}S^{i j} +
{1\over   4 } (2+3 {\bar \chi} +{\bar \chi}^2)
F_{i j k l}S^{k}S^{l}S^{i j}    - 
{1\over  2  }G_{i}G_{j k l}S^{i l}S^{j k}+
\cr &
{1\over 2} (2+ 3{\bar \chi} + {\bar \chi}^2) 
F_{i j k}G_{l}S^{j}S^{k}S^{i l}- 
G_{i}G_{j k}S^{i}S^{j k} -  
{1\over  2  }G_{i}G_{j}G_{k l}S^{i l}S^{j k}+ 
(1+{3\over 2} {\bar \chi} )F_{i j k}G_{l}S^{k}S^{l}S^{i j} -
\cr & 
{1\over   2 }F_{i j k}H_{l}S^{i l}S^{j k} + 
(1+ 2{\bar \chi} ) F_{i j k}G_{l}SS^{i l}S^{j k}  +  
F_{i j k}G_{l}G_{m}S^{m}S^{i l}S^{j k} + 
(1+ {1\over 2}{\bar \chi}) F_{i j k}G_{l m}S^{m}S^{i l}S^{j k} +
\cr & 
(1+ {1\over 2}{\bar \chi}) F_{i j k l}G_{m}S^{l}S^{i m}S^{j k} - 
{1\over   4 }G_{i j}G_{k l}S^{i k}S^{j l} + 
(1+ {1\over 2}{\bar \chi})F_{i j k}G_{l}G_{m}S^{k}S^{i m}S^{j l} - 
{1\over 8   }G_{i j k l}S^{i j}S^{k l}+
\cr &  
(1+ {1\over 2}{\bar \chi}) F_{i j k}G_{l m}S^{k}S^{i l}S^{j m}  - 
{1\over   4 } (2+3 {\bar \chi} +{\bar \chi}^2)F_{i j k}F_{l m n}S^{k}S^{n}S^{i 
l}S^{j m} + 
{1\over  4  }(1+2 {\bar \chi}) F_{i j k l}S\, S^{i j}S^{k l} +   
\cr &
{1\over   8 }(2+{\bar \chi})F_{i j k l m}S^{m}S^{i j}S^{k l} + 
{1\over   4 }F_{i j k l}G_{m}S^{m}S^{i j}S^{k l}+ 
{1\over  4  }(1+{\bar \chi}) F_{i j k}F_{l m n}F_{p q r}S^{r}S^{i l}S^{j p}S^{k 
n}S^{m q}+ 
\cr & 
{1\over  8  }(2+{\bar \chi})F_{i j k}F_{l m n}F_{p q r}S^{k}S^{i p}S^{j n}S^{l 
m}S^{q r}  +
{1\over   4 }F_{i j k l}G_{m n}S^{i m}S^{j n}S^{k l} + 
{1\over   2 }F_{i j k}G_{l}G_{m n}S^{i m}S^{j n}S^{k l} +  
\cr &
{1\over   6 }F_{i j k}G_{l m n}S^{i l}S^{j m}S^{k n} - 
{1\over   6 }(1+ 2 {\bar \chi}) F_{i j k}F_{l m n}S\ S^{i l}S^{j m}S^{k n} - 
{1\over   6 }(1+{\bar \chi}) F_{i j k}F_{l m n p} S^{p}S^{i l}S^{j m} S^{k n} - 
\cr &
{1\over   6 }F_{i j k}F_{l m n}G_{p}S^{p}S^{i l}S^{j m}S^{k n} - 
{1\over 2} (2+{\bar \chi}) F_{i j k}F_{l m n}G_{p}S^{n}S^{i l}S^{j m}S^{k p} + 
{1\over   2 }F_{i j k}G_{l}G_{m}S^{k}S^{i j}S^{l m} +
\cr &  
{1\over  16 }F_{i j k}F_{l m n}F_{p q r s}S^{i j}S^{k s}S^{l m}S^{n p}S^{q r}- 
{1\over   8 }(2+3 {\bar \chi}) F_{i j k}F_{l m n}S^{k}S^{n}S^{i j}S^{l m} - 
{1\over   2 }F_{i j k}F_{l m n}G_{p}S^{n}S^{i p}S^{j k}S^{l m} -
\cr & 
{1\over   4 }(2+3 {\bar \chi}+{\bar \chi}^2) F_{i j k}F_{l m n}S^{j}S^{k}S^{i 
n}S^{l m}- 
{1\over 2} (2+{\bar \chi}) F_{i j k}F_{l m n}G_{p}S^{k}S^{i p}S^{j n}S^{l m} +  
{1\over  8  }F_{i j k l m}G_{n}S^{i j}S^{k n}S^{l m} - 
\cr &
{1\over  16  }F_{i j k}F_{l m n}F_{p q r}F_{s t u}S^{i s}S^{j t}S^{k n}S^{l p}S^{m 
q}S^{r u}+  
{1\over  4  }F_{i j k}G_{l m n}S^{i j}S^{k n}S^{l m} +  
{1\over  48  }F_{i j k l m n}S^{i j}S^{k l}S^{m n} - 
\cr &
{1\over   4 }F_{i j k}F_{l m n}G_{p}S^{p}S^{i j}S^{k n}S^{l m}-  
{1\over  4  } (2+ {\bar \chi} ) F_{i j k}F_{l m n p} S^{p}S^{i j}S^{k n} S^{l m}+ 
{1\over  12 }F_{i j k}F_{l m n}F_{p q r s}S^{i p}S^{j q}S^{k r}S^{l s}S^{m n} -
\cr &   
{1\over   4 }F_{i j k}F_{l m n}G_{p q}S^{i p}S^{j q}S^{k n}S^{l m} - 
{1\over   48 }F_{i j k l}F_{m n p q}S^{i m}S^{j n}S^{k p}S^{l q} - 
{1\over  4  } (2+{\bar \chi}) F_{i j k}F_{l m n p}S^{k}S^{i l}S^{j p}S^{m n}- 
\cr &
{1\over  4  } (1+2 {\bar \chi} )F_{i j k}F_{l m n}S\ S^{i j}S^{k n}S^{l m}  - 
{1\over  4  }F_{i j k}F_{l m n p}G_{q}S^{i l}S^{j q}S^{k p}S^{m n}   - 
{1\over   4 }F_{i j k}F_{l m n}G_{p}G_{q}S^{i p}S^{j q}S^{k n}S^{l m} - 
\cr &
{1\over  4  }F_{i j k}F_{l m n}G_{p}G_{q}S^{i l}S^{j q}S^{k n}S^{m p} -  
{1\over  4  }F_{i j k}F_{l m n}G_{p q}S^{i l}S^{j p}S^{k n}S^{m q} -  
{1\over  12  }F_{i j k}F_{l m n p q}S^{i l}S^{j m}S^{k q}S^{n p} +
 \cr &
{1\over  4  }F_{i j k}F_{l m n}F_{p q r}G_{s}S^{i p}S^{j s}S^{k n}S^{l q}S^{m r} + 
{1\over  8  }F_{i j k}F_{l m n}F_{p q r s}S^{i p}S^{j q}S^{k n}S^{l r}S^{m s}  + 
{1\over   4 }F_{i j k l}G_{m}G_{n}S^{i m}S^{j n}S^{k l}  + 
\cr &
{1\over  2  }F_{i j k}G_{l}G_{m n}S^{i j}S^{k n}S^{l m} + 
{1\over   16 }F_{i j k}F_{l m n p q}S^{i j}S^{k q}S^{l m}S^{n p} -
{1\over   48 }F_{i j k}F_{l m n}F_{p q r}F_{s t u}S^{i p}S^{j s}S^{k n}S^{l m}S^{q 
r}S^{t u} -  
\cr &
{1\over  16  }F_{i j k l}F_{m n p q}S^{i j}S^{k m}S^{l q}S^{n p} + 
{1\over   4 }(2+{\bar \chi} )F_{i j k}F_{l m n}F_{p q r}S^{n}S^{i l}S^{j m}S^{k 
r}S^{p q}- 
{1\over   6 }F_{i j k}F_{l m n p}G_{q}S^{i l}S^{j m}S^{k p}S^{n q} +
\cr & 
{1\over   12 }F_{i j k}F_{l m n}F_{p q r}S^{r}S^{i l}S^{j m}S^{k n}S^{p q} -  
{1\over   4 }F_{i j k}F_{l m n p}G_{q}S^{i j}S^{k l}S^{m q}S^{n p} +
{1\over  8 }F_{i j k}F_{l m n}F_{p q r}G_{s}S^{i p}S^{j s}S^{k n}S^{l m}S^{q r} + 
\cr }
$$}}

\vbox{\ninepoint{
$$\eqalign{ & 
{1\over  4  }F_{i j k}F_{l m n}F_{p q r}G_{s}S^{i l}S^{j m}S^{k r}S^{n s}S^{p q} +  
{1\over  8  }F_{i j k}F_{l m n}F_{p q r}S^{r}S^{i j}S^{k n}S^{l m}S^{p q}+
{1\over   6 }F_{i j k}G_{l}G_{m}G_{n}S^{i m}S^{j n}S^{k l} +
\cr &
{1\over   8 }F_{i j k}F_{l m n}F_{p q r s}S^{i l}S^{j p}S^{k s}S^{m n}S^{q r} + 
{1\over   8}F_{i j k}F_{l m n}F_{p q r s}S^{i l}S^{j m}S^{k s}S^{n p}S^{q r}
+{1\over   2 }F_{i j k}G_{l m}S^{k}S^{i j}S^{l m} -
\cr &
{1\over  8  }F_{i j k}F_{l m n p}S^{k}S^{i j}S^{l m}S^{n p}-  
{1\over  24  }F_{i j k}F_{l m n}F_{p q r}F_{s t u}S^{i p}S^{j s}S^{k n}S^{l q}S^{m 
t}S^{r u} -
 {1\over  8  }F_{i j k}F_{l m n}G_{p q}S^{i j}S^{k q}S^{l m}S^{n p}-
\cr &
F_{i j k}HS^{k}S^{i j}- 
{1\over   16 }F_{i j k}F_{l m n} F_{p q r}F_{s t u}S^{i l}S^{j m}S^{k r}S^{n 
s}S^{p q}S^{t u}
- {1\over   8 }F_{i j k}F_{l m n}F_{p q r}F_{s t u}S^{i l}S^{j p}S^{k u}S^{m 
q}S^{n r}S^{s t} 
 }$$}}
$4780$ graphs contribute to $F_4$, which starts with
$$F_4={S^3\over 248832} (\chi-24)\chi(1728+168 \chi+5 \chi^2)+\ldots$$
This first  term comes from the graphs  
{\baselineskip=12pt \sl
\goodbreak\midinsert
\centerline{\epsfxsize 3 truein\epsfbox{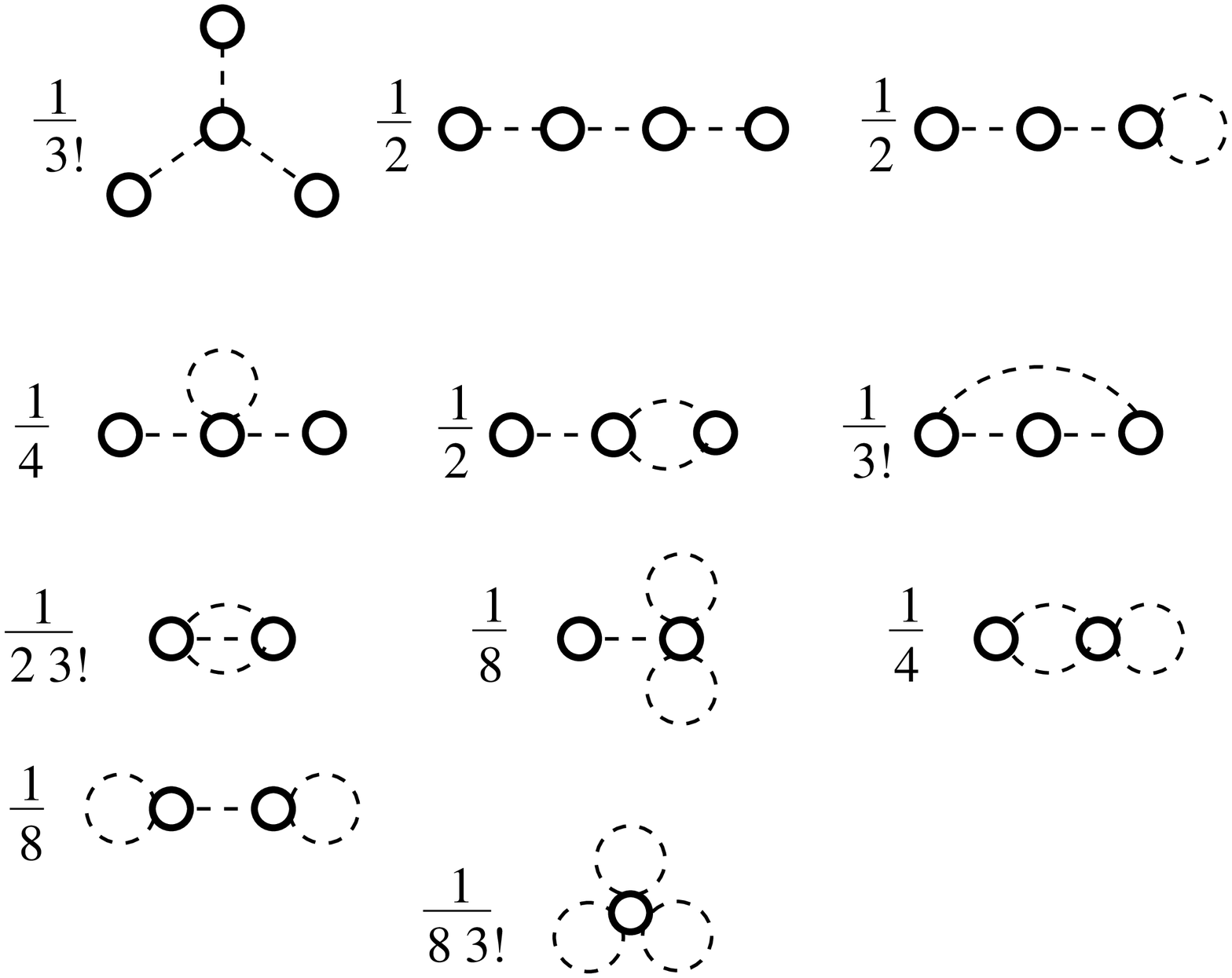}}
\leftskip 1pc\rightskip 1pc \vskip0.3cm
\endinsert}
The contributions of the remaining graphs have been calculated and 
used to evaluate the generating function for the genus 4 curves on 
the quintic see Table 10. $172631$ graphs contribute to $F_5$. 
These data are available on request.

We finally report a ``Ward-identity'' between the correlators at genus 
$3$ on the quintic. 
$$\eqalign{
&(S^{\psi\psi})^4 (134560 S^\Psi F_{\psi\psi\psi}^3- 
7305 (S^{\psi\psi})^2-8139 F_{\psi\psi\psi\psi}^2+3364 
F_{\psi\psi\psi}F_{\psi\psi\psi\psi\psi}-
\cr & 
3372 F_{\psi\psi\psi} F_{\psi\psi\psi\psi}G_\psi+1440 F_{\psi\psi\psi}^2 
G_\psi^2-4 S{\psi\psi} F_{\psi\psi\psi}^2(2877 F_{\psi\psi\psi}G_\psi
-1697 F_{\psi\psi\psi\psi})+\cr & 20184 F_{\psi\psi\psi}^2 G_{\psi\psi})=0.}$$

\listrefs
 
\bye